\documentclass[12pt]{article}
\usepackage{graphicx}
\usepackage{latexsym,amsmath,amsfonts,amssymb,color}
\usepackage[latin1]{inputenc}
\usepackage{cite}
\usepackage[american]{babel}
\usepackage{float}

\usepackage{hyperref}
\newcommand{\barO}{\overline {\Omega}}
\newcommand{\be}{\begin{equation}}
\newcommand{\ee}{\end{equation}}
\newcommand{\beq}{\begin{equation}}
\newcommand{\eeq}{\end{equation}}
\newcommand{\bea}{\begin{eqnarray}}
\newcommand{\eea}{\end{eqnarray}}

\newcommand{\ba}{\begin{eqnarray}}
\newcommand{\ea}{\end{eqnarray}}
\def\bemat{\left( \begin{array}}
\def\enmat{\end{array} \right)}

\def\boldsig{\mbox{\boldmath{$\sigma$}}}
\def\d{\partial}

\begin{document}
\baselineskip=15.5pt
\pagestyle{plain}
\setcounter{page}{1}

\def\ie{{\em i.e.},}
\def\eg{{\em e.g.},}
\newcommand{\rc}{\nonumber\\}
\newcommand{\bear}{\begin{eqnarray}}
\newcommand{\eear}{\end{eqnarray}}
\newcommand{\Tr}{\mbox{Tr}}    
\newcommand{\ack}[1]{{\color{red}{\bf Pfft!! #1}}}

\def\a{\alpha}
\def\b{\beta}
\def\c{\gamma}
\def\d{\delta}
\def\eps{\epsilon}           
\def\f{\phi}               
\def\vf{\varphi}  \def\tvf{\hat{\varphi}}
\def\vp{\varphi}
\def\g{\gamma}
\def\h{\eta}
\def\j{\psi}
\def\k{\kappa}                    
\def\l{\lambda}
\def\m{\mu}
\def\n{\nu}
\def\o{\omega}  \def\w{\omega}
\def\p{\pi}
\def\q{\theta}  \def\th{\theta}                  
\def\r{\rho}                                     
\def\s{\sigma}                                   
\def\t{\tau}
\def\u{\upsilon}
\def\x{\xi}
\def\z{\zeta}
\def\pt{\hat{\varphi}}
\def\tt{\hat{\theta}}
\def\lab{\label}
\def\6{\partial}
\def\wg{\wedge}
\def\atanh{{\rm arctanh}}
\def\bpsi{\bar{\psi}}
\def\bt{\bar{\theta}}
\def\bvf{\bar{\varphi}}
\def\W{\Omega}

\numberwithin{equation}{section}

\renewcommand{\theequation}{{\rm\thesection.\arabic{equation}}}


\begin{flushright}
\end{flushright}

\begin{center}

\centerline{\Large {\bf Holographic Floquet states in low dimensions (I)}}

\vspace{8mm}

\renewcommand\thefootnote{\mbox{$\fnsymbol{footnote}$}}
Ana Garbayo\footnote{ana.garbayo.peon@usc.es},
Javier Mas\footnote{javier.mas@usc.es}
and  Alfonso V. Ramallo\footnote{alfonso@fpaxp1.usc.es}

\vspace{4mm}

{\small \sl Departamento de  F\'\i sica de Part\'\i  culas} \\
{\small \sl Universidade de Santiago de Compostela} \\
{\small \sl and} \\
{\small \sl Instituto Galego de F\'\i sica de Altas Enerx\'\i as (IGFAE)} \\
{\small \sl E-15782 Santiago de Compostela, Spain} 
\vskip 0.2cm

\end{center}

\vspace{8mm}
\numberwithin{equation}{section}
\setcounter{footnote}{0}
\renewcommand\thefootnote{\mbox{\arabic{footnote}}}

\begin{abstract}

We study the response of a (2+1)-dimensional gauge theory to an external rotating electric field.  In the strong coupling regime such system is formulated holographically in a top-down model constructed by intersecting D3- and D5-branes along 2+1 dimensions, in the quenched approximation, in which the D5-brane is a probe in the $AdS_5\times {\mathbb S}^5$ geometry.
The system has a non-equilibrium phase diagram with conductive and insulator phases. The external driving induces a rotating current due to vacuum polarization (in the insulator phase) and to Schwinger effect (in the conductive phase).    
For some  particular values of the driving frequency  the external field resonates with the vector mesons of the model and a rotating current can be produced even in the limit of vanishing driving field. These features are in common with the (3+1) dimensional setup based on the D3-D7 brane model \cite{Hashimoto:2016ize,Kinoshita:2017uch} and hint on some interesting universality.  We  also compute the conductivities paying special attention to the photovoltaic induced Hall effect, which is only present for massive charged carriers. In the vicinity of the Floquet condensate the optical Hall coefficient persists at zero driving field, signalling time reversal symmetry breaking.

\end{abstract}

\newpage
\tableofcontents

\newpage
\section{Introduction}

The physics of periodically driven quantum systems has been the subject of intense study in recent years (see \cite{Bukov_2015,Eckardt:2017kqr, Weinberg_2017,Holthaus_2016} for reviews with tons of citations). Several reasons back up this interest, a very relevant one being technological in origin: the possibility of manipulating  quantum systems in a controlled way by using time-periodic external fields. This approach goes under the name of Floquet engineering \cite{Oka_2019,Rudner_2020}, following from  the Floquet theorem,  a temporal analogue of the Bloch theorem. The artificial setup involves mainly irradiating the system with a circularly polarised laser, or shaking it around. With  appropriate periodic drivings, new phases of quantum materials have been created and  non-equilibrium phenomena have emerged. Examples include light-induced superconductivity \cite{Fausti_2011,Mitrano_2016} and Floquet topological insulators \cite{Oka_2009,Kitagawa_2011,Cayssol_2112,Lindner,Rechtsman_2013,Jotzu_2014,Wang_2013}. Moving on to  (3+1) dimensions, artificial Weyl semimetals have also been reported  \cite{Zhang_2016,Hubener_2016,Bucciantini_2017}.

In this paper we study the behavior of a strongly-coupled $(2+1)$-dimensional gauge theory under the influence of a  external electric field  $\vec {\cal E}$ rotating in the $xy$ plane as
\beq
{\cal E}_x\,+\,i\,{\cal E}_y\,=\,E\,e^{i\Omega t}\,
\,.
\label{rotating_E}
\eeq
We will address this problem in the context of the holographic AdS/CFT correspondence \cite{Maldacena:1997re} (see \cite{AdS_CFT_reviews} for reviews),  and will follow closely the pioneering work in  Refs. \cite{Hashimoto:2016ize,Kinoshita:2017uch} in their study of $(3+1)$-dimensional systems.  In the first of these two references, the authors studied the massless D3-D7 flavour system and computed the artificially induced Hall conductivity.  In the second they extend part of the analysis to massive flavours where an interesting phase space shows up. Our paper streamlines the same analysis in (2+1) dimensions,  extending the analysis to encompass the optical and static conductivities also for massive flavours. The  $(2+1)$-dimensional theory in place is realised  as a codimension-one defect in ${\cal N}=4$ supersymmetric $SU(N_c)$ super Yang-Mills theory in four dimensions. The field content and lagrangian   was explicitly constructed in \cite{DeWolfe:2001pq}. It contains  a matter hypermultiplet living in the defect which couples to the fields of the ambient four-dimensional theory. 
In String Theory, this model can be realized  as the intersection of D3- and D5-branes along $2+1$ dimensions. In this setup the  4d gauge theory on the D3-branes is holographically dual to the $AdS_5\times {\mathbb S}^5$ geometry, while the D5-branes provide $(2+1)$-dimensional flavors, \ie\ fields living in the fundamental representation of the gauge group \cite{Karch:2002sh,Erdmenger:2002ex,Skenderis:2002vf}. We will work in the quenched  probe approximation and will hence neglect the backreaction of the  D5-branes on the $AdS_5\times {\mathbb S}^5$ geometry. 
The fluctuation of the fields living on the brane are dual to the mesonic excitations of the gauge theory. In ref. \cite{Arean:2006pk} the complete analysis of these fluctations was performed for the D3-D5 system and the exact spectrum of mesons was found (see \cite{Kruczenski:2003be} for a similar analysis in the D3-D7 case). 

When an external electric field acts on a medium, a vacuum polarization due to virtual charged particles (quarks) is produced.  As a consequence, an oscillating polarization current is induced. If the external field is weak enough, the medium will remain in an insulator gapped phase in which the induced current is always perpendicular to the applied field. There is no Joule heating and the system is dissipation-less.  When the external field is large enough the vacuum is unstable against the creation of quark-antiquark pairs (Schwinger effect). In the conductive  phase the current becomes dissipative, and the driving induces a  Joule heating.

For an external oscillating electric field as in (\ref{rotating_E}) the critical value $E_c$ at which the insulator-conductor transition takes place depends on the frequency $\Omega$, \ie\  $E_c=E_c(\Omega)$ \cite{Takayoshi:2020}. Remarkably, for some values $\Omega=\Omega_c$ of the frequency, the transition occurs for vanishing external electric field, \ie\ $E_c(\Omega_c)=0$.  At this point of the phase space the rotating current $j$ is not zero, in spite of the fact that the external driving field vanishes. Physically, for these frequencies the driving field enters in resonance with the vector meson excitations of the gauge theory. Actually, this resonating states with zero electric field and non-zero current also occur in the conductive pase in a finite range of frequencies $\Omega_c<\Omega\le \Omega_m$.   Following \cite{Kinoshita:2017uch}, where this behaviour was first found in the D3-D7 model, we will call a configuration with $E=0$ and $j\not=0$ a zero field Floquet condensate of vector mesons, or simply a vector meson  Floquet condensate. Such stable rotating state is a fixed point of time evolution, namely a non-thermal fixed point \cite{Berges:2004ce}.

In our holographic  setup the flavor degrees of freedom are modeled as the excitations of a  probe D5-brane embedded in the $AdS_5\times {\mathbb S}^5$ ten-dimensional background. The probe brane extends along an  $AdS_4\times {\mathbb S}^2$ submanifold of the ten-dimensional background. The holographic duality in the D5-brane worldvolume relates the fields at the boundary of  $AdS_4$ to those of the dual 
$(2+1)$-dimensional gauge theory. In particular, to model the rotating field (\ref{rotating_E}) we have to switch on an electric gauge field on the worldvolume attaining the value  (\ref{rotating_E})  as we approach the $AdS_4$ boundary (the subleading term near the boundary determines the current $j$). The D5-brane with this world-volume gauge field has a non-trivial profile  which can be found by solving the equations of motion of the probe brane. The leading value of the profile function at the boundary determines mass of the quarks, while the subleading term is related to the quark condensate. 

A probe brane with an electric field in his worldvolume can develop an event horizon in the effective (open string) metric on its worldvolume. This is analogous, but not  the same, to the case in which the brane is embedded in a black hole background geometry  with a non-zero Hawking temperature\cite{Frolov:2006tc,Mateos:2006nu,Mateos:2007vn}. Accordingly, we can  borrow the 
terminology of the non-zero temperature case and classify the configurations of the brane depending on whether or not it crosses or not the effective horizon. In the so-called Minkowski embeddings the brane reaches the origin of $AdS_4$ without developing the effective horizon. These Minkowski configurations are dual to the insulating phase of the defect gauge theory. On the contrary, in the conductive phase the induced horizon forms at a finite radial distance from the D3-brane system. The flavour mesons are deconfined and a current can set up in response to an external electric field.  At the interface between these two cases we find the critical embeddings, in which the probe  develops the effective horizon right at its IR endpoint. In this work we study both types of embeddings in order to determine the stable non-equilibrium phases of the theory. In this task we will employ both numerical and analytic techniques.

The rest of this paper is organized as follows. In section \ref{setup} we present our setup, find the equations of motion for the probe D5-brane and study the boundary behavior of the different functions and their relation to the observables of the gauge theory.  In section \ref{types_of_embeddings} we analyze the different types of embeddings and determine the IR boundary conditions that regular solutions must satisfy. We also determine in this section the effective metric and the effective Hawking temperature.  The numerical integration of the equations of motion and the structure of the phase diagram of the theory is presented in section 
\ref{phase_diagram}.  

When either  the mass is zero or the driving  frequency  is large, the equations of motion can be solved analytically in the linearized approximation. The exact analytical solutions are presented in section \ref{sec:analsol}. Section 
\ref{Conductivities} deals with the analysis of the AC and DC conductivities of the model, which are obtained by studying the response of the brane to an additional  probe electric field which is treated in a linearized approximation. We end up with a summary and some concluding remarks in section \ref{conclu}.  The paper is supplemented with some appendices.  In appendix \ref{app_A} we give details about the derivation of the regularity conditions of the brane embeddings at the pseudo-horizon. In appendix \ref{app_B}, we find approximate  analytic solutions of the equations of motion  when the mass  is small. The different embeddings when the equations of motion are linearized are studied in 
appendix \ref{linearized_embeddings}. The  exact conductivities for the massless case  are obtained  in appendix \ref{app_C}.

\section{Setup and ansatz}
\label{setup}

In order to analyze the Floquet states in the strong coupling regime, we will engineer a holographic brane setup in which two sets of branes intersect along $2+1$ dimensions. 
More concretely we will consider the  intersection of two stacks of  D3- and D5-branes according to the following array
\beq
\begin{array}{ccccccccccl}
 &1&2&3& 4& 5&6 &7&8&9 &  \\
D3: & \times &\times &\times &\_ &\_ & \_&\_ &\_ &\_ &      \\
D5: &\times&\times&\_&\times&\times&\times&\_&\_&\_ &
\end{array}
\label{D3D5intersection}
\eeq
In (\ref{D3D5intersection}) the D3- and D5-branes share the spatial directions $1$ and $2$, whereas the direction $3$ is parallel to the D3-brane and orthogonal to the D5-brane.  The field theory dual to the brane setup (\ref{D3D5intersection})  is well-known \cite{Karch:2002sh,DeWolfe:2001pq,Erdmenger:2002ex,Skenderis:2002vf}. It consists of a supersymmetric  theory with $N_f$ matter hypermultiplets (flavors) living on the $(2+1)$-dimensional defect and coupled to the ambient ${\cal N}=4$ theory realized by the stack of $N_c$ color D3-branes.  We will adopt the approximation in which the D5-branes are probes in the geometry generated by the D3-branes, so $N_f\ll N_c$. The latter is the standard holographic dual of four-dimensional ${\cal N}=4$ super Yang-Mills, namely  $AdS_5\times {\mathbb S}^5$  with flux. Let us start by recalling this geometry
\bear
&&ds^2_{10}\,=\,{\rho^2+w_1^2+w_2^2+w_3^2\over R^2}\,\Big(-dt^2+dx^2+dy^2+dz^2\Big)\,+\, \\ 
& \rule{0mm}{8mm} &\qquad\qquad\qquad\qquad
+\,{R^2\over \rho^2+w_1^2+w_2^2+w_3^2}\,\Big(d\rho^2+\rho^2\,d\Omega_2^2+dw_1^2+dw_2^2+dw_3^2\Big)\,\,, \nonumber
\eear
where $d\Omega_2^2=d\theta^2+\sin^2\theta d\phi^2$ is the metric of a unit two-sphere and $R$ is the common radius of the $AdS_5$ and ${\mathbb S}^5$ factors  (in what follows we will take $R=1$).   In order  to embed a D5-brane  wrapping a $AdS_4\times {\mathbb S}^2$ submanifold    we choose the following set of worldvolume coordinates
\beq
\xi^{a}\,=\,(t, x,y, \rho, \theta, \phi)\,\,,
\eeq
and the following ansatz  for the transverse scalars
\beq
z\,=\,{\rm constant}\,\,,
\qquad\qquad
w_1=w(t,\rho)\,\,,
\qquad\qquad
w_2\,=\,w_3\,=\,0\,\,.
\eeq
The induced metric on the D5-brane woldvolume now takes the form
\bear
&&ds_6^2\,=\,-\Big(\rho^2+w^2\Big)\,\Bigg(1\,-\,{\dot w^2\over \big(\rho^2+w^2\big)^2}\Bigg)\,dt^2\,+\,(\rho^2+w^2)(dx^2+dy^2)\,+\,\rc\rc
&&\qquad\qquad\qquad
+{1\over \rho^2+w^2}\,\Big((1+w'^{\,2}) d\rho^2\,+\,2\,\dot w\,w'\,dt d\rho\,+\,\rho^2\,d\Omega_2^2\Big)\,\,,
\label{induced_metric}
\eear
where the dot denotes derivative with respect to $t$ and the prime  with respect to the holographic coordinate $
\rho$.
The main objective to study the response of this system to the driving of an external circularly polarized electric fied
\beq
 {\vec {\cal  E}}(t) = 
\bemat{c} {\cal E}_x(t)\\{\cal  E}_y(t)\enmat\, =  \bemat{cc} \cos \Omega t  & -\sin \Omega t  \\ \sin \Omega t & \cos \Omega t \enmat  \bemat{c} { E}_x\\{  E}_y\enmat \equiv  O(t)\vec E\,\,,
\label{rotating_electric_field}
\eeq
with $\vec E\,=\, {\vec {\cal  E}}(t=0)$. This will  source of a worldvolume gauge field with one-form potential 
\beq
2\pi\,\alpha'\,{\cal A}\,=\,a_x(t,\rho)\,dx\,+\,a_y(t,\rho)\,dy\,\,,\label{vec_pot}
\eeq
whose field strength is
\beq
2\pi\,\alpha'\,{\cal F}\,=\,\dot a_x\,dt\wedge dx\,+\,a_x'\,d\rho\wedge dx\,+\,
\dot a_y\,dt\wedge dy\,+\,a_y'\,d\rho\wedge dy\,\,.
\eeq
The action of the probe brane is given by the DBI action
\beq
S\,= - N_f T_5\,\int d^6\xi\,\sqrt{-\det \big(g_6+2\pi\alpha'\,{\cal F}\big)}\,\,,
\label{DBI_action}
\eeq
where $T_5$ is the tension of the D5-brane. In the following  we will adhere to the notation and reasoning advocated in  \cite{Hashimoto:2016ize,Kinoshita:2017uch}, with slight modifications. To start with, the following switch from  vector to complex notation is useful
\beq
 E = E_x + i E_y ~~~~, ~~~~ a\,=\,a_x+ia_y\,\,.
\eeq

Now,  computing the  determinant $-\det(g_6+2\pi\alpha'\,F)$,  the  DBI action becomes
\bear
&&S \sim -  \int dt\,d\rho\,\,{\rho^2\over \rho^2+w^2}\,\,
\Bigg[(1+w'^{\,2})\big( (\rho^2+w^2)^2\,-\,|\dot a|^2\big)\,-\,
\dot w^2\big(1+|a'|^2\big)\,+\,\rc\rc
&&\qquad\qquad\qquad\qquad
+\,|a'|^2\, (\rho^2+w^2)^2\,+\,2\,\dot w\,w' (\,{\rm Re}(\dot a\,a'^{\,*})\,-\,\big({\rm Im}(\dot a\,a'^{\,*})\big)^2\,\Bigg]^{{1\over 2}}\,\,.
\label{DBI_complex}
\eear
Another convenient switch is to represent the complex gauge field $a$ in the rotating frame
\beq
{\cal E} = E e^{i\Omega t} ~~~~~, ~~~~ a(t,\rho)\,=\,b(t,\rho)\,e^{i(\Omega t+\chi(t,\rho))}\,\,.
\label{complexified_potential}
\eeq
 In the new variables, $(b, \chi)$  the DBI action (\ref{DBI_complex}) becomes
\bear
&&S  \sim -  \int dt\,d\rho\,\,{\rho^2\over \rho^2+w^2}\,\,
\Bigg[(1+w'^{\,2})\big( (\rho^2+w^2)^2\,-\,\dot b^2\,-\,(\Omega+\dot\chi)^2\,b^2\big)\,-\,\rc\rc
&&\qquad\qquad
-\dot w^2\big(1+b'^{\,2}\,+\,b^2\,\chi'^{\,2}\big)\,+\,(\rho^2+w^2)^2\big(b'^{\,2}\,+\,b^2\,\chi'^{\,2}\big)\,+\,\rc
&&\qquad\qquad
+2\,\dot w\,w'\,\big(\dot b\,b'\,+\,(\Omega+\dot\chi)\,\chi'\,b^2\big)\,-\,b^2\Big(\dot b\,\chi'\,-\,(\Omega+\dot\chi)\,b'\Big)\Bigg]^{{1\over 2}}\,\,.
\eear
Since the action does not depend on $t$ explicitly,  taking the functions $b$, $\chi$ and $w$ as independent of the time is a consistent ansatz\footnote{this of course does not preclude the existence of time dependent solutions. For example,  at special points in phase space, instabilities may trigger first order phase transitions that can occur between Black Hole and Minkowski embeddings}
\beq
b\,=\,b(\rho)\,\,,
\qquad\quad
\chi\,=\,\chi(\rho)\,\,,
\qquad\quad
w\,=\,w(\rho)\,\,.
\eeq
By removing the terms with the time derivative from the action, we can rewrite it in a much more simplified fashion as $
S\sim \int d\rho\,{\cal L}\,\,,$
where now
\beq
{\cal L}\,=\,{\rho^2\over \rho^2+w^2}\,
\sqrt{\Big( (\rho^2+w^2)^2\,-\,\Omega^2\,b^2\Big)\,\Big(
1+b'^{\,2}+w'^{\,2}\Big)\,+\,(\rho^2+w^2)^2\,b^2\,\chi'^{\,2}}\,\,.
\label{cal_L_0}
\eeq
We first notice that $\chi$ is a cyclic variable which means that $q$, defined as
\beq
q\,=\,\Omega\,{\partial {\cal L}\over \partial \chi'}\,=\,\Omega\,{\rho^4\over {\cal L}}\,b^2\,\chi'\,\,,
\label{q_def}
\eeq
is independent of $\rho$. The Euler-Lagrange equations derived from ${\cal L}$ are
\bear
&&\rho(\rho^2+w^2)\,\Big[(\rho^2+w^2)^2\,-\,\Omega^2\,b^2\Big]\,b''\,=\,\rc\rc
&&\qquad
=-(1+b'^{\,2}+w'^{\,2})\Big[2(\rho^2+w^2)^3\,b'\,+\,\Omega^2\,b\Big(\rho(\rho^2+w^2)-2w^2\,b\,b'\Big)\Big]\,+\,\rc\rc
&&\qquad\qquad\qquad\qquad\qquad\qquad+
b(\rho^2+w^2)^3(\rho-2 b b')\,\chi'^{\,2}\,\,,\qquad\qquad\rc\rc
&&\rho(\rho^2+w^2)\,\Big[(\rho^2+w^2)^2\,-\,\Omega^2\,b^2\Big]\,w''\,=\,\rc\rc
&&\qquad-2\,(1+b'^{\,2}+w'^{\,2})\Big[(\rho^2+w^2)^3\,w'\,-\,\Omega^2\,b^2\,w\,(\rho+w\,w')\Big]\,-\,2\,b^2\,(\rho^2+w^2)^3\,w'\, \chi'^{\,2}\,\,,
\qquad\qquad\rc\rc
&&b\rho(\rho^2+w^2)\,\Big[(\rho^2+w^2)^2\,-\,\Omega^2\,b^2\Big]\,\chi''\,=\,-2\,\chi'\,\,\Bigg[
(\rho^2+w^2)^3\,\Big(b^3\,\chi'^{\,2}\,+\,\rho\,b'\,+\,b(1+b'^{\,2}+w'^{\,2})\Big)\,-\,\rc\rc
&&\qquad\qquad\qquad\qquad
-\Omega^2\,b^3\Big(2\rho^2+2\rho\,w\,w'\,+\,w^2(1+b'^{\,2}+w'^{\,2})\Big)\Bigg]\,\,.
\label{eoms}
\eear
The last equation in (\ref{eoms}) is redundant, since we can use the first integral (\ref{q_def}) to obtain the value of $\chi'$ in terms of the other functions
\beq
\chi'^{\,2}\,=\,{q^2\,\big[(\rho^2+w^2)^2\,-\,\Omega^2\,b^2\big]\over
(\rho^2+w^2)^2\,b^2\,\big[\Omega^2\,\rho^4\,b^2\,-\,q^2\big]}\,\,(1+b'^{\,2}+w'^{\,2})\,\,.
\label{chi_p}
\eeq
We can next use this last equation to eliminate $\chi'$ in the first two equations in (\ref{eoms}). The resulting equations for $w$ and $b$ depend on $q$ (alternatively one could Legendre transform and obtain the corresponding Routhian). 
\subsection{Boundary conditions}

Let us now study the equations of motion of the gauge field at the UV $\rho\to\infty$. It is convenient to define the  new complex combination 
\beq
c(\rho)\,=\,b(\rho)\,e^{i\chi(\rho)}\,\,,\label{bdec}
\eeq
hence $a(\rho, t) = c(\rho) e^{i\Omega t}$.
In terms of $c(\rho)$ the lagrangian density ${\cal L}$ gives
\bear
{\cal L}\,\sim\,{\rho^2\over \rho^2+w^2}\,\Bigg[
(1+w'^{\,2})\,\Big[ (\rho^2+w^2)^2\,-\,\Omega^2\,|c|^2\Big]\,+\,(\rho^2+w^2)^2\,|c'|^2\,-\,\Omega^2\,
\Big[{\rm Re}\,(c\,c'^{\,*})\Big]^2\Bigg]^{{1\over 2}}\,\,,\qquad
\label{Lagrangian_complex_c}
\eear
and the equation of motion for $c(\rho)$ derived from ${\cal L}$ is
\bear
&&\partial_{\rho}\,\Big[{\rho^4\over (\rho^2+w^2)^2\,{\cal L}}\,\Big( (\rho^2+w^2)^2\,c'\,-\,\Omega^2\,{\rm Re}\,(c\,c'^{\,*})\,c\Big)\Big]\,+\,\\
&&\qquad\qquad\qquad\qquad
+\,  \Omega^2\,{\rho^4\over (\rho^2+w^2)^2\,{\cal L}}\Big((1+w'^{\,2})\,c\,+\,{\rm Re}\,(c\,c'^{\,*})\,c'\Big)\,=\,0\,\,. \nonumber
\eear
Near the UV at $\rho\to\infty$ we can show  that ${\cal L}\approx \rho^2$ and the equation of motion reduces to
$\partial_{\rho}(\rho^2\,c')  = 0$, whose general solution is of the form $c(\rho) = a_1 + a_2/\rho$.
Redefining the constants $a_1$ and $a_2$, we can write asymptotically
\beq
c(\rho)\,=\,b(\rho)\,e^{i\chi(\rho)}\,=\,{i\,E\over \Omega}\,+\,{j\over \rho}\,+\,\cdots\,\,,
\label{asymp_c}
\eeq
where $E$ is the complexified electric field defined in (\ref{complexified_potential}), and $j$ the related  current.  Proceeding similarly, the embedding function $w(\rho)$  behaves near the UV boundary $\rho\to \infty$ as
\beq
w(\rho)\sim m+{{\cal C}\over \rho}\,+\,\cdots\,\,,
\label{asymp_w}
\eeq
where $m$ and ${\cal C}$ are  constants related to the quark mass and quark condensate respectively. We will actually refer to them as such. The relation between $E$, $j$, $m$ and ${\cal C}$ and the electric field ${\cal E}_{YM}(t)$, the electric current ${\cal J}_{YM}(t)$, quark mass $m_q$ and quark condensate $\langle O_m\rangle $ in the boundary theory are
\bear
&&{\cal E}_{YM}(t)\,=\,\sqrt{{\lambda\over 2\pi^2}}\,e^{i\Omega\,t}\,E\,\,,
\qquad\qquad
{\cal J}_{YM}(t)\,=\,{N_f\,N_c\over \pi^2}\,e^{i\Omega\,t}\,j\,\,,\label{dictionaryE}\\
&& m_q\,=\,{\sqrt{\lambda\over 2\pi^2}}\,m\,\,,\qquad\qquad
\langle O_m\rangle\,=\,-{N_f\,N_c\over \pi^2}\,c\,\,,
\label{dictionarym}
\eear
where $\lambda=g^2_{YM}\,N_c$ is the 't Hooft coupling of the ${\cal N}=4$ theory.  The integration constant
$q$  defined in (\ref{q_def}) can be written in terms of the   field $c$ and $c^*$
\beq
q\,=\,i\,{\Omega\,\rho^4\over 2\,{\cal L}}\,\big(c\,c'^{\,*}\,-\,c^*\,c'\big)\,\,,
\label{q_c_c*}
\eeq
and can be easily related  to UV data. As $w(\rho)\sim m+{\cal C}/\rho$ when $\rho\to\infty$, it follows that ${\cal L}\approx \rho^2$ for large $\rho$. Moreover, from (\ref{asymp_c}) $c'\sim -j/\rho^2$ and, therefore
\beq
q\,=\,{\rm Re}(E\,j^*)\,=\,j_x\,E_x\,+\,j_y\,E_y\,\,,
\eeq
which endows  $q$ with the physical interpretation of the Joule heating. If we assume that the driving effectively heats the system we should expect that  $q\ge 0$, and from (\ref{q_def})  this implies $\Omega\,\chi{\,}'\geq 0$. We can see that this result is consistent with the  ingoing boundary condition at the pseudohorizon, which can be argued as follows.  The wavefront of the complexified gauge potential $a(t,\rho)$ in (\ref{complexified_potential}) is the surface $\chi(\rho)+\Omega\,t\,=\,{\rm constant}$ (we are in the case in which $\chi$ does not depend on $t$). By performing a generic variation of this condition, 
it follows that
\beq
\delta\,\rho\,=\,-{\Omega\over \chi{\,'}}\,\delta\,t\,\,.
\label{variation_wavefront}
\eeq
Requiring that our gauge potential represents an infalling wave at the pseudo-horizon implies 
 $\delta\rho<0$ for $\delta t>0$ which, following (\ref{variation_wavefront}), only occurs when $\Omega\,\chi{\,'}\ge 0$ and, therefore, 
$q\ge 0$.

The equations of motion (\ref{eoms}) and the UV boundary conditions (\ref{asymp_c}) and  (\ref{asymp_w}) have the following scaling symmetry
\bear
&&
t\to t/\alpha\,\,,\qquad
\rho\to\alpha\,\rho\,\,,\qquad
w\to \alpha\,w\,\,,\qquad
b\to\alpha b\,\,,\qquad
\chi\to\chi\,\,,\rc\rc
&&\Omega\to\alpha\,\Omega\,\,,\qquad
E\to\alpha^2\,E\,\,,\qquad
j\to\alpha^2\,j\,, \label{scaling} \\ \rc
&&m\to\alpha \,m\,\,,\qquad
{\cal C}\to\alpha^2\,{\cal C}\,\,,\qquad
q\to\alpha^4\,q\,\,.\nonumber
\eear
The Lagrangian and the action transform homogeneously  ${\cal L} \to \alpha^2 {\cal L}$ and $S\to \alpha^2 S$.
By choosing $\alpha=1/m$ in  (\ref{scaling}) we can make $m=1$ and deal with the remaining quantities in units of (the  appropriate powers of) $m$.

\section{Types of embeddings}
\label{types_of_embeddings}

\begin{figure}[!ht]
\center
 \includegraphics[width=0.50\textwidth]{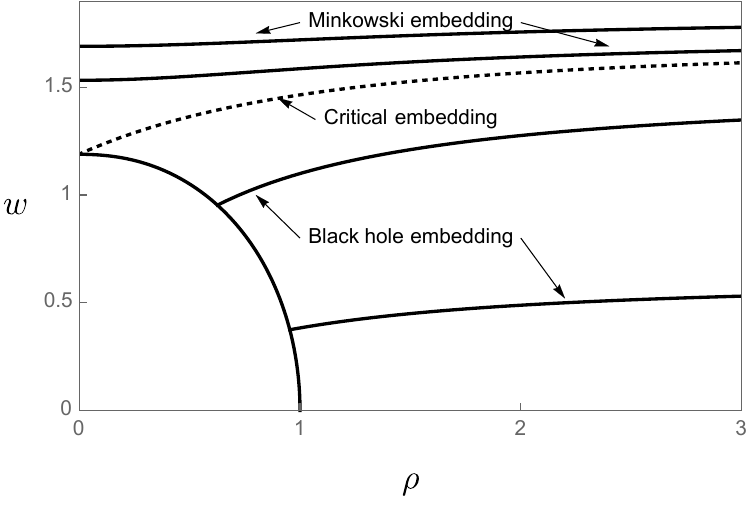}
  \caption{Profiles of the different types of embeddings for $E=\Omega=1$. }
\label{embeddings}
\end{figure}

The equations of motion written in (\ref{eoms}) are potentially singular when
\beq
b_0\,=\,{w_0^2+\rho_c^2\over \Omega}\,\,,
\label{b0_w0_rho_c}
\eeq
where $b_0=b(\rho=\rho_c)$ and $w_0=w(\rho=\rho_c)$. 
The point $\rho=\rho_c$  where the condition (\ref{b0_w0_rho_c}) holds will be referred to as the pseudohorizon. Indeed, we will show below that $\rho=\rho_c$ is the  event horizon of the induced open string metric on the D5-brane. The IR  behaviour  of the bulk fields when approaching this pseudohorizon determines three types of embeddings (see figure \ref{embeddings}).  We can have black hole embeddings (which cross the horizon), Minkowski embeddings (which do not intersect the pseudohorizon and reach the point $\rho=0$) and critical embeddings (for which  the brane reaches the $\rho_c$ at  $\rho_c=0$).  In order to avoid the singularity at $\rho=\rho_c$  the first derivatives of  the functions $w(\rho)$, $b(\rho)$ and $\chi(\rho)$ have to be adjusted properly
for black hole and critical embeddings. 

\subsection{Black hole embeddings}

Expanding the functions $w(\rho)$, $b(\rho)$ and $\chi(\rho)$  around $\rho=\rho_c$ as
\bear
&&w(\rho)\,=\,w_0\,+\,w_1\,(\rho-\rho_c)\,+\,\cdots\,\,, \nonumber \\
\rule{0mm}{5mm}
&&b(\rho)\,=\,b_0\,+\,b_1\,(\rho-\rho_c)\,+\,\cdots\,\,,\\
\rule{0mm}{5mm}
&&\chi(\rho)\,=\,\chi_0\,+\,\chi_1\,(\rho-\rho_c)\,+\,\cdots\,\,,\nonumber
\eear
we obtain coefficients $w_1$, $b_1$ and $\chi_1$ whose explicit expressions are given in appendix \ref{app_A}. Notice that $\chi_1$ appears in the expression of the on-shell lagrangian density ${\cal L}_0$  evaluated at the pseudohorizon $\rho=\rho_c$
\beq
{\cal L}_0(\rho_c)\,=\,\rho_c^2\,b_0\,\chi_1\,\,.
\label{cal_L_0_chi_1}
\eeq
We can use this expression to obtain the Joule heating $q$ in (\ref{q_def}). Indeed as $q$ is constant, we can evaluate it at the pseudohorizon, with the result
\beq
q\,=\,\Omega\,\rho_c^2\,b_0\,\,.
\label{qJH}
\eeq
which only depends on $b_0$. Using (\ref{b0_w0_rho_c}),  we get
\beq
q\,=\,\rho_c^2\big(w_0^2+\rho_c^2\big)\,\,.
\label{q_w0_rho_c}
\eeq

\subsection{Minkowski embeddings}
In this case there is no $\rho_c$  that solves (\ref{b0_w0_rho_c}) and the brane reaches the origin $\rho=0$. We can therefore evaluate the Joule heating constant $q$ by taking $\rho=0$ in (\ref{chi_p}) and, since the right-hand side of this equation contains a $\rho^4$ factor we get $q=0$. Thus, for these  embeddings there is no net energy flux on the boundary theory. Moreover, from (\ref{chi_p}) one readily concludes that  the phase function $\chi(\rho)$ is in fact a constant. Expanding $b(\rho)$ and 
$w(\rho)$ around $\rho=0$
\bear
&&b(\rho)\,=\,b_0\,-\,{b_0\,\Omega^2\over 6\big(w_0^4-\Omega^2\,b_0^2\big)}\,\rho^2\,+\,\cdots\,\,,\rc\rc
&&w(\rho)\,=\,w_0\,+\,{b_0^2\,\Omega^2\over 3\,w_0\big(w_0^4\,-\,\Omega^2\,b_0^2\big)}\,\rho^2\,+\,\cdots\,\,.
\eear
\begin{figure}[!ht]
\center
 \includegraphics[width=0.47\textwidth]{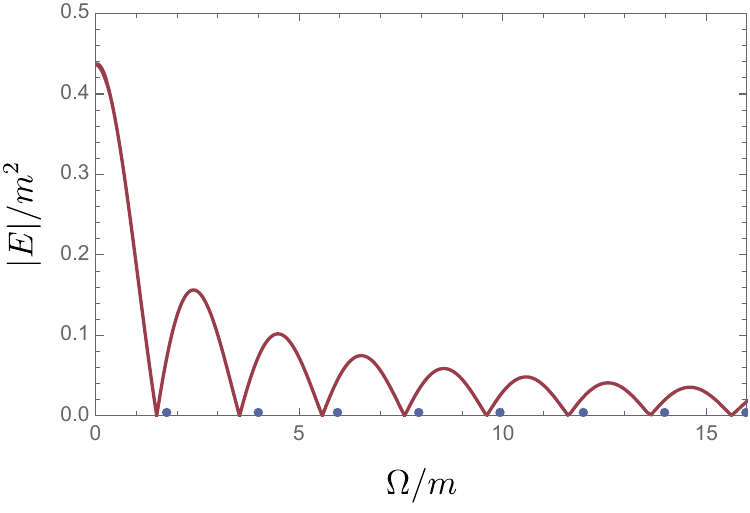} \includegraphics[width=0.5\textwidth]{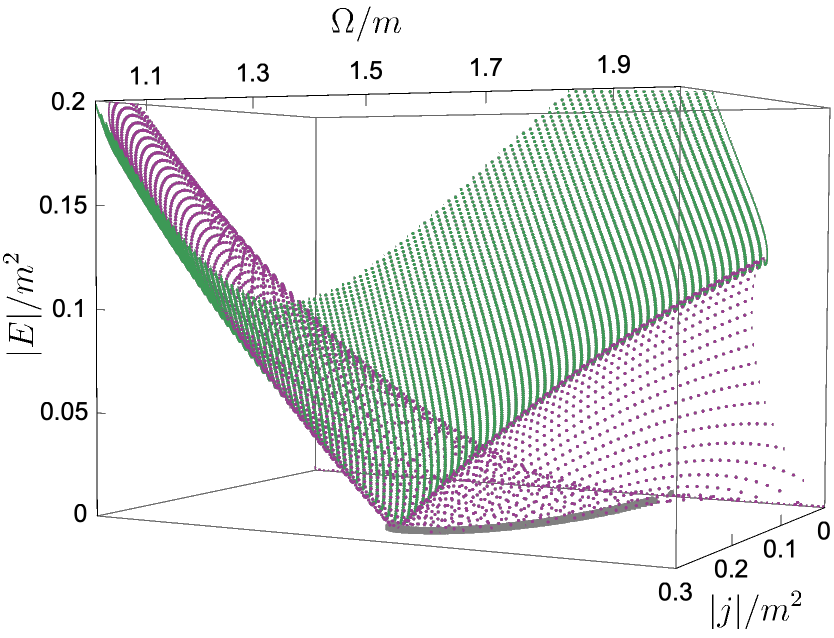}
  \caption{Electric field of the critical embeddings versus driving frequency. The frequencies for which $E$ vanish are the critical frequencies $\Omega_c$ of the vector meson Floquet condensates. The blue dots signal the frequencies (\ref{Omega_meson}) of the vector meson spectrum  for different values of the principal quantum number $n$. On the right hand side, the three dimensional version of the left plot, focussing close to the first critical point, which stretches into a segment $\Omega/m \in ( 1.496, 1.732)$. 
   }
\label{E_Omega}
\end{figure}

\subsection{Critical embeddings}
\label{Critical_bc}
When the  effective  horizon reaches the origin, $\rho_c=0$, the embedding is critical. The value of the 
$b$ field at the horizon is $b_0=w_0^2/\Omega$, $\chi(\rho)$ is also here constant and we can expand 
$b(\rho)$ and $w(\rho)$ around $\rho=0$, with the result
\bear
&&b(\rho)\,=\,{w_0^2\over \Omega}\,-\,{\Omega\over 2\sqrt{\Omega^2+4\,w_0^2}}\,\rho\,+\,\cdots\rc\rc
&&w(\rho)\,=\,w_0\,+\,{w_0\over \sqrt{\Omega^2+4w_0^2}}\,\rho\,+\,\cdots\,.
\eear

\begin{figure}[!ht]
\center
 \includegraphics[width=0.5\textwidth]{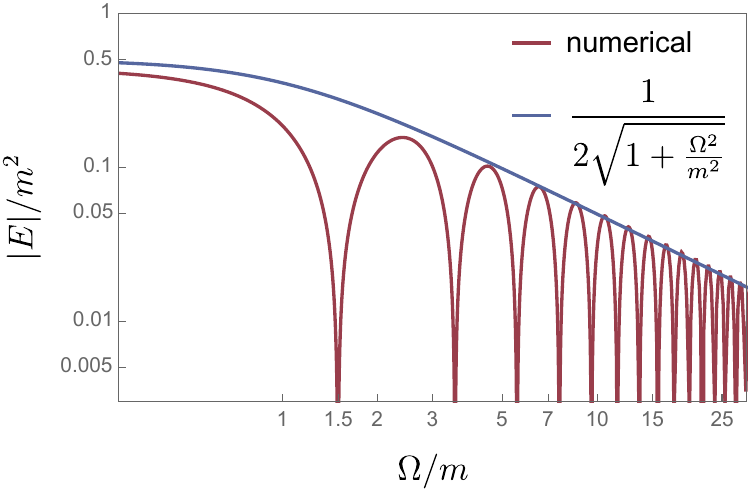}
  \caption{Logarithmic version of the plot in Fig.\,\ref{E_Omega}.    The high $\Omega/m$ fitting is derived in Sec. \ref{sec:critembed} (see eq. (\ref{ECMinkSol})). }
\label{E_Omega_log}
\end{figure}

\subsection{Effective metric}
The effective open string metric $\gamma_{ab}$ is defined as \cite{Seiberg:1999vs}
\beq
\gamma_{ab}\,=\,h_{ab}\,+\,(2\pi \alpha')\,F_{ac}\,F_{bd}\,h^{cd}\,\,,
\eeq
where we now denote by $h_{ab}$ the induced metric (\ref{induced_metric}). Explicitly,   for our ansatz, we have
\bear
&&\gamma_{ab}\,d\xi^a\,d\xi^b=-F(\rho)\,dt^2+{1+w'^{\, 2}+|c'|^2\over \rho^2+w^2}\,d\rho^2-
{2\Omega\,{\rm Im}(c\,c'^{\,*})\over  \rho^2+w^2}\,dt\,d\rho
+\,F(\rho)\,e_+\,e_-\,+\,
\qquad\qquad\qquad
\rc\rc
&&
\qquad\qquad
+{\rho^2+w^2\over 4(1+w'^{\, 2})}\,\big(e_-\,c'+e_+\,c'^{\,*})^2\,+\
{\Omega^2\over 4(\rho^2+w^2)}\big(e_-\,c+e_+\,c^{*})^2+{\rho^2\over \rho^2+w^2}d\Omega_2^2\,\,,
\eear
with $F(\rho)$ given by the following function
\beq
F(\rho)\equiv {(\rho^2+w^2)^2-\Omega^2\,|c|^2\over \rho^2+w^2}\,\,,
\eeq
and $e_{\pm}$ are  complex 1-forms
\beq
e_{\pm}\,=\,e^{\mp\,i\Omega t}\big(dx\pm i dy\big)\,\,.
\eeq
The $(t,\rho)$ part of the metric gets diagonalized  by means of  the following change of coordinates
\beq
d\tau\,=\,dt\,-\,A(\rho)\,d\rho\,\,,
\qquad\qquad
d\rho_*\,=\,B(\rho)\,d\rho\,\,,
\label{tortoise_coord}
\eeq
where $A(\rho)$ and $B(\rho)$ are the following functions:
\beq
A(\rho)\,=\,{\Omega\,{\rm Im}\big(c\,c'^{\,*}\big)\over 
(\rho^2+w^2)^2-\Omega^2\,|c|^2}\,\,,
\qquad\qquad
B(\rho)\,=\,{(\rho^2+w^2)\,{\cal L}_0\over \rho^2\big[(\rho^2+w^2)^2-\Omega^2\,|c|^2\big]}\,\, .
\label{A_B_tortoise}
\eeq
 The new coordinate $\rho_*$ is just the Eddington-Finkelstein tortoise coordinate.  
In these coordinates, the transformations  in (\ref{scaling}) act as a rescaling  $ (\tau, \rho_*)\to \lambda^{-1} (\tau, \rho_*)$, and  the effective metric takes the form
\bear
&&\gamma_{ab}\,d\xi^a\,d\xi^b=F(\rho)\,(\,-d\tau^2+d\rho_*^2\,)\,+\,
F(\rho)\,e_+\,e_-\,+\,{\rho^2+w^2\over 4(1+w'^{\, 2})}\,\big(e_-\,c'+e_+\,c'^{\,*})^2\,+\rc\rc
&&
\qquad\qquad\qquad\qquad
+{\Omega^2\over 4(\rho^2+w^2)}\big(e_-\,c+e_+\,c^{*})^2+{\rho^2\over \rho^2+w^2}d\Omega_2^2\,\,.
\label{open_string_metric}
\eear
The function $F(\rho)$ vanishes at $\rho=\rho_c$ and, therefore, this function acts as a blackening factor for the effective metric (\ref{open_string_metric}), with $\rho=\rho_c$ playing the role on an event horizon. The effective Hawking temperature $T_H$ is given by
\beq
T_H\,=\,{\kappa\over 2\pi}\,=\,-{\gamma_{tt}'\over 4\pi\,\gamma_{t\rho}}\Bigg|_{\rho=\rho_c}\,\,.
\eeq
Using the values of $\gamma_{tt}$ and $\gamma_{t\rho}$ found above, we get the $T_H$ in terms of near horizon data
\beq
T_H\,=\,{2(\rho_c+w\,w')-\Omega\,b'\over 2\pi\,\chi'\,b}\Bigg|_{\rho=\rho_c}\,=\,
{2(\rho_c+w_0 w_1)\,-\,\Omega\,b_1\over 2\pi\chi_1\,b_0}\,\,.
\label{TH_final}
\eeq

\begin{figure}[!ht]
\center
 \includegraphics[width=0.47\textwidth]{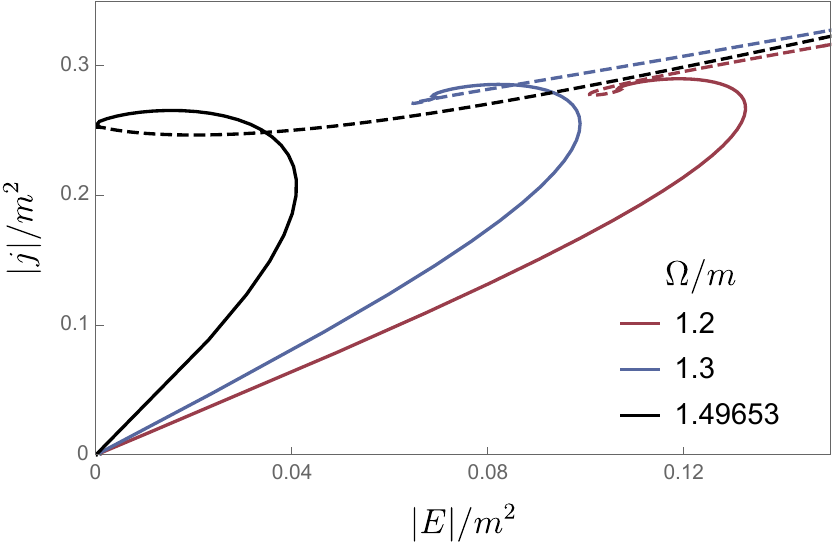}
 \qquad
 \includegraphics[width=0.47\textwidth]{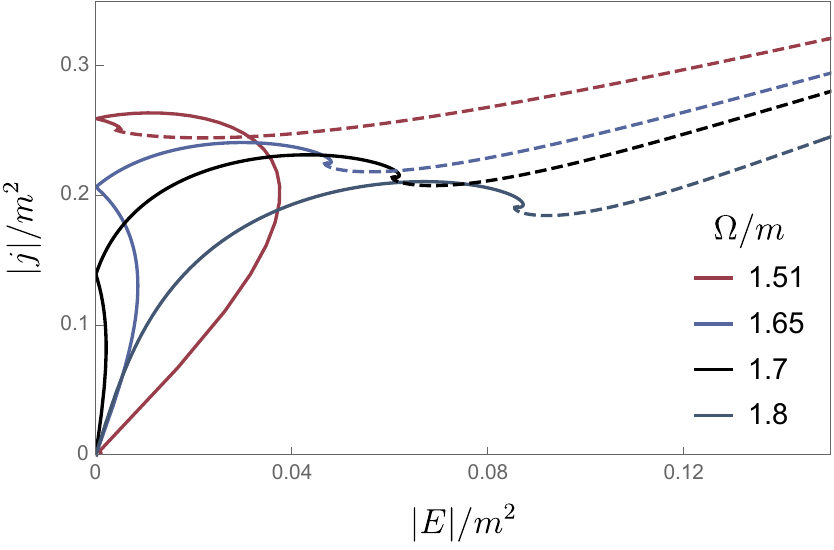}
  \caption{Electric current versus electric field for $\Omega\le \Omega_c$ (left) and 
  $\Omega>\Omega_c$ (right) around the first vector meson resonance. The solid curves correspond to the insulator (Minkowski) phase, whereas the dotted curves represent the conductive (black hole) phase. On the right curve we show that $E$ can only vanish for $j\not=0$ for Minkowski embeddings with $\Omega_c= 1.4965\leq \Omega\leq 1.732 = \Omega_{meson}$. }
\label{J_vs_E}
\end{figure}

\section{Phase diagram}
\label{phase_diagram}

In the previous section we have addressed the different types of embedding according to the IR boundary conditions. On the other hand, the phase diagram is constructed out of the UV boundary data that contain information about the sources and the responses. 
The analysis of the generic case proceeds by  numerically integrating  the system of equations  (\ref{eoms}). In this section we will obtain the precise boundary of the phase space of solutions presented in Figs. \ref{E_Omega} and \ref{E_Omega_log}.   Although the qualitative results are the same as the ones found in \cite{Kinoshita:2017uch}, we need the precise values of these boundaries in order to locate interesting places where to look for conductivity properties in the next section.  
As usual, integration proceeds from the IR to the UV. Indeed, enforcing regularity conditions found in section \ref{types_of_embeddings} as initial conditions at the pseudohorizon $\rho=\rho_c$ and integrating out up to the boundary yields a unique solution. For a given frequency $\Omega$ we can specify  $b_0=b(\rho=\rho_c)$ and $w_0=w(\rho=\rho_c)$  and use the values of $w_1$, $b_1$ and $\chi_1$ computed in appendix \ref{app_A}(for Minkowski embeddings  $\rho_c=0$ and $\chi(\rho)$ is constant).  From the UV behavior of the solutions $b(\rho)$, $\chi(\rho)$ and $w(\rho)$ when $\rho\to\infty$  one can  extract 
$E$, $j$, $m$ and ${\cal C}$.  Due to the scaling symmetry (\ref{scaling}) these quantities are not independent. It is natural to measure all quantities in terms of the quark mass $m$. Accordingly, we will present our results for $\Omega/m$,  $|E|/m^2$, $j/m^2$ and ${\cal C}/m^2$.

\begin{figure}[!ht]
\center
 \includegraphics[width=0.47\textwidth]{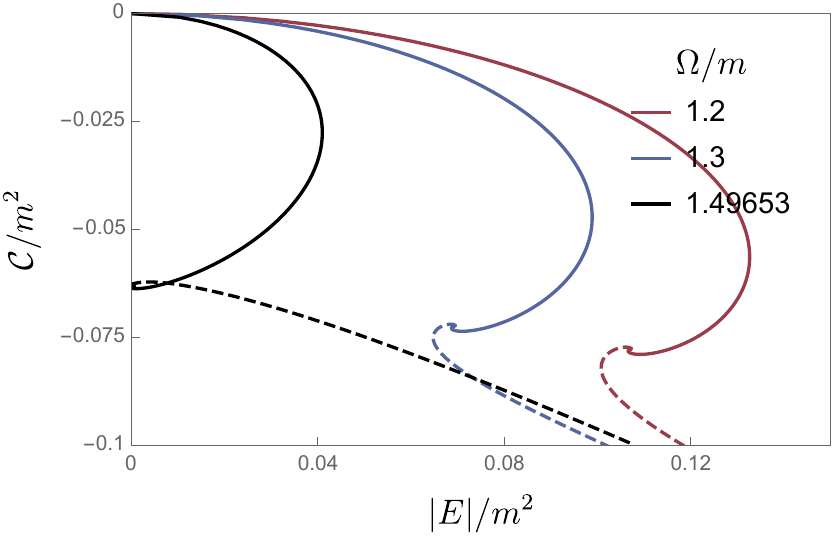}
 \qquad
 \includegraphics[width=0.47\textwidth]{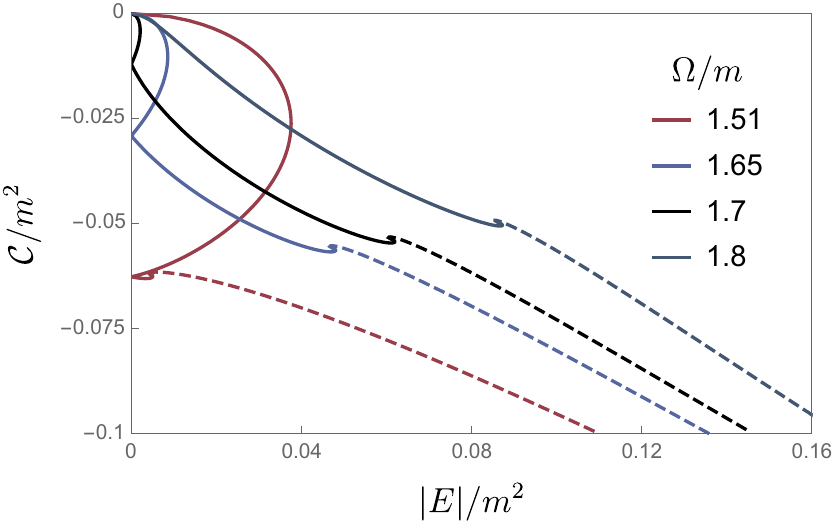}
  \caption{Quark condensate as a function of the electric field for several $\Omega$ around the first resonance. The curves on the left (right) are for $\Omega\le \Omega_c$ ($\Omega>\Omega_c$). Solid (dashed) curves correspond to Minkowski (black hole) embeddings. }
\label{c_vs_E}
\end{figure}

\begin{figure}[!ht]
\center
 \includegraphics[width=0.47\textwidth]{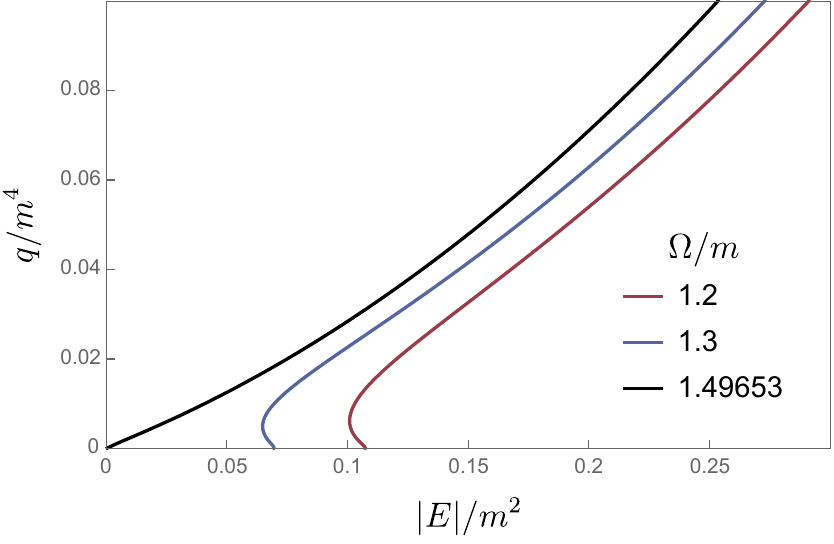}
 \qquad
 \includegraphics[width=0.47\textwidth]{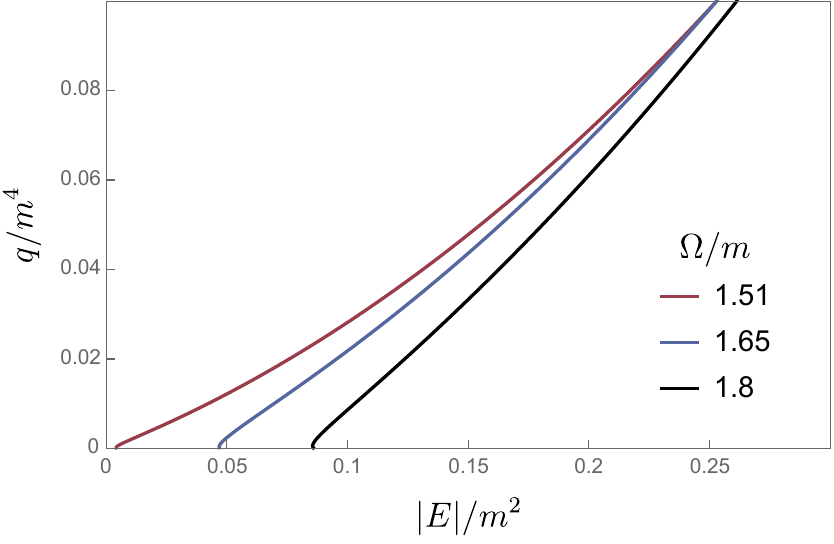}
  \caption{The Joule heating $q$ as a function of the electric field $E$ for black hole embeddings and several frequencies around the first resonance. On the left (right) plots $\Omega\le \Omega_c$ ($\Omega>\Omega_c$).  }
\label{fig:q_vs_E}
\end{figure}
\begin{figure}[H]
\center
  \includegraphics[width=0.55\textwidth]{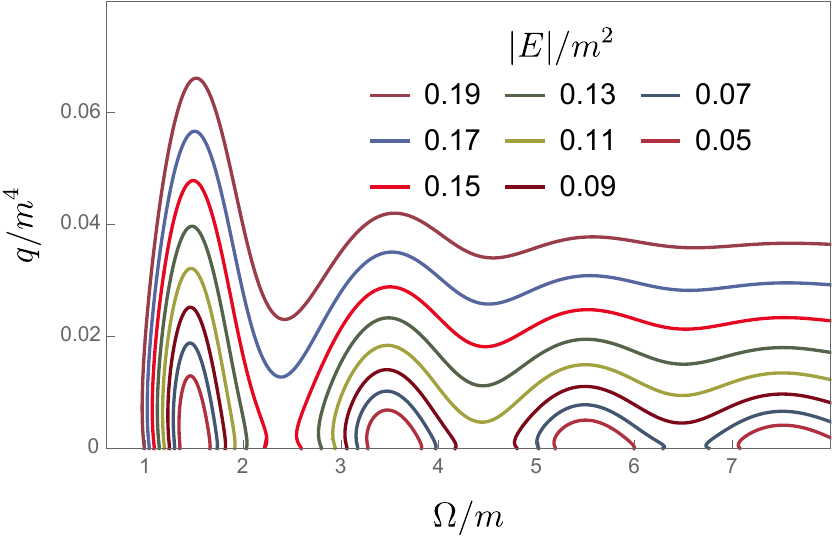}
  \caption{The Joule heating $q$ as a function of the frequency $\Omega$ for black hole embeddings and several values of the electric field.  }
\label{fig:q_vs_Omega}
\end{figure}

In figure \ref{E_Omega} we show the phase diagram in the parameter space 
$(\Omega/m, |E|/m^2)$. The solid curve is the locus of critical embeddings. We can  think of it roughly as a  boundary between the conductive phase (above) made of  black hole embeddings,  and insulator phase (below) made of Minkowski embeddings\footnote{Although it is pretty clear from figure \ref{embeddings} that the critical embeddings lie in between  Minkowski and Black Hole embeddings, the spiralling multivaluedness in the vicinity of the solid line (see figure (\ref{J_vs_E}) below) implies that one can  find Black Hole embeddings closely below it, and also Minkowski embedding above.}. In figure \ref{J_vs_E} we plot the electric current versus the electric field for driving frequencies around the firs resonance. In figure \ref{c_vs_E} we represent the quark condensate for Minkowski and black hole embeddings and different driving frequencies.  We notice in Figures  \ref{E_Omega} and \ref{E_Omega_log}
 that $|E|/m^2$ has a series of maxima whose values decrease as $\Omega/m$ increases. Moreover, for some discrete values $\Omega_c/m$ of the frequency the electric field vanishes and, as shown in figure \ref{J_vs_E}, the corresponding rotating current is non-zero. 

 Thus, for these frequencies we have a vector meson Floquet condensate of the type mentioned in the introduction. Actually, these $E=0$ Floquet states exist  also for Minkowski embeddings in a finite range of frequency  $\Omega_c< \Omega \le\Omega_{meson}$, where $\Omega_{meson}$ is the mass of a  vector meson in the defect theory. This mass spectra was computed in \cite{Arean:2006pk} and is given by a tower of values depending on a principal quantum number $n$

\beq
\Omega_{meson}/m\,=\,2\sqrt{
\Big(n+{1\over 2}\Big)\Big(n+{3\over 2}\Big)}\,\,,
\qquad\qquad
n\,=\,0,1,2,\cdots\,\,.
\label{Omega_meson}
\eeq
In section \ref{Large_Omega}  and appendix \ref{linearized_embeddings} we obtain these frequencies by means of a linear analysis of the equations of motion. The first four values of the critical resonant frequencies and of the meson masses are
\bear
&&\Omega_c/m\,=\,1.4965\,,\,3.5308\,,\,5.5676\,,\,7.5851\,\cdots\,  \label{floqcondfreqs}\\ \rule{0mm}{6mm}
&&\Omega_{meson}/m\,=\,1.7320\,,\,3.8730\,,\,5.9161\,,\,7.9372\,,\cdots\,\,.  \label{mesonfreqs}
\label{Omega_c_Omega_meson}
\eear

In figure \ref{fig:q_vs_E} we plot the Joule heating $q$ as a function of the electric field, whereas in figure \ref{fig:q_vs_Omega} the Joule heating for black hole embeddings is plotted versus the driving frequency for several values of the electric field.

To get further insight on the structure of the phase diagram of the model, it is interesting to analyze in detail the relation between the current $j$ and the electric field $E$. Since $j$ and $E$ are vector quantities with two components in the $xy$ plane, we expect to have  a relation of the type:
\beq
\begin{pmatrix}
j_x\cr j_y
\end{pmatrix}\,=\,
\begin{pmatrix}
\gamma_{xx}&&\gamma_{xy}\cr
-\gamma_{xy}&&\gamma_{xx}
\end{pmatrix}\,
\begin{pmatrix}
E_x\cr E_y
\end{pmatrix}\,\,,
\label{gamma_def}
\eeq
where the form of the matrix is dictated by the rotational symmetry in the $xy$ plane.  The matrix elements $\gamma_{xx}$ and $\gamma_{xy}$ are susceptibilities, defined in analogy with the longitudinal and transverse (Hall) conductivities. 
From the definitions (\ref{gamma_def}) it is immediate to find $\gamma_{xx}$ and $\gamma_{xy}$ in terms of the components of $E$ and $j$:
\beq
\gamma_{xx}\,=\,{E_x\,j_x+E_y\,j_y\over E_x^2+E_y^2}\,=\,{|j|\over |E|}\,\cos\delta\,\,,
\qquad\qquad
\gamma_{xy}\,=\,{E_y\,j_x-E_x\,j_y\over E_x^2+E_y^2}\,=\,{|j|\over |E|}\,\sin\delta\,\,,
\eeq
where $\delta$ is the angle formed by the vectors $(j_x, j_y)$ and $(E_x, E_y)$ in the $xy$ plane. The quantity $\gamma_{xx}$ is related to the Joule heating $q$. Indeed, one has $\gamma_{xx}=q/|E|^2$. Given this, and the fact that we have profusely  illustrated the behaviour of $q$ in Figs. \ref{fig:q_vs_E} and \ref{fig:q_vs_Omega}, we shall  focus on the study of $\gamma_{xy}$.

\begin{figure}[!ht]
\center
 \includegraphics[width=0.47\textwidth]{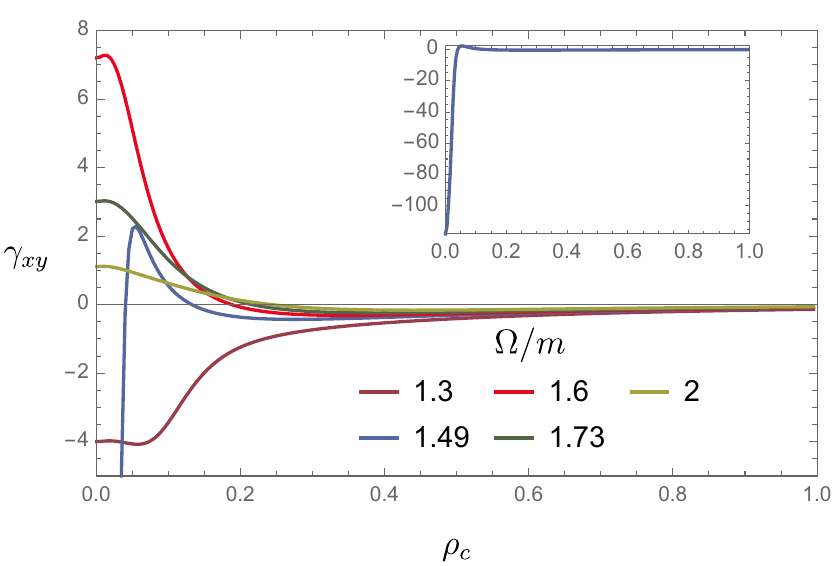}
 \qquad
  \includegraphics[width=0.47\textwidth]{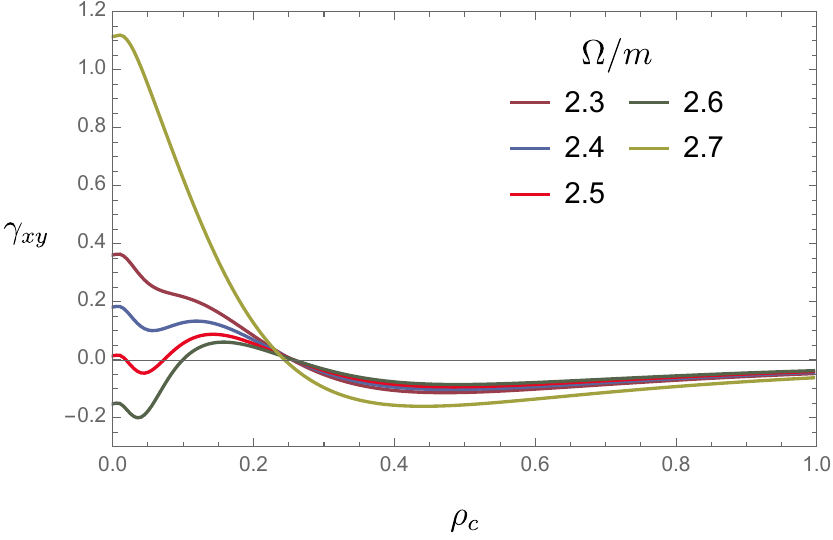}
   \caption{The susceptibility coefficient $\gamma_{xy}$ as a function of $\rho_c$ for black hole embeddings and different driving frequencies. On the left we plot the values for frequencies close to the first critical resonant frequency $\Omega_c/m=1.49$. On the right we represent $\gamma_{xy}$ for driving frequencies around $\Omega/m=2.5$, which is frequency  of the first zero of $j$ for critical embeddings.
}
    \label{gamma_xy}
\end{figure}

For massless black hole embeddings $\gamma_{xy}$ vanishes, since in this case $j$ and $E$ are colinear, as we demonstrate analytically in section \ref{massless_embd} below. On the other hand $\gamma_{xy}\not=0$ for Minkowski embeddings since there is no Joule heating in this case and the vectors $j$ and $E$ are orthogonal. This behavior can be verified by studying $\gamma_{xy}$ for black hole embeddings as a function of $\rho_c$ (see figure \ref{gamma_xy} ). When $\rho_c$ is large enough $\gamma_{xy}\to 0$, whereas it reaches a non-vanishing value as $\rho_c\to0$. Interestingly, the sign of $\gamma_{xy}(\rho_c\to 0)$ changes at a discrete set of values of the driving frequency. We have identified these values as the critical resonant frequencies $\Omega_c$ of (\ref{Omega_c_Omega_meson}), as well as those for which the current $j$ vanishes for critical embeddings. These last frequencies with $j=0$ are very close to the maxima of the lobes in the $|E|/m^2$ versus $\Omega/m$ plot of figure \ref{E_Omega}. The first numerical values of the frequencies which make $E$ maximal and $j$ zero for critical embeddings are:
\bear
&& {\Omega\over m}(j=0)\,=\,2.5080\,,\,4.5489\,,\,6.5769\,,\,8.5960\,,\cdots\,\,,\rc\rc
&& {\Omega\over m}(E\,{\rm maximal})\,=\,2.4008\,,\,4.4772\,,\,6.5270\,,\,8.5638\,,\,\cdots
\eear
\begin{figure}[!ht]
\center
 \includegraphics[width=0.45\textwidth]{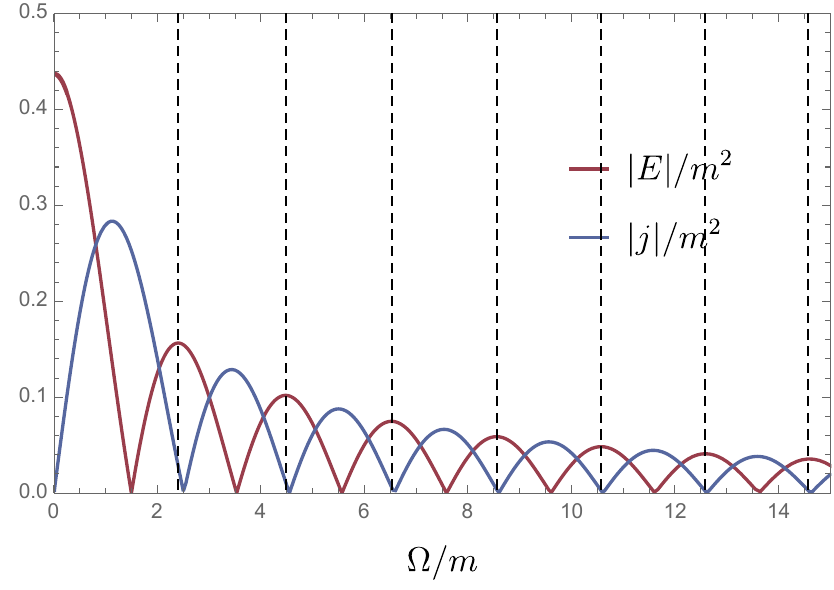}
 \qquad
  \includegraphics[width=0.45\textwidth]{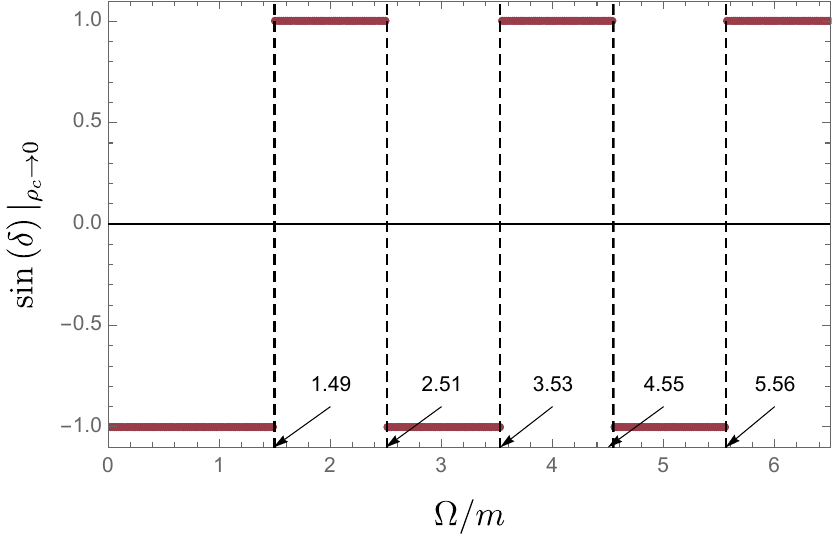}
   \caption{On the left we plot  $|E|/m^2$ and $j/m^2$ versus the driving frequency for critical embeddings. On the right we depict the value of $\sin\delta$  for embeddings with $\rho_c=0$.  }
    \label{E_j_lobes_sin_delta}
\end{figure}
In figure \ref{E_j_lobes_sin_delta} we plot together the $|E|/m^2$ and $j/m^2$ curves as a function of $\Omega/m$. We notice that the current curve also displays a lobe-shaped structure, displaced with respect to the $|E|/m^2$ curve, whose maxima (zeros) are very close to the zeros (maxima) of the $|E|/m^2$ curve. We have also checked that $\sin \delta(\rho_c\to 0)=\pm 1$ and changes discontinuously its sign at the discrete set of frequencies described above (see  figure \ref{E_j_lobes_sin_delta}).

\section{Analytic Solutions}\label{sec:analsol}

There are two limits where the equations are amenable to analytic solution. These are the low mass limit $m\to 0$, and the high frequency limit $\Omega \to \infty$. In the first case all embeddings are of black hole type. In the second, the three types of embeddings can occur. 

\subsection{Massless embeddings }
\label{massless_embd}

Let us consider the particular case of massless black hole embeddings, which correspond to putting the embedding function $w(\rho)=0$. By inspection  one easily concludes that $w=0$ is a consistent truncation of the equations of motion in (\ref{eoms}). In this massless case,  the lagrangian density (\ref{cal_L_0})  takes the form
\beq
{\cal L}_0\,=\,\sqrt{
(\rho^4-\Omega^2\,b^2)(1+b'^{\,2})\,+\,
\rho^4\,b^2\,(\chi')^2}\,\,.
\eeq
As in the general case, $\chi(\rho)$ is a cyclic variable which, according to (\ref{chi_p}), is now given by
\beq
\chi'\,=\,{q\over \rho^2 b}\,
{\sqrt{
(\rho^4-\Omega^2\,b^2)(1+b'^{\,2})}\over 
\sqrt{\Omega^2\,\rho^4\,b^2\,-\,q^2}}\,\,.
\label{chi_p_massless_general}
\eeq
Let us now eliminate $\chi(\rho)$ by introducing the Routhian ${\cal R}_0$, defined as
\beq
{\cal R}_0\,=\,\chi'\,{\partial {\cal L}_0\over \partial\,\chi'}\,-\,{\cal L}_0\,\,.
\eeq
After some calculation, ${\cal R}_0$ as a function of  $b(\rho)$ and $q$ becomes
\beq
{\cal R}_0\,=\,-\sqrt{
\Big(1\,-\,{q^2\over \Omega^2\,\rho^4\,b^2}\Big)\,(\rho^4-\Omega^2\,b^2)}\,\,\,
\sqrt{1+b'^{\,2}}\,\, .
\eeq
As usual in this flavour brane setup, demanding reality of the Routhian fixes  transport coefficients in terms of horizon data \cite{Karch:2007}. Indeed, at $\rho_c$ the two factors under the square root must change sign, and therefore vanish,   simultaneously. 
This enforces $b_0= b(\rho_c)$ and $q$ to 
satisfy both equations  (\ref{b0_w0_rho_c}) with $w_0=0$,  and (\ref{qJH}). Combined together, these equation entail, in the massless case 
\be
q = \Omega^2 b_0^2\, .\label{qdeb0}
\ee

Let us look for solutions to the  Euler Lagrange equations  in which $b(\rho)$ is constant. Writing these equations in the following form 
\beq
{\partial\over \partial b}
\Bigg[\Big(1\,-\,{q^2\over \Omega^2\,\rho^4\,b^2}\Big)\,(\rho^4-\Omega^2\,b^2)
\Bigg]\,=\,-{2\over b}\,\Big(\Omega^2\,b^2\,-\,{q^2\over \Omega^2\,b^2}\Big)=0\,\,,
\label{condition_constant_b}
\eeq
it's clear that there is a chance to have $b(\rho)$ constant  precisely if it is equal to $b_0$ related to $q$ as in (\ref{qdeb0}).
 Plugging this value of   $b=b_0$ into the right-hand side of (\ref{chi_p_massless_general}) we find
\beq
\chi'^{\,2}\,=\,{\Omega^2\over \rho^4}\,\,,
\label{chip_square}
\eeq
that can be integrated right away
\beq
\chi\,=\,-{\Omega\over \rho}\,+\,\chi_0\,\,,
\eeq
with $\chi_0$  constant.  The minus sign is consistent with a positive Joule heating $q$, see (\ref{q_def}), and infalling boundary conditions, see  (\ref{variation_wavefront}).
 Then $c(\rho)=b e^{i\chi}$ ends up being
\beq
c(\rho)\,=\,{\rho_c^2\over \Omega}\,
e^{i\chi_0}\,e^{-i\,{\Omega\over \rho}}\,\,.
\eeq
The extreme simplicity of this solution must be contrasted with the need for numerical integration in the D3-D7 setup 
\cite{Hashimoto:2016ize}.
By expanding  near the UV as in (\ref{asymp_c}),  the electric field and current  at the boundary are equal  and given by
\beq
E\,=\,j\,=\,-i\,\rho_c^2\,e^{i\chi_0}\,\,.
\eeq
Hence, in the massless case, the rotating electric field and the induced current are  aligned. 
The effective Hawking temperature for this massless case can be obtained from (\ref{TH_final})
\beq
T_H\,=\,{\rho_c\over \pi\,b_0\,\chi'(\rho_c)}\,\,.
\eeq
Using (\ref{chip_square}) and (\ref{b0_w0_rho_c}) this simplifies into

\beq
T_H\,=\,{\rho_c\over \pi}\,=\,{q^{{1\over 4}}\over \pi}\,\,.
\eeq

Black hole embeddings with a small non-vanishing mass can be studied as perturbations of the exact massless solutions of this section. This analysis is carried out in detail in appendix \ref{app_B}. 

\subsection{High frequency limit}
\label{Large_Omega}

If the frequency $\Omega$ is large, the boundary condition at the UV (\ref{asymp_c}) fixes that $|c(\rho\to\infty)|=E/\Omega\ll 1$ and this  enforces $c(\rho)$ to be small for all values of $\rho$.   The gauge field  can then be treated as a perturbation around the state with $c=0$ and, as we shall see, admits an analytical solution.  To start with, using the scaling symmetry (\ref{scaling}) we can choose to  linearize the  equations of motion derived from (\ref{Lagrangian_complex_c})
near the solution with $c=0$ and $w=1$. Writing  $w=1+\delta w$ we get
\bear
&& c''\,+\,{2\over \rho}\, c'\,+\,{\Omega^2\over (1+\rho^2)^2}\, c\,=\,0\,\,  ,\\ \rule{0mm}{8mm}
&&\delta w''\,+\,{2\over \rho}\,\delta w'\,=\,{2\Omega^2\over (1+\rho^2)^3}\,| c|^2\,\,.
\label{c_deltaw_eqs}
\eear
where $c$ and $\delta w$ are small.  Notice that, although at linear order the equations decouple, as in \cite{Kinoshita:2017uch} we have included the lowest order contribution  to the coupling of modes, ${\cal O}(|\delta c|^2)$, in the equation of $\delta w$.  
The general solution of the equation of $\delta c$ is
\beq
 c(\rho)\,=\,c_1\,g_1(\rho)\,+\,c_2\,g_2(\rho)\,\,,
\label{general_c_linear}
\eeq
where $c_1$ and $c_2$ are constants and $g_1(\rho)$ and $g_2(\rho)$ are the functions
\bear
&&g_1(\rho)\,=\,{\sqrt{1+\rho^2}\over \rho}\,\sin\Big(\sqrt{1+\Omega^2}\,\arctan(\rho)\Big)\,\,,\rc\rc
&&g_2(\rho)\,=\,{\sqrt{1+\rho^2}\over \rho}\,\cos\Big(\sqrt{1+\Omega^2}\,\arctan(\rho)\Big)\,\,.
\eear
with asymptotic  behaviour   as $\rho\to\infty$ given by
\bear
&&g_1(\rho)\,=\,\sin\Big({\pi\over 2}\,\sqrt{1+\Omega^2}\Big)\,-\,
\sqrt{1+\Omega^2}\,\cos\Big({\pi\over 2}\,\sqrt{1+\Omega^2}\Big)\,{1\over \rho}\,+\,\cdots\,\,,\rc\rc
&&g_2(\rho)\,= \,\cos\Big({\pi\over 2}\,\sqrt{1+\Omega^2}\Big)\,+\,
\sqrt{1+\Omega^2}\,\sin\Big({\pi\over 2}\,\sqrt{1+\Omega^2}\Big)\,{1\over \rho}\,+\,\cdots\,\,.
\label{g12_UV}
\eear
Similarly  near $\rho=0$,  $g_1$ and $g_2$ behave as
\bear
&&g_1(\rho)\,=\,\sqrt{1+\Omega^2}\,-\,{\Omega^2\,\sqrt{1+\Omega^2}\over 6}\,\rho^2\,+\,\cdots\rc\rc
&&g_2(\rho)\,=\,{1\over \rho}\,-\,{\Omega^2\over 2}\,\rho\,+\,\cdots
\label{g12_IR}
\eear

At the linearized level the Joule heating $q$ of Eq. (\ref{q_c_c*}) is proportional to the Wronskian of $ c$ and $ c^{*}$
\beq
q\,=\,i\,{\Omega\,\rho^2\over 2}\,\Big( c\, c'^{\,*}\,-\, c^{*}\, c'\Big)\,\,.
\label{q_linear}
\eeq
Since $q$ is independent of $\rho$, we could compute it by using the expansions (\ref{g12_UV}) around $\rho=\infty$ or (\ref{g12_IR}) around $\rho=0$. The result one obtains is the same and given by
\beq
q\,=\,\Omega\,\sqrt{1+\Omega^2}\,\,{\rm Im}\,(c_1\,c_2^{*})\,\,.
\eeq

The solution for $w(\rho)$ can be written as 
\beq
\delta w(\rho)\,=\,-{2\Omega^2\over \rho}\int_{0}^{\rho}\,{s^2\over (s^2+1)^{3}}\,| c(s)|^2\,-\,
2\Omega^2\,\int_{\rho}^{\infty}\,{s\over (s^2+1)^{3}}\,| c(s)|^2\,+\,c_3\,\,,
\label{delta_w_linear}
\eeq
with $c_3$ being a new integration constant. 

Since we obtained the general solution of our equations, we can now analyze the different types of embeddings. The 
details of this study are deferred to appendix \ref{linearized_embeddings}. Let us summarize here the results that we obtain there. (i) Regularity of the  Minkowski embeddings at $\rho=0$ enforces that  the constant $c_2$ in (\ref{general_c_linear}) has to vanish.  (ii) By imposing the additional condition $E=0$ to these Minkowski solutions, we obtain the resonant frequencies (\ref{Omega_meson}), corresponding to the meson masses of the D3-D5 model.  (iii) To construct black hole embeddings we have to build up a pseudohorizon in our linearized theory and impose the corresponding infalling boundary conditions. Although the linear approximation breaks down for the critical embeddings, it turns out that  one can use the linear Minkowski solution to represent the critical configuration in the UV region of large $\rho$ and use this result to represent features of the phase diagram, such as the envelope curve of figure 
\ref{E_Omega_log}.

\section{Conductivities}
\label{Conductivities}

One of the main uses of the Floquet driving  is the artificial engineering of materials exhibiting transport phenomena with topological character. In this sense, the obtention of a Hall conductivity without a magnetic field  is an important hallmark. In the present context, this has been termed photovoltaic Hall effect\cite{Oka_2009}  where it's origin has  been shown to be  topological, like the Thouless pumping. The time-reversal symmetry breaking and hence, the nontrivial band topology,  is induced  by the circularly polarized light.  This proposal has been the subject of both theoretical and experimental studies  during the last decade \cite{Morimoto:2010, Ikebe:2010, Wang:2013, Kumar:2016, McIver:2020}.
In \cite{Hashimoto:2016ize} the photovoltaic Hall conductivity for massless carriers was obtained in the D3-D7 model,   and speculated to behave consistently with the picture of a strongly coupled version of a  (3+1) Weyl semimetal.  
Our findings show similarities and  departures from their results. For example, the massless case will be shown to be  trivial. However, in the massive case, we will indeed find an intricate Hall effect. 

Following the proposal in \cite{Oka_2009b} we proceed to study the response of our system to a small additional electric field on the boundary pointing in a fixed direction. From this response we will be able to extract the AC and DC conductivities.
We will closely mimic the strategy in \cite{Hashimoto:2016ize} whose main steps  will be reviewed and extended here for completeness, as our case involves  three coupled field perturbations. 

For the analysis it is most convenient to stick to the cartesian basis used in (\ref{rotating_electric_field}).
  On top of this background  field, the idea is to place another   fixed direction electric field 
\beq
 {\vec  {\cal E}}(t) = O(t)\vec E + \vec \epsilon(t)   \label{perturbed_E}\, ,
 \eeq
    with $|\vec \epsilon| \ll  |\vec E|$, and a  harmonic (AC) time dependence 
 \beq
   \vec \epsilon(t) = \vec\epsilon\,\,e^{-i\omega\,t}\ \, . \label{perturbed_E2}
 \eeq
Our task is to extract the response of the system out of the non equilibrium steady state (NESS) solutions spelled out in the previous sections. In particular,  the currents will also suffer a perturbation $\vec {\cal  J}(t) \to  O(t)\vec j + \delta \vec  j(t)$ and one of our targets will be to extract the {\em effective} conductivity response matrix $\boldsig$, such that $\delta \vec j(t) = \boldsig \cdot  \vec \epsilon(t)$.

From (\ref{perturbed_E}) we expect a perturbation of the bulk gauge field, which we also write now in cartesian components, $\vec a(t,\rho) \to \vec a(t,\rho) + \delta \vec a(t,\rho)$. 
The idea is now to refine the ansatz, taking into account  the new  driving  data (\ref{perturbed_E}) and (\ref{perturbed_E2}).
 First of all, let us reconsider the boundary conditions of the unperturbed gauge field, $\vec a(t,\rho)$ (\ref{vec_pot}),  now in cartesian coordinates.  The near boundary expansion will be
\bear
\vec a(t,\rho) &=&
-\int^t\vec {\cal E}(t) dt \, + \frac{\vec{\cal J}}{\rho} + ... \nonumber\\
&=& \,-{1\over \Omega}\, O(t)\,
\mbox{\boldmath{$\varepsilon$}}\,\vec E\, + \frac{\vec{ \cal J}}{\rho} + ... \label{boundcona}
\eear
where  $\mbox{\boldmath{$\varepsilon$}}$ stands for  the  $2\times 2$ antisymmetric matrix
\beq
\mbox{\boldmath{$\varepsilon$}}\,=\,
\bemat{cc}0 & 1 \\ -1& 0\enmat\,\,.
\eeq
The leading term matches the vector potential of  the rotating electric field through   $\vec {\cal E}(t)\,=\,-\partial_t\, \vec a(t,\rho=\infty)$.
This easily follows from the relation   $\partial_t\,O\,\mbox{\boldmath{$\varepsilon$}}=-\Omega\,O$. 
The boundary condition (\ref{boundcona}) suggests   the following ansatz for the bulk solution
\beq
\vec a\,(t,\rho)\,=\,O(t)\,\vec c\,(\rho)\,\,.
\eeq
which is equivalent to (\ref{complexified_potential}) and (\ref{bdec}) in cartesian basis. 
Near the boundary
\beq
\vec c(\rho)\approx \vec c_0\,+\,{\vec c_1\over \rho}\,+\,...\,\, ~
\eeq
with\footnote{This replaces (\ref{asymp_c}) with $j = c_{1x}  + i c_{1y}$} 
\beq
\vec c_0\,=\,-{1\over \Omega}\mbox{\boldmath{$\varepsilon$}}\,\vec E\, ~~~,~~~
\vec{\cal J}(t)\,=\,O(t)\,\vec c_1\,\,.
\label{cejota}
\eeq
Considering the perturbations,  now the bulk gauge potential $\vec a+\delta \vec a$   has to match the full electric field (\ref{perturbed_E}) at the boundary, $-\partial_t (\vec a +\delta a) = \vec{\cal E}(t)$
\beq
\vec a(t,\rho=\infty)\,+\delta\vec a(t,\rho=\infty)
\,=\,-{1\over \Omega}\, O(t)\,
\mbox{\boldmath{$\varepsilon$}}\,\vec E\,-\,{i\over \omega}\,
\vec\epsilon\,\,e^{-i\omega\,t}\,\,. \label{pertura}
\eeq
Induced by this new boundary conditions, the ansatz will assume that the static field, $c(\rho)$, develops a time dependent perturbation
\beq
\vec a\,(t,\rho)\,+\delta\vec a\,(t,\rho)
=\,O(t)\,\Big(\vec c\,(\rho)+\delta \vec c\,(t,\rho)\Big)\,\,. \label{defdeltac}
\eeq
When the flavour branes are massive, these fluctuations couple to a perturbation of the brane embedding functions $w(\rho) \to w(\rho) + \delta w(t,\rho)$.  In this case we will be dealing with a 3 component vector of fluctuations
$\delta \vec \xi(t,\rho) = (\delta c_x,\delta c_y,\delta \omega)$. 
The perturbed  equations of motion, linear in  $\delta \vec \xi$, will assume the following form,
in terms of the $(\tau, \rho_*)$ coordinates  (see Eq. (\ref{tortoise_coord}))
\beq
\big(\partial^2_{\tau}\,-\,\partial_{\rho_*}^2\,+\,{\bf A}(\rho)\,\partial_{\tau}\,+\,{\bf B}(\rho)\,\partial_{\rho_*}\,+\,
{\bf C}(\rho)\,\big)\,\delta\vec \xi\,=\,0\,\, . \label{pertequ}
\eeq
Here, ${\bf A}$,  ${\bf B}$ and  ${\bf C}$ are $3\times 3$ matrices depending on the radial coordinate $\rho(\rho^*)$ and, parametrically, on the rotating frequency $\Omega$. 
At the pseudo-horizon $\rho=\rho_c\,  (\rho^* \to -\infty)$ one finds
\beq
{\bf A}(\rho=\rho_c)\,=\,-{\bf B}(\rho=\rho_c)\,\equiv\,{\bf A}_c\, 
\qquad\qquad
{\bf C}(\rho=\rho_c)\,=\,0\,\,.
\eeq
Thus, in this limit, $\rho^*\to -\infty$, the fluctuation equations become
\beq
\big(\partial^2_{\tau}\,-\,\partial_{\rho_*}^2\big)\delta\vec \xi\,+\,{\bf A}_c\,(\partial_{\tau}\,-\,\partial_{\rho_*})\delta\vec \xi\,=\,0\,\,,
\eeq
whose general solution takes  the form
\beq
\delta\vec \xi\,=\,\vec f (\tau+\rho_*)\,+\,e^{-{\bf A}_c\,\rho_*}\,\vec g (\tau-\rho_*)\,\,,
\eeq
with $\vec f$  and $\vec g$ arbitrary vector functions.  Imposing that  $\vec g=0$ selects the ingoing wave boundary condition at the pseudohorizon. 

Let us now work out the boundary conditions for the fluctuations. First, we want to keep the mass of the flavor brane fixed. After inserting (\ref{defdeltac}) into (\ref{pertura}) and multiplying by ${\cal O}^{-1}$,  the boundary conditions for $\delta \vec \xi(t,\infty)$ turn out to be
\beq
\delta\vec c\,(t,\rho=\infty)\,=\,
-{i\over \omega}\,O(-t)\,e^{-i\omega\,t}\,\vec \epsilon\,\, ~~~, ~~~\delta \omega(t,\rho=\infty) = 0\, .
\label{UV_bc_deltac}
\eeq
Following \cite{Hashimoto:2016ize}, we expand the matrix ${\cal O}(t)$, defined in (\ref{rotating_electric_field}) as
\beq
 O(t) = {\bf M}_+ e^{i\Omega t} + {\bf M}_-  e^{-i\Omega t}\,\,, \label{odete}
\eeq
where ${\bf M}_{\pm} $ are orthogonal projectors given by
\beq
{\bf M}_{\pm} = \frac{1}{2} \bemat{cc} 1 & \pm i \\ \mp i & 1 \enmat \,\, .
\label{M_pm_def}
\eeq
Finally, defininig $
\omega_{\pm}\,=\,\omega\,\pm\,\Omega\,\,,
$
the boundary  UV condition  (\ref{UV_bc_deltac}) of $\delta\vec c$ can be written as
\beq
\delta\vec c (t,\rho=\infty)\,=\,-{i\over \omega}\,\Big({\bf M}_+ e^{-i\omega_+ t} \,+\,{\bf M}_-e^{-i\omega_- t} \Big)\,
\vec \epsilon\,\, .
\label{delta_c_bd}
\eeq
It is natural to assume that from these boundary condutions the  bulk fluctuations  $\delta\vec c\,(t,\rho)$ will oscillate with frequencies 
\beq
\delta\vec c\,(t,\rho)\,=\,\vec\beta_+(\rho)\,e^{-i\omega_+ t}\,+\,\vec\beta_-(\rho)\,e^{-i\omega_- t}\,\,.
\label{timedepbeta}
\eeq
As a consequence of the coupling in the linearized  equations, we will assume likewise that the embedding fluctuations will resonate with the same frecuencies
\be
\delta w(t,\rho) = \gamma_+(\rho) e^{-i\omega_+ t} + \gamma_- (\rho)e^{-i\omega_- t} \, .
\ee
In summary, setting $\delta \vec \xi(t,\rho) = \vec \xi_\pm(\rho) e^{-i\omega_\pm t}$,  the 3 component vectors $\vec\xi_{\pm}(\rho)= (\beta_{\pm,x} ,\beta_{\pm,y} ,\gamma_\pm )$ will satisfy the linearized ordinary differential system 
\beq
\left[\frac{d^2}{d\rho_*^2} - {\bf B}(\rho) \frac{d}{d\rho^*} + \omega_\pm^2 + i\omega_\pm {\bf A}(\rho) - {\bf C}(\rho) \right] \vec\xi_{\pm} = 0\, .
\label{difeqsys}
\eeq
From (\ref{UV_bc_deltac}), the brane fluctuations have  vanishing boundary source $
\gamma_\pm(\infty) = 0 \, .
$
Inserting  the  boundary expansions for the fields in \eqref{timedepbeta}
\beq
\delta\vec c(t,\rho)\approx \delta\vec c^{(0)}(t)\,+\,{\delta\vec  c^{(1)}(t)\over \rho}\,+\,...
~~~~~, ~~~~~ \vec \beta_\pm (\rho) = \vec \beta_\pm^{(0)} + \frac{\vec\beta_\pm^{(1)}}{\rho} + ...
\eeq
and comparing we get
\beq
\delta \vec c^{(0)}(t)\,=\,\vec\beta^{(0)}_+\,e^{-i\omega_+ t}\,+\,\vec\beta^{(0)}_-\,e^{-i\omega_- t}\, ~~~\,,
\qquad
\delta \vec c^{(1)}(t)\,=\,\vec\beta^{(1)}_+\,e^{-i\omega_+ t}\,+\,\vec\beta^{(1)}_-\,e^{-i\omega_- t}\,\,.
\eeq
Comparing the first of these equations with (\ref{delta_c_bd}) we conclude that  
\beq
\vec\beta^{(0)}_{\pm}\,=\,-{i\over \omega}\,{\bf M}_{\pm}\,\vec \epsilon\,\,.
\label{beta_0_M_epsilon}
\eeq

The subleading vectors $\vec\beta^{(1)}_{\pm}$  determine the variation of the current $\delta \vec {\cal J}(t) = O(t)\,\delta \vec c^{(1)}(t) $.
Using (\ref{cejota}), (\ref{odete}) and (\ref{delta_c_bd}) 
\bear
\delta \vec{\cal J}(t)\,= \,e^{-i\omega\,t}\,
\Big({\bf M}_+\,\vec\beta^{(1)}_+\,+\,{\bf M}_-\,\vec\beta^{(1)}_-\,+\,
{\bf M}_+\,\vec\beta^{(1)}_-\,e^{2\,i\Omega\,t}\,+\,{\bf M}_-\,\vec\beta^{(1)}_+\,e^{-2\,i\Omega\,t}
\Big)\,\,.
\label{delta_J_beta1} 
 \rule{0mm}{7mm}
\eear
As usual in the AdS/CFT program, given a regular  solution, the vectors $\vec \beta^{(1)}_{\pm}$ and $\vec \beta^{(0)}_{\pm}$ will no more be independent. From the linearity of the equations of motion it follows that this relation is also linear
\beq
\vec \beta^{(1)}_{\pm}\,=\,{\bold X_{\pm}}\,\vec \beta^{(0)}_{\pm}\, = 
\,-{i\over \omega}\,{\bold X_{\pm}}\,{\bold M_{\pm}}\,\vec \epsilon \, ,
\label{X_pm}
\eeq
where ${\bold X_{\pm}}$ are $2\times 2$ matrices to be computed and  (\ref{beta_0_M_epsilon}) has been used.
Plugging this into (\ref{delta_J_beta1}) allows to find a relation between the current  $\delta{\cal J}$ and the applied electric field $\vec \epsilon$
which defines $\boldsig$, $\boldsig_+$ and $\boldsig_-$ as the $2\times 2$ conductivity matrices of the three current modes, one  with frequency $\omega$ and two  with $\omega\pm 2\Omega$ (the heterodyning mixing modes \cite{OcaBuccian})
\beq
\delta \vec{\cal J} =
 \,\Big[
\boldsig(\omega)\,e^{-i\omega\,t}\,+\,\boldsig^+(\omega)\,
e^{-i(\omega+2\Omega)\,t}\,+\,\boldsig^-(\omega)\,
e^{-i(\omega-2\Omega)\,t}\Big]\,\vec\epsilon\,\,,
\label{delta_J_epsilon}
\eeq
where
\bear
 \boldsig(\omega) &=&-\frac{ i}{\omega}   \left( {\bf M}_+ {\bf X}_+ {\bf M}_+   + {\bf M}_-  {\bf X}_- {\bf M}_- \right)\,   \nonumber\\
 \boldsig^+(\omega) &=& -\frac{ i}{\omega}  \, {\bf M}_- {\bf X}_- {\bf M}_+ \,\,  \label{conduct} \\
 \boldsig^-(\omega) &=&-\frac{ i}{\omega} \,  {\bf M}_+  {\bf X}_+ {\bf M}_- \,\,. \nonumber
\eear

The procedure to compute ${\bold X}_\pm$ starts by solving the differential equation system (\ref{difeqsys}) from the pseudohorizon $\rho_* \to -\infty$
to the AdS boundary at $\rho_*=0$. At the horizon we impose the ingoing boundary condition $\vec \xi_\pm(\rho_*) \simeq \vec c \, e^{-i\omega_\pm \rho_*}$, where $\vec c$ are constant vectors $(1,0,0), (0,1,0)$ and $(0,0,1)$. Due to linearity of the equations  we obtain in this way two complex $3\times 3$ matrices $P_\pm$ and $Q_\pm$ such that $\vec\xi_\pm^{(0)} = P_\pm \vec c_\pm$, and  $\vec\xi_\pm^{(1)} = Q_\pm \vec c_\pm$. In this way we can solve  for $\vec \xi^{(1)}_\pm = Q_\pm P^{-1}_\pm \vec\xi_\pm^{(0)} $. Since we will be setting the source for the embedding fluctuations to zero, i.e. $\xi^{(0)}_{\pm,3} \equiv  \gamma_\pm^{(0)} = 0$ and
we only care about $\xi^{(1)}_{\pm,\, 1,2}$, the searched for matrices ${\bf X}_\pm $ in (\ref{X_pm}) will be given by the $2\times 2$ submatrix of the product $Q_\pm P^{-1}_\pm$
$$
{\bf X}_{\pm\, ij} =  [Q_\pm P^{-1}_\pm]_{ij=1,2} \, .
$$
From them, the conductivities are extracted using (\ref{conduct}). 
\begin{figure}[!ht]
\center
\includegraphics[width=0.47\textwidth]{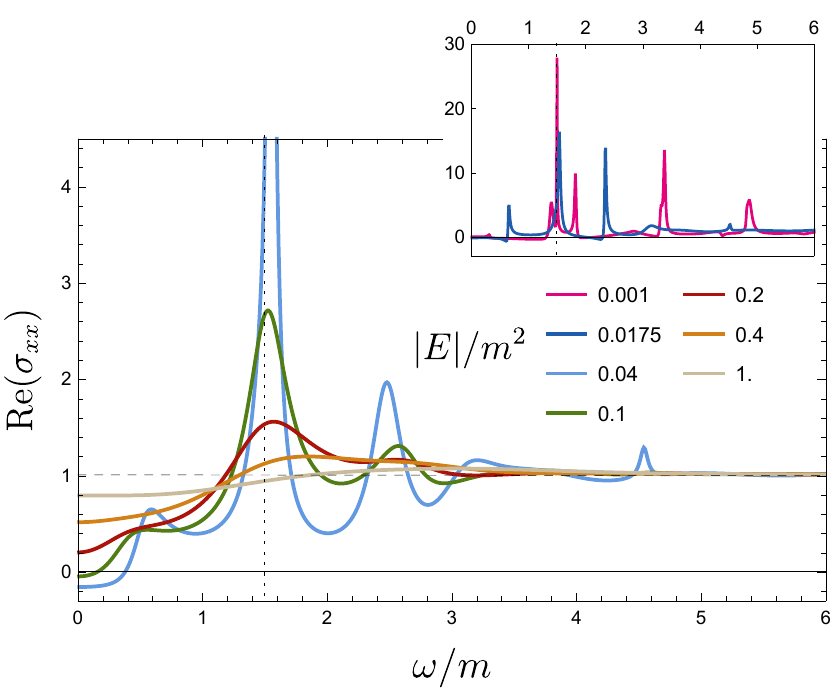} ~ \includegraphics[width=0.48\textwidth]{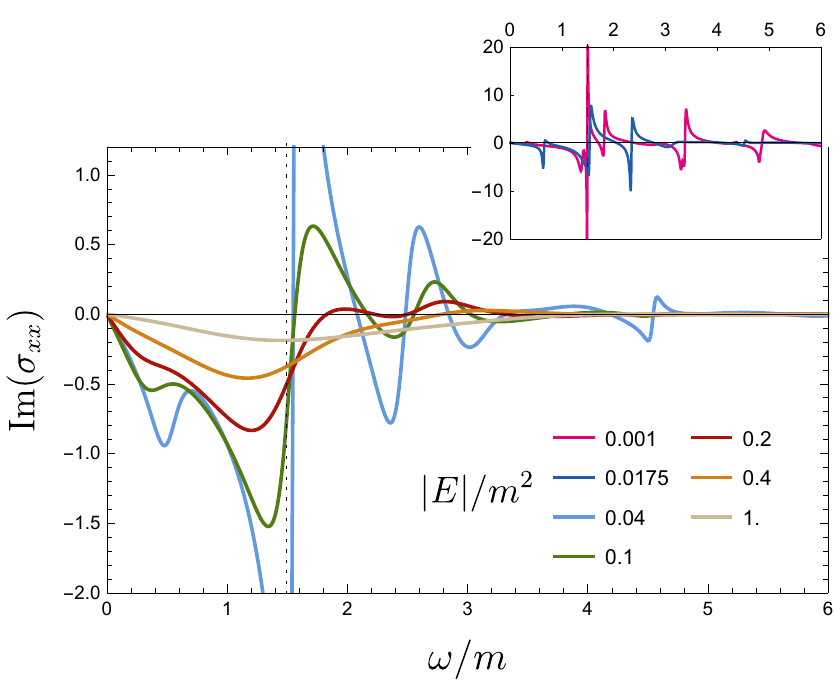} \\
\includegraphics[width=0.47\textwidth]{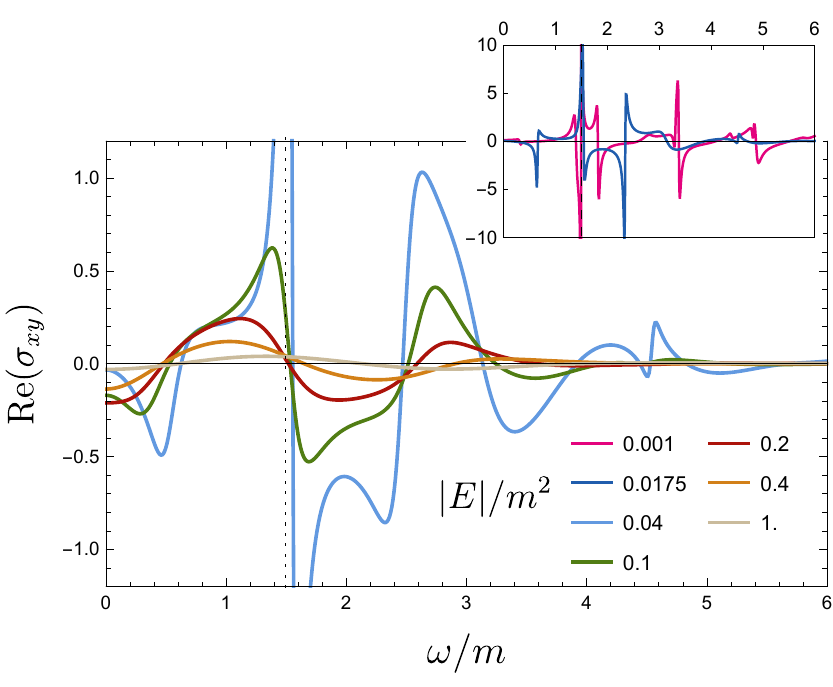} ~ \includegraphics[width=0.47\textwidth]{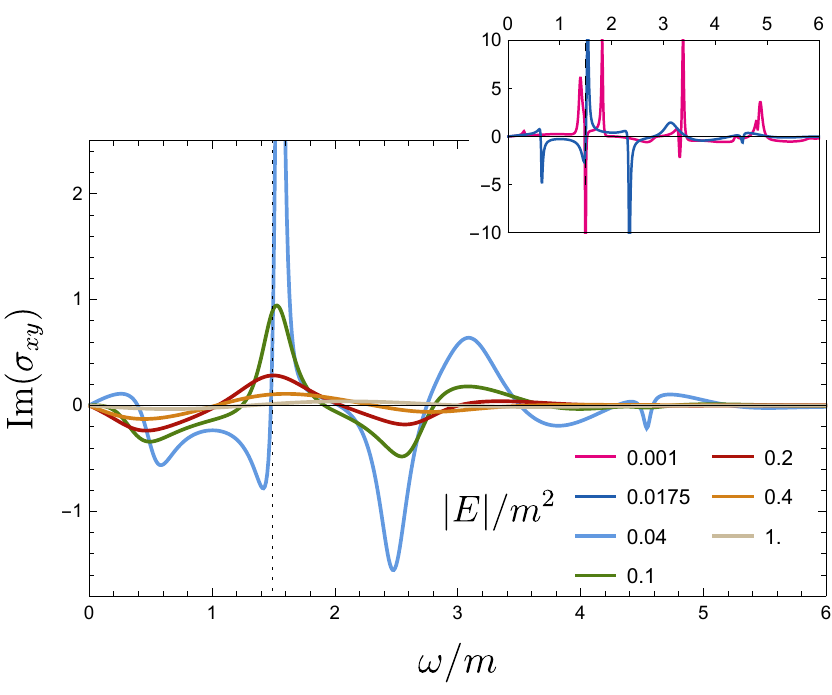} \\
\caption{AC conductivities for fixed value of $\Omega/m = 1.496$ and different values of $|E|/m^2$. They correspond to the vertically aligned dots in Fig.\,\ref{fig:PhSpMap} with the same color code.    In the insets, the curves at low values of $|E|/m^2\to 0$ deep into the critical wedge, where they develop peak resonances.  The value $\omega/m = \Omega/m = 1.496$ is signalled with a dashed vertical line. }
\label{fig:AC_conductivities}
\end{figure}

\subsection{AC conductivities}\label{sec:ACcond}

The results for the photovoltaic optical Hall conductivity, $\sigma_{xy}$, as well as for the photovoltaic optical  absortion spectrum, $\sigma_{xx}$,  can be seen in Fig.\,\ref{fig:AC_conductivities}.
We have plotted everything in units of the carrier mass $m$.\footnote{Numerically we have performed the integration for 2 different sets of  values of $(E,\Omega,m)$ in order to verify the scaling symmetry of Eq. (\ref{scaling}) }\, The driving laser  has been set to  the first resonant frequency  $\Omega/m=1.496$  for a discrete set of values of  $|E|/m^2$ (see the coloured dots along the vertical dashed line in Fig.\,\ref{fig:PhSpMap}). 

The  behaviour in all components of $\boldsig$  shows oscillations whose amplitude grows  for decreasing  $|E|/m^2$. All conductivities oscillate around $0$ except for Re$(\sigma_{xx})$ which  does it around a saturation value  $ 1$.

The departure in (2+1) dimensions,  from the analog (3+1) dimensional holographic system is significant (see Fig.\,7 and 8 in  \cite{Hashimoto:2016ize}).
In both cases, in the large frequency limit,  the conductivity  becomes proportional to the identity matrix  $\boldsig (\omega\to \infty) = c\, \mathbb{I}$. In the D3/D7 case,  $c$ is a complex linearly growing function of  $\omega$, whereas in our case $c$ is a constant  (equal to 1 in our normalization in this section, see later however).  This asymptotic  behaviour is also attained in the limit of high  $|E|/m^2\gg 1$, as controlled by the massless limit $m\to 0$ which is analytically solvable.
Not only the amplitude of the maxima, but also their movement is different. In the D3/D7 case their positions shift to lower frequency for decreasing external field. Here we see that they tend to approach the critical $\omega=1.496$.

As $|E|/m^2$ becomes smaller, strong oscillations develop, with the most prominent one sharpening around the resonance value for the optical and driving frequencies $\omega = \Omega$ (signalled with a dotted vertical line in all the plots). This feature bears  resemblance with the one obtained in \cite{Oka_2009b} (Fig 1b), although the behaviour of the peak amplitude with the intensity is the opposite. 

In the limit $|E|/m^2 \to 0$ these oscillations become very sharp  peaks as shown in the insets in Figs.\ref{fig:AC_conductivities}. As we move down towards $E/m^2\to 0$ the shape and  positions of these peaks change rapidly. This can be appreaciated in the insets, where the curves for $E/m^2 = 0.0175$ and $0.001$ look decorrelated. This last magenta curve shows a subseries of  peaks at frequencies   $\omega=1.496, 3.38, 4.87,...$ which have some overlap with the vector meson Floquet condensate spectrum  $\omega=1.496, 3.53, 5.56,...$ at shifted values.

  The fact that $\boldsig$ is dimensionless  in (2+1)  dimensions eases the comparison with results in the literature.  For the Hall component, $\sigma_{xy}(\omega)$ there is a striking similarity with the plot  in Figs.\,8  of \cite{Deghani2015}, where also the resonance peak is visible as well as secondary peaks.
     Importantly, in the limit $E\to 0$ the photoelectric Hall  conductivity does not vanish. This signals spontaneous time reversal symmetry breaking and might be a consequence of the underlying vector meson Floquet condensate state.\footnote{We thank T. Oka for this observation.}

\begin{figure}[!ht]
\center
\includegraphics[width=0.47\textwidth]{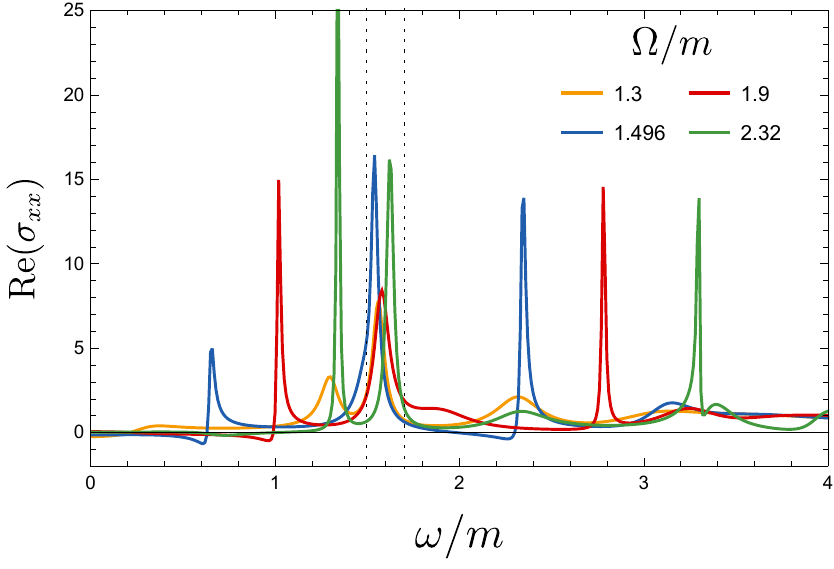} ~~~ \includegraphics[width=0.47\textwidth]{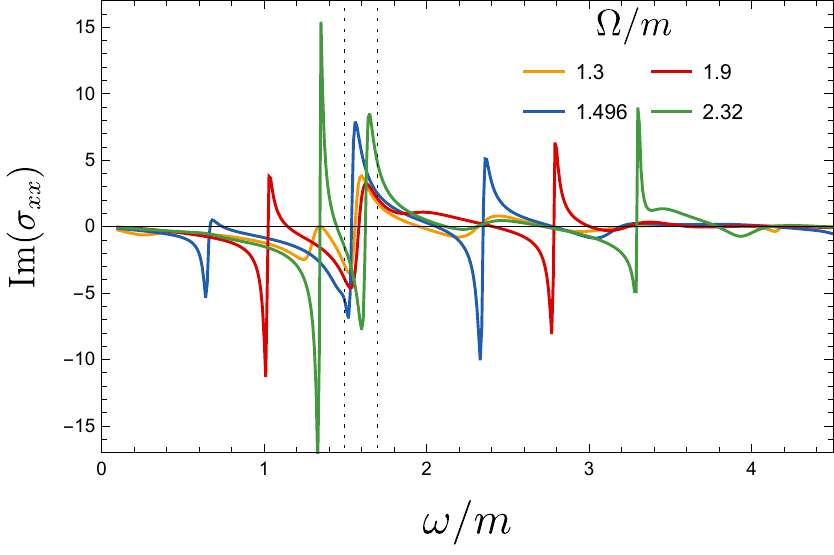} \\
\includegraphics[width=0.47\textwidth]{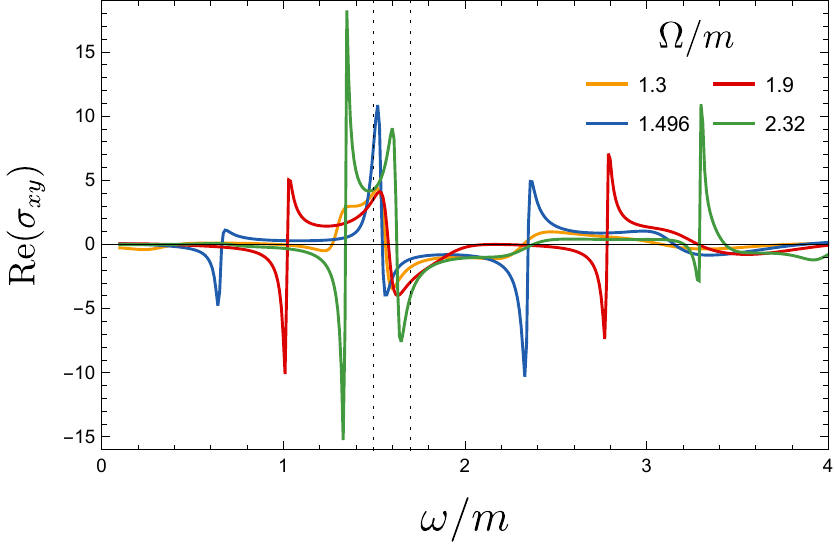} ~ \includegraphics[width=0.47\textwidth]{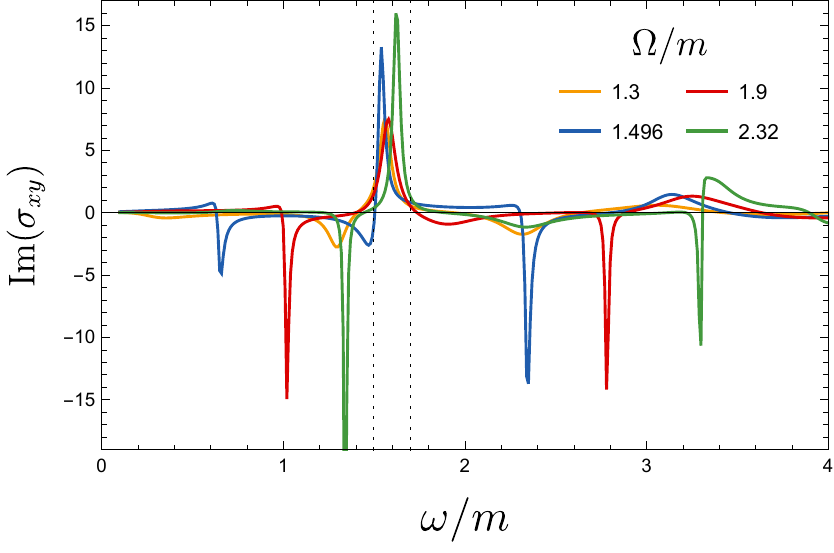} \\
\caption{Peaks in the AC conductivities closely above along the critical line. The exact position of these curves in phase space can be seen in Fig.\,\ref{fig:PhSpMap} as coloured circles. The vertical dashed lines signal the vector meson  Floquet condensate and the  meson  mass frequencies respectively.  The central peaks bunch in this region whereas the two lateral peaks move with $\omega$ while keeping their separation almost constant, $\Delta \omega /m \sim 1.7$.  
}
\label{fig:criticalcond}
\end{figure}

\begin{figure}[!ht]
\center
 \includegraphics[width=0.5\textwidth]{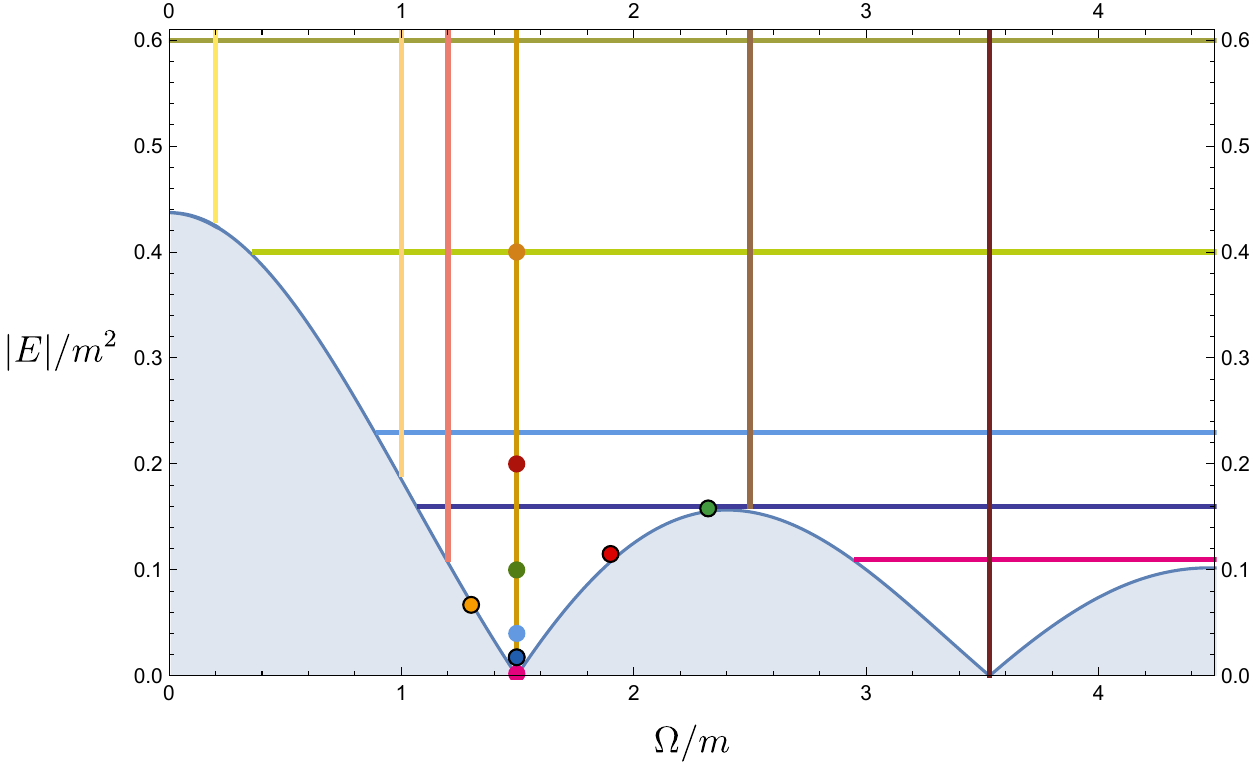} 
\caption{ The dots and circles signal the points where the AC conductivities in Figs.\,\ref{fig:AC_conductivities} and \ref{fig:criticalcond} have been plotted respectively.  Also shown are  the lines with either $\Omega/m$ or  $|E|/m^2$ fixed, used to plot the DC conductivities in Figures .\,\ref{fig:DC_Eovm2} and \ref{fig:DC_Omovm}.  In all cases, the same color code has been used.  }
\label{fig:PhSpMap}
\end{figure}

The enhancement of the oscillations  and peak formation observed in Fig.\ref{fig:AC_conductivities} is a universal feature as we approach  the line of critical embeddings.   We have followed the movement of the peaks for a sequence of configurations very closely above the critical line, for background frequencies in the range $\Omega/m \in (1.3, 2.32)$ (see Fig.\,\ref{E_Omega}). 
The resulting curves are plotted  in Fig.\,\ref{fig:criticalcond}. A remarkable fact is  the presence of one peak whose position barely changes and is confined to  the interval $\omega/m \in  (1.4965, 1.73) $, i.e. among the vector meson Floquet condensate frequency and the meson mass resonance (see Fig.\,\ref{E_Omega} and eqs. \eqref{floqcondfreqs}\eqref{mesonfreqs}).
Very slightly above the critical line there are two sidewise peaks which  appear within a  short interval in $E/m^2$. They all can be seen also in Fig.\,\ref{fig:criticalcond}. The remarkable fact about these lateral peaks is that  they shift position with the driving in such a way that their mean value coincides with the background frequency $\Omega$, while their splitting is almost constant $\Delta \omega /m \sim 1.7$ . This curious result is very neatly seen on embeddings  to the right of the vector meson Floquet condensate, i.e. in the segment $\Omega/m \in(1.496,2.32)$ and, very much damped, nearby  to its  left. The blue curve at $\Omega/m=1.496$ is the same one that can be observed in the insets of Fig.\,\ref{fig:AC_conductivities}, which has $E/m^2=0.0175$. At this frequency $\Omega=\Omega_c$, there are embedding solutions down to $E=0$ and this gives the magenta curves in those insets at values of $E/m^2 = 0.001$ and even lower. 

We also computed the heterodyning conductivities $\boldsig^{\pm}(\omega)$.  They bear the same features as for the normal conductivity, and the reader can found further information in Appendix
\ref{app_E}. Roughly stated, the curves for these exotic conductivities share, each of them, half of the peaks visible in $\boldsig(\omega)$.  

\begin{figure}[!ht]
\center
\includegraphics[width=0.43\textwidth]{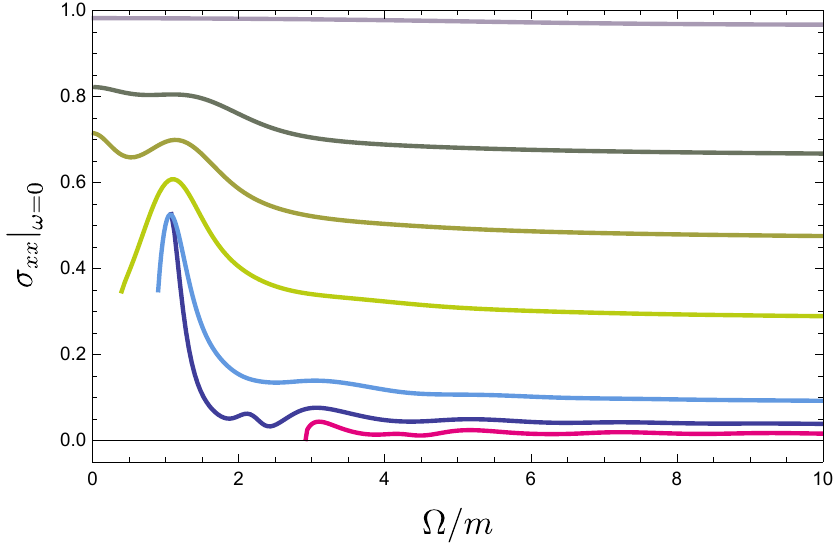} ~ \includegraphics[width=0.45\textwidth]{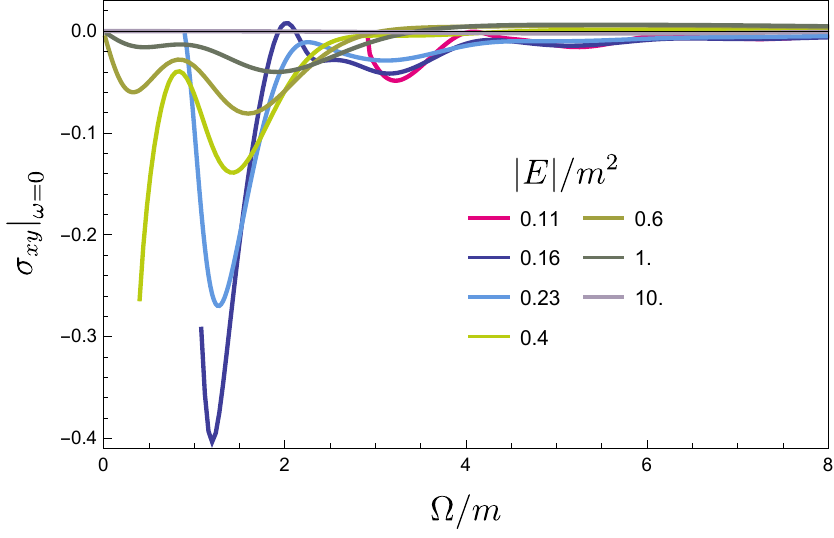} 
\caption{DC Conductivities as a function of $\Omega/m$ for various values of $|E|/m^2$.    }
\label{fig:DC_Eovm2}
\end{figure}
\begin{figure}[!ht]
\center
\includegraphics[width=0.43\textwidth]{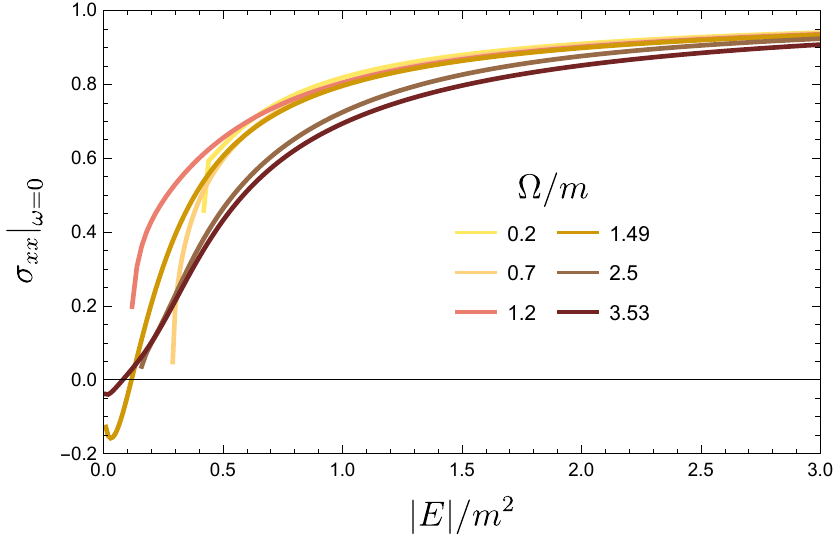} ~ \includegraphics[width=0.45\textwidth]{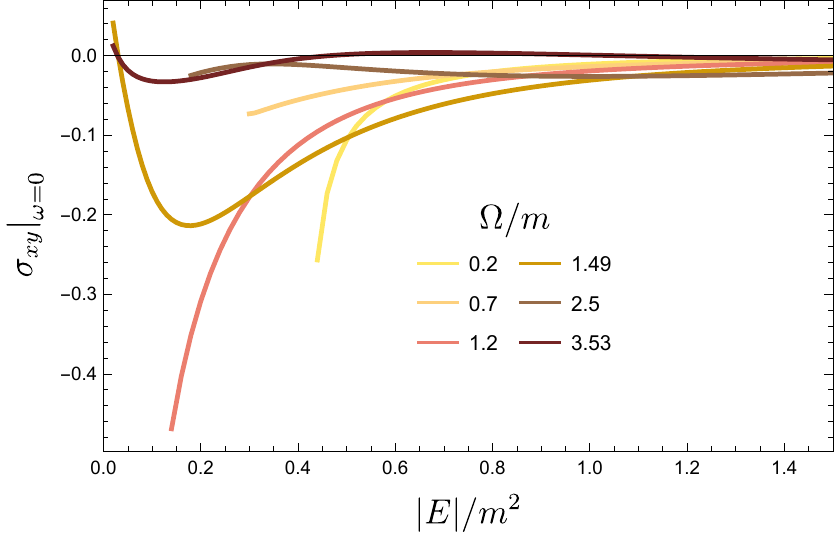} 
\caption{DC Conductivities as a function of $|E|/m^2$ for various values of $\Omega/m$.    }
\label{fig:DC_Omovm}
\end{figure}

\subsection{DC conductivities}
The zero frequency limit  $\omega\to 0$  is formally singular  (c.f.(\ref{UV_bc_deltac})). However the  numerical value of the AC conductivity is well behaved and stable down  to values of $\omega/\Omega$ well below $10^{-3}$. Hence we will use this value  to define our DC conductivity. The authors of   \cite{Hashimoto:2016ize} performed
 a numerical analysis in the time domain  and showed consistent  agreement between the two prescriptions.  From the plots in Figs.\,\ref{fig:AC_conductivities}, it is apparent that the imaginary parts vanish in the limit $\omega\to 0$. Hence
 the DC conductivities are real, as follows from the relation  $\boldsig(-\omega)=\boldsig(\omega)^*$.
  
 Figs.\ref{fig:DC_Eovm2} and  \ref{fig:DC_Omovm}  summarize our numerical results.
 In Fig.\,\ref{fig:DC_Eovm2} we exhibit  $\boldsig(\omega=0)$ for varying $\Omega/m$  along lines of constant $|E|/m^2$.
For high enough $E/m^2$ the limit $\Omega\to 0$ stays in the conducting phase and the direct conductivity $\sigma_{xx}$ has the correct static limit obtained in \cite{Karch:2007} where it becomes a constant that only depends on $m$, and saturates to a maximum value equal to one  (see section \ref{sec:massless} below for the correct normalization).

In Fig.\,\ref{fig:DC_Omovm}  we plot $\boldsig(\omega=0)$, varying $|E|/m^2$  along lines of constant $\Omega/m$. 
 Our results should be compared with the condensed matter calculation in \cite{Oka_2009} (see the discussion after Fig.2). For the Hall conductivity we see qualitative agreement in that, in absolute value,  $|\sigma_{xy}|$ first increases with $|E|/m^2$ and then decreases. For high $\Omega/m (\sim 3.53)$ we find values where this conductivity has zeros, this being  a remarkable feature.   For $\sigma_{xx}$ we find   a monotonic growth that  reaches a  limiting value for high $|E|/m^2$. In contrast, authors of \cite{Oka_2009} (see Fig.2b) obtained an intermediate  valley  which they proposed  to be a reflection of the  gap opening at the Dirac points. If so, our model is too crude to capture this feature.

The comparison with the (3+1) case  is complicated because the authors of  \cite{Hashimoto:2016ize} only considered the massless case, which in our setup turns out to be analytic and trivial.   Indeed, taking  $(|E|/m^2,\Omega/m)\to \infty$
the Hall conductivity vanishes whereas the DC conductivity asymptotes to a constant value.  One important difference is the sign of $\sigma_{xy}$ which in our  case comes out negative, hence opposite to the one in the (3+1) dimensional setup. This is so, despite the fact that we have been careful in keeping  all the conventions the same, and the sign in $\sigma_{xx}$ is positive as well.\footnote{yet, notice a  change of sign for very low values of $|E|/m^2$ approaching the Floquet condensates. It would be interesting to know if the configuration is unstable in this region and would decay to Minkowski embeddings.  }

The region in phase space where a significant photoelectric Hall conductivity  is obtained is rather small. It happens for values of the parameters in the region of  $\Omega/m \leq 2$ and also small  $|E|/m^2  \leq 0.6$. This corresponds roughly to  the close vicinity of the first dome to the left of  the critical  value $\Omega_c = 1.496$ in Fig.\,\ref{fig:PhSpMap}.


\subsection{Massless limit}\label{sec:massless}

When the unperturbed embedding of the probe brane in massless, \ie\ when $w=0$, the equations for $\delta w$ and $\delta \vec c$ decouple and we can take safely 
$\delta w=0$. The unperturbed configuration in this case is the analytic solution discussed in section \ref{massless_embd}. The equations of motion for $\delta c_x$ and $\delta c_y$ greatly simplify for this solution. Amazingly, the general solution of these coupled  equations can be obtained analytically, as shown in detail in appendix \ref{app_C}. Using these results one can extract the exact value of the conductivity matrices, which turn out to be
\beq
\boldsig(\omega)\,= c\,{\mathbb I}\,\,,
\qquad\qquad
\boldsig^+(\omega)\,=\,\boldsig^-(\omega)\,=\,0\,\,.
\label{massless_cond}
\eeq
with $c=1$.
As mentioned before, this analytical result controls the limits in the previous sections where  $|E|/m^2$ and $\Omega/m$ both become very large. 

Of course, the coefficient $c$  in Eq. \eqref{massless_cond} is not equal to 1.  Using \eqref{dictionaryE} it is easy to derive the correctly normalised physical conductivity, relating the the physical electric field and the physical current, as 
$$
\boldsig(\omega)\,=\frac{\sqrt{2}N_f N_c}{\pi\sqrt{\lambda}}~{\mathbb I} \, .
$$
Up to a convention factor of $\sqrt{2}$ this result matches the one obtained in \cite{Karch:2007}, see eq.\,(5.8),  were the observation is made that this conductivity is nothing else than $1/g^2$ in terms of the effective defect field theory gauge coupling on the D5-brane. This normalisation should affect all  the conductivities obtained previously in this paper.

\section{Summary and outlook}
\label{conclu}

In this paper we employed holographic techniques to analyze the response of a (2+1)-dimensional  gauge theory to an external rotating electric field. The system can be in two phases (insulator and conductor) depending on the type of embedding of the probe brane.  In agreement with the (3+1)-dimensional results of \cite{Hashimoto:2016ize,Kinoshita:2017uch} we have found that for certain frequencies the driving field resonates with the vector meson excitations of the gauge theory and gives rise to zero field Floquet resonances. The lobbed structure shown in Fig.\,\ref{E_Omega} may be a universal feature.  

Transport properties are an important benchmark, and they exhibit substantial differences in the two dimensionalities.  In  (2+1) dimensions, the massless case turns out to be trivial. In the massive situation, however, the conductivity tensor exhibits a rich structure. We have investigated the AC and DC conductivities, paying special attention to the first resonant frequency $\Omega = \Omega_c = 1.4965$ which is close to the meson mass frequency. In this situation the system creates efficiently charged carriers and holes. In experimental setups, the asymmetry in the electron-hole pairs for different chiralities gives rise to the so called ``optically induced valley polarization". This might be the closest physical realization to our setup at the resonant point. 
Weather the double peak structure observed in Fig: \ref{fig:criticalcond}  may have some meaning in this direction is something that deserves further scrutiny. Otherwise, it would probably mean this model is too crude to capture the chiral asymmetry in the splitting of the Dirac point.

Many distinguishing features are present for configurations approaching the critical line separating the conducting from the isolating phase. The strong enhancement of the oscillations deforming into sharp peaks  is one such. These peaks seem to point to bound states that are shifted from the meson masses. At this point, a very relevant question is that of stability of the embeddings. The exact point where, say, Black Hole embeddings decay into Minkowski embeddings and viceversa is a problem that can be addressed in equilbrium thermodynamics, and has a practical answer in terms of the equal area law. Here however, the similar tools are not available from a would be periodic thermodynamics \cite{Kohn14}, and this certainly an important aspect that will need further study in the future. 

Let us now discuss possible related lines of  research for the future. An interesting avenue would be to complete the phase diagram of the D3-D5 model by studying the system at (real) non-zero temperature and/or finite chemical potential. Moreover, we have treated the system in the quenched approximation. Dealing with the backreaction of the D5-branes is, in general, a tough problem. However, we could adopt some approximation, as it is done in the smearing approach  (see \cite{Nunez:2010sf} for a review). Interestingly, the smeared backreacted D3-D5 background has been obtained in \cite{Conde:2016hbg}
(see also \cite{Penin:2017lqt,Jokela:2019tsb,Gran:2019djz}) and one could try to generalize these results to the case in which an external rotating electric field is acting on the flavor brane. 

The comparison of our results with the D3-D7 model points towards some universality in the phase space. To strengthen this point,  a natural strategy is to extend the study to other brane setups. Most prominently, the (2+1) dimensional ABJM model is an example of superconformal QFT with a well stablished gravity dual \cite{Aharony:2008ug}.  The flavor branes in this case are D6-branes \cite{Hohenegger:2009as,Gaiotto:2009tk} (see \cite{Jokela:2012dw} for an analysis of the thermodynamics of the flavor branes in ABJM). The ABJM supergravity background has a non-trivial topological structure which allows to switch on fluxes of the gauge fields and realize the quantum Hall effect \cite{Bea:2014yda}. This suggests that it could be the right place to find non-trivial topological effects induced by an external driving in an holographic setup like the ones we are missing here.

Another direction worth pushing is  to consider other types of drivings. The brane picture of gauge theories allow for several possibilities. For example, instead of an electric field we could consider a magnetic field, or let the brane oscillate periodically. Actually, this last scenario has been considered  in \cite{Hoyos:2011us} in connection with the realization of the chiral magnetic effect in holography. It would be good to revisit this setup and search for possible stable non-equilibrium phases and condensates. The natural expectation is to observe resonances with the other type of mesons of the theory, like scalar mesons.

\vspace{1cm}
\noindent 

\section*{Acknowledgements}
We are  indebted to Keiju Murata for many patient clarifications about his work. Also we would like to thank  Yago Bea and Javier Tarr\'io for discussions and an early  collaboration in the initial stages of this project. Special thanks also go to Takaaki Ishii inspiring conversations, and  to Takashi Oka for  insightful remarks on the physical interpretation of our results.

This work was supported by MINECO FPA2017- 84436-P, Xunta de Galicia ED431C 2017/07, Xunta de Galicia (Centro singular de investigaci\'on de Galicia accreditation 2019-2022) and the European Union (European Regional Development Fund - ERDF ``Mar\'ia de Maeztu" Units of Excellence MDM- 2016-0692, and the Spanish Research State Agency. The work of A.G. has been supported by the Xunta de Galicia action
under the grant ED481A-2019/054.

\appendix

\section{Regularity conditions}
\label{app_A}

In this appendix we derive the regularity conditions for black hole embeddings. 
The right-hand side of the equations (\ref{eoms}) must vanish at the pseudohorizon
$\rho=\rho_c$  in order to avoid the potential  singularity at this point. These conditions yield the following three equations
\bear
&&-\rho_c\,(\Omega+2\rho_c\,b_1)\,(1+b_1^2+w_1^2)\,+\,\Omega\,b_0^2\,(\rho_c-2 b_0 b_1)\,\chi_1^2\,=\,0\,\,,\rc\rc
&&\rho_c\,(\rho_c\,w_1-w_0)\,(1+b_1^2+w_1^2)\,+\,\Omega\,b_0^3\,w_1\,\chi_1^2\,=\,0\,\,,\rc\rc
&& \rho_c^2\,(-1+b_1^2+w_1^2)\,-\,\rho_c\,(2 w_0 w_1-\Omega b_1)\,+\,\Omega\,b_0^3\chi_1^2\,=\,0\,\,. 
\label{regularity_eqs}
\eear
From the first equation in (\ref{regularity_eqs}), we get
\beq
\chi_1^2\,=\,{\rho_c\,(\Omega+2\rho_c\,b_1)\,(1+b_1^2+w_1^2)\over 
\Omega\,b_0^2\,(\rho_c-2b_0\,b_1)}\,\,.
\label{ch_1_b_1_w_1}
\eeq
Using this result in the second equation in (\ref{regularity_eqs}) we arrive at the following relation between $b_1$ and $w_1$
\beq
b_1\,=\,{\Omega\over 2 w_0}\,
{\rho_c\,w_0\,-\,(2\rho_c^2+w_0^2)\,w_1\over \rho_c^2+w_0^2}\,\,.
\label{b_1_w_1}
\eeq
Let us next plug  (\ref{ch_1_b_1_w_1}) into the third condition in (\ref{regularity_eqs}). We get
\beq
-w_0 b_1^2\,+\,\big(4\rho_c b_0+\Omega\rho_c+4 w_0 b_0 w_1)b_1+
w_0^2-2\rho_c w_0 w_1+2(\rho_c^2+w_0^2) w_1^2\,=\,0\,\,.
\eeq
Using now (\ref{b_1_w_1}) we arrive at the following quadratic equation for $w_1$
\beq
\Big[1+{\Omega^2 (\rho_c^2+w_0^2) \over 4(\rho_c^2+w_0^2)^2+\Omega^2\,\rho_c^2}\Big]\, w_1^2\,+\,
{2\rho_c\over w_0}\,w_1\,-\,1\,=\,0\,\,.
\label{quadratic_eq_w1}
\eeq
Let us write the solution of (\ref{quadratic_eq_w1}). First of all  we define the quantities $s$ and $D$ as
\bear
&&s\equiv \sqrt{
\big[4(\rho_c^2+w_0^2) +\Omega^2\big]\,\big[4(\rho_c^2+w_0^2)^2+\Omega^2\rho_c^2\big]}\,\,,\rc\rc
&&D\equiv  4(\rho_c^2+w_0^2)^2\,+\,\Omega^2\,(2\rho_c^2+w_0^2)\,\,.
\label{s_d_def}
\eear
Then, we can write
\beq
w_1\,=\,{1\over w_0\,D}\Big[-4\rho_c\,(\rho_c^2+w_0^2)^2\,-\,\Omega^2\,\rho_c^3\,+\,
(\rho_c^2+w_0^2)\,s\Big]\,\,.
\label{w_1_expr}
\eeq

Let us now analyze some limiting cases of the  relations found above.

\subsection{Large frequency}

Let us expand in powers of $1/\Omega$ for large frequency $\Omega$. Taking the limit $\Omega\to\infty$ in 
(\ref{s_d_def}) we get that $s\approx \rho_c\,\Omega^2$. and  $D\approx (w_0^2+2\rho_c^2)\,\Omega^2$. 
Plugging these values in (\ref{w_1_expr}) we get that $w_1$ is given at leading order by
\beq
w_1\,=\,{\rho_c\,w_0\over 2\rho_c^2+w_0^2}\,\,.
\eeq
Using this value of $w_1$ to evaluate the rhs of (\ref{b_1_w_1}) we get that $b_1\to 0$ as $\Omega\to\infty$. Actually, one can prove that
\beq
b_1\,=\,-{w_0^2 (\rho_c^2+w_0^2)\over \rho_c\,(2\rho_c^2+w_0^2)}\,\,{1\over \Omega}\,\,.
\eeq
Taking into account the values of $b_1$ and $w_1$ found in this large $\Omega$ limit, we get the following value of $\chi'$ at leading order
\beq
\chi_1\,=\,{\Omega \over \sqrt{\rho_c^2+w_0^2}}\,{\sqrt{4\rho_c^2+w_0^2}\over 2\rho_c^2+w_0^2}\,\,.
\eeq

\subsection{Small frequency}
When $\Omega\to 0$ the quantities $s$ and $D$ attain the values
\beq
s\approx 4\,(\rho_c^2+w_0^2)^{{3\over 2}}\,\,,
\qquad\qquad
D\approx 4\,(\rho_c^2+w_0^2)^{2}\,\,,
\qquad\qquad 
(\Omega\to 0)\,\,,
\eeq
and one can verify readily that
\beq
w_1\to {1\over w_0}\,\Big[-\rho_c\,+\,\sqrt{\rho_c^2+w_0^2}\Big]\,\,.
\eeq

\section{Small mass  solutions}
\label{app_B}

We now consider a small perturbation around the  analytic massless solution of  section \ref{massless_embd}. We write  $b(\rho)$ and $w(\rho)$ as
\beq
b(\rho)\,=\,b_0\,+\,\beta(\rho)\,\,,
\qquad\qquad
w(\rho)\,=\,\lambda(\rho)\,\,,
\eeq
and we consider the equations of motion (\ref{eoms}) at first-order in $\beta$ and $\lambda$. Using (\ref{chi_p}) to eliminate $\chi'$, we arrive at the following decoupling linear equations for $\lambda(\rho)$ and $\beta(\rho)$
\bear
&&\rho^2\,(\rho^4\,-\,q)\,\lambda''\,+\,2\rho\,(\rho^4\,+\,q)\,\lambda'\,-\,2q\,\lambda\,=\,0\,\,,\\
\rule{0mm}{6mm}
&&\rho^2\,(\rho^4\,-\,q)\,\beta''\,+\,2\rho\,(\rho^4\,+\,q)\,\beta'\,+\,
4\,\Omega^2\,{\rho^6\over \rho^4\,-\,q}\,\,\beta\,=\,0\,\,.
\label{eom_lambda_beta}
\eear
(We are assuming that $q>0$).  Let us consider from now on the equation for $\lambda$ in (\ref{eom_lambda_beta}). This equation can be solved analytically in terms of hypergeometric functions
\beq
\lambda(\rho)\,=\,c_1\,\rho\,F\Big({1\over 4}\,,\,{1\over 2}\,;\,{3\over 4}\,;\,{\rho^4\over q}\Big)\,+\,
c_2\,\rho^2\,F\Big({1\over 2}\,,\,{3\over 4}\,;\,{5\over 4}\,;\,{\rho^4\over q}\Big)\,\,,
\label{general_sol_lambda}
\eeq
where $c_1$ and $c_2$ are constants. The hypergeometric functions written in (\ref{general_sol_lambda}) have a logarithmic singularity as we approach the pseudohorizon at $\rho= q^{{1\over 4}}$. Indeed, from the general equation
\beq
F\Big(\alpha, \beta;\alpha+\beta;z\Big)\approx -{\Gamma(\alpha+\beta)\over \Gamma(\alpha)\Gamma(\beta)}\,\,\log (1-z)\,\,,
\qquad\qquad {\rm as}\,\,z\to 1^{-}\,\,,
\eeq
we get
\beq
\lambda(\rho)\,\approx\,-{q^{{1\over 4}}\over \sqrt{\pi}}\,
\Bigg[c_1\,{\Gamma\big({3\over 4}\big)\over \Gamma\big({1\over 4}\big)}\,+\,
c_2\,q^{{1\over 4}}\,{\Gamma\big({5\over 4}\big)\over \Gamma\big({3\over 4}\big)}\,\Bigg]\,
\log\Big(1-{\rho^4\over q}\Big)\,\,,
\qquad\qquad {\rm as}\,\,\rho\to q^{{1\over 4}}\,\,.
\eeq
The absence of this logarithmic singularity imposes the following ratio between the constants $c_1$ and $c_2$
\beq
{c_1\over c_2}\,=\,-{q^{{1\over 4}}\over 4}\,\,
\Bigg[{\Gamma\big({1\over 4}\big)\over \Gamma
\big({3\over 4}\big)}\Bigg]^2\,\,.
\label{c_ratio_IR_regular}
\eeq
Let us now look at the UV  behavior $\rho\to\infty$. In general, for large $z$, one has
\beq
F(\alpha,\beta;\alpha+\beta; z)\,\approx\,
\Gamma(\alpha+\beta)\,\Bigg[\,
e^{-i\pi\alpha}\,{\Gamma(\beta-\alpha)\over \Gamma^2(\beta)}\,z^{-\alpha}\,+\,
e^{-i\pi\beta}\,{\Gamma(\alpha-\beta)\over \Gamma^2(\alpha)}\,z^{-\beta}
\,\Bigg]\,\,.
\eeq
Applying this last equation to our case, we get the UV behavior of the two functions in (\ref{general_sol_lambda})
\bear
&&\rho\,F\Big({1\over 4}\,,\,{1\over 2}\,;\,{3\over 4}\,;\,{\rho^4\over q}\Big)\,\approx\,
q^{{1\over 4}}\,\Gamma\Big({3\over 4}\Big)\Bigg[ {e^{-{i\pi\over 4}}\over \pi}\,\Gamma\Big({1\over 4}\Big)
\,+\,q^{{1\over 4}}\,
{e^{-{i\pi\over 2}}\,\Gamma\big(-{1\over 4}\big)\over \Gamma^2\big({1\over 4}\big)}\,\,{1\over \rho}\Bigg]\,\,,\rc\rc
&&\rho^2\,F\Big({1\over 2}\,,\,{3\over 4}\,;\,{5\over 4}\,;\,{\rho^4\over q}\Big)\,\approx
q^{{1\over 4}}\,\Gamma\Big({5\over 4}\Big)\Bigg[ q^{{1\over 4}}\,
e^{-{i\pi\over 2}}\,{\Gamma\Big({1\over 4}\Big)\over \Gamma^2
\Big({3\over 4}\Big)}\,+\,q^{{1\over 2}}\,{e^{-{3i\pi\over 2}}\over \pi}\Gamma\Big(-{1\over 4}\Big)\,{1\over \rho}\,
\Bigg]\,\,.
\qquad\qquad
\eear
Therefore,  the coefficient of the leading term  in $\lambda(\rho)$ as $\rho\to\infty$ is
\beq
q^{{1\over 4}}\,c_1\,e^{-{i\pi\over 4}}
{\Gamma\big({3\over 4}\big)\Gamma\big({1\over 4}\big)\over \pi}\,+\,q^{{1\over 2}}\,c_2\,
e^{-{i\pi\over 2}}\,\Gamma\Big({5\over 4}\Big)\,{\Gamma\big({1\over 4}\big)\over \Gamma^2\big({3\over 4}\big)}\,\,.
\eeq
The imaginary part of this leading term in the UV is zero if the ratio of $c_1$ and $c_2$ is
\beq
{c_1\over c_2}\,=\,-{q^{{1\over 4}}\over 4}\,
{\sqrt{2}\,\Gamma\Big({5\over 4}\Big)\over \Gamma^3
\big({3\over 4}\big)}\,\,,
\eeq
which can be seen to be equivalent to (\ref{c_ratio_IR_regular}). If this ratio holds, the  leading term $\lambda(\rho=\infty)$ is finite and given by
\beq
m\,\equiv\, q^{{1\over 4}}\,c_1\,{\Gamma\big({3\over 4}\big)\,\Gamma\big({1\over 4}\big)\over \sqrt{2}\,\pi}\,=\,
q^{{1\over 4}}\,c_1\,\,.
\label{c_1_m}
\eeq
Let us now look at the  UV subleading term, which is of the form ${\cal C}/\rho$. The coefficient ${\cal C}$ (\ie\ the condensate), is given by
\beq
{\cal C}\,=\,
q^{{1\over 2}}\,c_1\,e^{-{i\pi\over 2}}
{\Gamma\big({3\over 4}\big)\,\Gamma\big(-{1\over 4}\big)\over \Gamma^2\big({1\over 4}\big)}\,+\,
q^{{3\over 4}}\,c_2\,e^{-{3i\pi\over 4}}
{\Gamma\big({5\over 4}\big)\,\Gamma\big(-{1\over 4}\big)\over \pi}\,\,.
\eeq
The imaginary part of ${\cal C}$ vanishes if
\beq
{c_1\over c_2}\,=\,-q^{{1\over 4}}\,
{\Gamma\big({5\over 4}\big)\,\Gamma^2\big({1\over 4}\big)\over \sqrt{2}\,\pi\,\Gamma\big({3\over 4}\big)}\,\,,
\eeq
which again can be proven to be equivalent to  (\ref{c_ratio_IR_regular}). Then, the condensate ${\cal C}$ is given by
\beq
{\cal C}\,=\,- q^{{3\over 4}}\,c_2\,
{\Gamma\big({5\over 4}\big)\,\Gamma\big(-{1\over 4}\big)\over \sqrt{2}\, \pi}\,=\,q^{{3\over 4}}\,c_2\,\,.
\label{c_2_c}
\eeq
By combining (\ref{c_2_c}), (\ref{c_1_m}) and (\ref{c_ratio_IR_regular}) we can obtain the relation between ${\cal C}$ and $m$
\beq
{\cal C}\,=\,-q^{{1\over 4}}\,
\Bigg[{2\,\Gamma\big({3\over 4}\Big)\over \Gamma
\big({1\over 4}\big)}\Bigg]^2\,\,m\,\,.
\label{c_m_rel_low_mass}
\eeq
Therefore, in this small mass regime, the condensate is proportional to the mass with a coefficient which  depends on 
$q^{{1\over 4}}\,=\,\sqrt{\Omega\,b_0}$. Notice that we can write the solution for $w=\lambda$ in terms of the mass and condensante as
\beq
w(\rho)\,=\,{m\over |q|^{{1\over 4}}}\,
\rho\,F\Big({1\over 4}\,,\,{1\over 2}\,;\,{3\over 4}\,;\,{\rho^4\over |q|}\Big)\,+\,
{{\cal C}\over |q|^{{3\over 4}}}\,\,\rho^2\,F\Big({1\over 2}\,,\,{3\over 4}\,;\,{5\over 4}\,;\,{\rho^4\over |q|}\Big)\,\,,
\eeq
where ${\cal C}$ is related to $m$ as in (\ref{c_m_rel_low_mass}). We can also obtain the value of the embedding function at the pseudohorizon in terms of the mass
\beq
w(\rho=\sqrt{\Omega b_0})\,=\,{\Gamma^2\big({3\over 4}\big)\over \sqrt{2\pi}}\,\,m\,\,.
\eeq

Let us now integrate the equation in (\ref{eom_lambda_beta}) for the fluctuation $\beta$ of the electric field. This equation can be solved analytically,. In order to write this solution,  let us define the  following hatted quantities
\beq
\hat\rho={\rho\over q^{{1\over 4}}}\,\,,
\qquad\qquad
\hat\Omega={\Omega\over q^{{1\over 4}}}\,\,,
\eeq
and the two functions $F(\hat\rho)$ and $G(\hat\rho)$ as
\beq
F(\hat\rho)\,\equiv\,\hat\Omega\arctan(\hat\rho)\,\,,
\qquad\qquad
G(\hat\rho)\,\equiv\,{\hat\Omega\over 2}\,\,\log{\hat\rho-1\over \hat\rho\,+\,1}\,\,.
\eeq
Then, the general solution for the function $\beta$ is
\beq
\beta(\hat\rho)\,=\,c_1\,\cos\big[F(\hat\rho)\,+\,G(\hat\rho)\big]\,+\,
c_2\,\sin\big[F(\hat\rho)\,+\,G(\hat\rho)\big]\,\,,
\label{general_sol_beta}
\eeq
where $c_1$ and $c_2$ are  real constants. In order to find $E$ and $j$ for this solution, we must study the UV behavior of the complex potential $c=b\,e^{i\chi}$. The equation for the perturbation $\delta c\,=\,c-b_0\,e^{-i\Omega/\rho}$ at first order is
\beq
\rho^2\,(\rho^4\,-\,q)\,\delta c{\,''}\,+\,2\,\Big(\rho(\rho^4+q)+i q\,\Omega\Big)\,\delta c{\,'}\,+\,
{\Omega^2\over \rho^2}\,\Big(\rho^4\,+\,q\,-\,4i\,\rho\,{q\over \omega}\,\Big)\delta c\,=\,0\,\,.
\label{delta_c_massless}
\eeq
It is difficult to solve this equation directly. However, the equation for $\delta \chi=\chi+{\Omega\over \rho}$ can be obtained form (\ref{chi_p}) and  is rather simple
\beq
\delta\,\chi{\,'}\,=\,-{2\,\rho^2\,\Omega\over b_0\,(\rho^4\,-\,q)}\,\beta\,\,.
\label{delta_chi_eq_massless}
\eeq
Notice that $\delta c$  is related to $\beta$ and $\delta \chi$ as
\beq
\delta c\,=\,e^{-i\,{\Omega\over \rho}}\,\big(\beta\,+\,i\,b_0\,\delta\chi\big)
\label{c_beta_chi}
\eeq
One can check that the second equation in (\ref{eom_lambda_beta}), together with (\ref{delta_chi_eq_massless}), imply (\ref{delta_c_massless}). Moreover, since we know  the solution for $\beta$ we can directly integrate (\ref{delta_chi_eq_massless}). Indeed, let us write the solution (\ref{general_sol_beta}) in terms of complex exponentials as
\beq
\beta\,=\,d \,e^{-i(F(\hat\rho)\,+\,G(\hat\rho))}\,+\,d^*\,e^{i(F(\hat\rho)\,+\,G(\hat\rho))}\,\,,
\eeq
with $d$ being a complex constant. Then, we can show that $\delta \chi$ is
\beq
\delta \chi\,=\,-{i\,d\over b_0}\,e^{-i(F(\hat\rho)\,+\,G(\hat\rho))}\,+\,{i\,d^*\over b_0}\,e^{i(F(\hat\rho)\,+\,G(\hat\rho))}\,+\,\lambda\,\,,
\eeq
with $\lambda\in {\mathbb R}$. Plugging $\beta$ and $\delta \chi$ on the right-hand side of (\ref{c_beta_chi}), we get
\beq
\delta c\,=\,2\,d\,e^{-i\,{\Omega\over q^{1/4}}\big({q^{1/ 4}\over \rho}\,+\,
\arctan(\rho/ q^{1/ 4})\,+\,{1\over 2}\log {\rho-q^{1/4}\over \rho+q^{1/4}}\big)}\,+\,i\,b_0\,\lambda\,e^{-i\,{\Omega\over \rho}}\,\,.
\eeq
Notice that the function multiplying $d$ is an outgoing wave, whereas that multiplying $\lambda$ is an incoming wave. It is also interesting to notice that the first order variation of the Joule heating $q$ is zero for this solution, as it can be checked by using  (\ref{q_def}).

\section{Embeddings in the linearized approximation}
\label{linearized_embeddings}
In this appendix we study in detail the three types of embeddings in the linearized approximation of section \ref{Large_Omega} (valid for large driving frequency $\Omega$ or small electric field $E$).

\subsection{Minkowski embeddings}
To obtain the Minkowski embeddings that reach $\rho=0$ we  impose regularity of the function $c(\rho)$  at $\rho=0$. This implies that the constant $c_2$ must vanish and, therefore the gauge field $\,c(\rho)$ is given by
\beq
 c(\rho)\,=\,c_1\,g_1(\rho)\,=\,c_1\,{\sqrt{1\,+\,\rho^2}\over \rho}\,\,\sin\Big(\sqrt{1+\Omega^2}\,\arctan(\rho)\Big)\,\,.
\label{delta_c_sol_Min}
\eeq

From the  boundary expansion (\ref{asymp_c}) we find the electric  field and current
hence
\bear
 E\,&=&\,-i\,c_1\,\Omega\,\sin\Big({\pi\over 2}\,\sqrt{1+\Omega^2}\Big)\,\,,
 \label{E_Min}
\\
\rule{0mm}{7mm}
j\,&=&\,-c_1\,\sqrt{1+\Omega^2}\,\cos \Big({\pi\over 2}\,\sqrt{1+\Omega^2}\Big)\,\,.
\eear
Eliminating the constant $c_1$ we obtain the current in terms of the electric field $E$
\beq
j\, =\,-{\,i\,E\over \Omega}\,\sqrt{1+\Omega^2}\,\cot \Big({\pi\over 2}\,\sqrt{1+\Omega^2}\Big)\,\,.
\label{j_Min}
\eeq
We can also eliminate $c_1$ in (\ref{delta_c_sol_Min}) using (\ref{E_Min})
\beq
 c(\rho)\,=\,{i\,E\over \Omega\,\sin\Big[{\pi\over 2}\,\sqrt{1+\Omega^2}\Big]}\,\,
{\sqrt{1+\rho^2}\over \rho}\,\sin\Big[\sqrt{1+\Omega^2}\arctan \rho\Big]\,\,.
\label{delta_c_E}
\eeq

Let us now look at the embedding function $\delta w$ given in (\ref{delta_w_linear}). Imposing  $\delta w(\rho=0)=0$ fixes the value of the integration constant $c_3$
\beq
c_3\,=\,2\,\Omega^2\,|c_1|^2\,\int_{0}^{\infty}\,ds\,{s\over (s^2+1)^{3}}\,g_1^2(s)\,\,.
\eeq
Then, the resulting $\delta w(\rho)$ can be written as
\bear
&&\delta w(\rho)\,=\,2\,\Omega^2\,|c_1|^2\,
\int_{0}^{\rho} \,ds\,\,{\sin^2\big[\sqrt{1+\Omega^2}\,\arctan(s)\big]\over s(s^2+1)^2}\,-\,\rc\rc
&&\qquad\qquad\qquad\qquad\qquad
-\,{2\,\Omega^2\,|c_1|^2\over \rho}\,\int_{0}^{\rho} \,ds\,\,{\sin^2\big[\sqrt{1+\Omega^2}\,\arctan(s)\big]\over (s^2+1)^2}\,\,.
\label{delta_w_Min_sol}
\eear
From the UV expansion (\ref{asymp_w}) we find  the mass and the condensate 
\bear
&&m\,=\,1\,+\,2\,\Omega^2\,|c_1|^2\,
\int_{0}^{\infty} \,ds\,\,{\sin^2\big[\sqrt{1+\Omega^2}\,\arctan(s)\big]\over s(s^2+1)^2}\,\,,\rc\rc
\rule{0mm}{8mm}&&{\cal C}\,=\,-2\,\Omega^2\,|c_1|^2\,
\int_{0}^{\infty} \,ds\,\,{\sin^2\big[\sqrt{1+\Omega^2}\,\arctan(s)\big]\over (s^2+1)^2}\,\,.
\eear
In terms of the electric field at the boundary, using (\ref{E_Min}) we obtain
\bear
&&m\,=\,1\,+\,{2\,|E|^2\over \sin^2\Big({\pi\over 2}\,\sqrt{1+\Omega^2}\Big)}\,\,
\int_{0}^{\infty} \,ds\,\,{\sin^2\big[\sqrt{1+\Omega^2}\,\arctan(s)\big]\over s(s^2+1)^2}\,\,,\rc\rc
&&{\cal C}\,=\,-
{|E|^2\over 4\,\sin^2\Big({\pi\over 2}\,\sqrt{1+\Omega^2}\Big)}\,\,
\left(\pi+{\sin\left(\pi\sqrt{1+\Omega^2}\right)\over\Omega^2\sqrt{1+\Omega^2}}\right)\,\,.
\eear

To find the resonant frecuencies  of the Floquet condensates we impose that $ E=0$ in (\ref{E_Min}), which  means that the $\Omega$'s must satisfy 
\beq
\sqrt{1+\Omega_n^2}\,=\,2(n+1)\,\,,
\qquad\qquad
\eeq
for $n\,=\,0,1,2,\cdots\,\,.$ This gives the following discrete set
\beq
\Omega_n\,=\,2\sqrt{
\Big(n+{1\over 2}\Big)\Big(n+{3\over 2}\Big)}\,\,,
\eeq
which are the same as the masses of the vector mesons in the D3-D5 model \cite{Arean:2006pk}.

\subsection{Black hole embeddings}
We now determine $c(\rho)$ for a black hole embedding by imposing the appropriate in-falling boundary condition at the pseudohorizon. Now the phase $\chi(\rho)$ is not constant and, in fact, $\chi'$ is related to the Joule heating $q$. To obtain this relation in the linearized approximation, let us  take into account that $c\,c\,'^{\,*}\,-\,c\,^{*}\,c\,'=-2i\,\chi'\,b^2$. Since  the lagrangian density ${\cal L}=\rho^2$ in the linear approximation, we get from (\ref{q_def}) that
\beq
\chi'(\rho)\,=\,{q\over \Omega}\,{1\over \rho^2\,b^2(\rho)}\,\,.
\eeq
Let us now take $\rho=\rho_c$ in this last expression and use (\ref{b0_w0_rho_c}) and (\ref{q_w0_rho_c}) to relate $b_0=b(\rho_c)$ and $q$ to 
$w_0=w(\rho_c)$ and $\rho_c$. We immediately get
\beq
\chi'(\rho_c)\,=\,{\Omega\over w_0^2+\rho_c^2}\,\,,
\label{chi_prime_initial_linear}
\eeq
which is an initial condition for $\chi(\rho)$. We begin by introducing
 a new variable $y$, related to $\rho$ as\footnote{
The coordinate $y$ is nothing but the tortoise coordinate $\rho_*$ in this linear approximation. Indeed, in this case the function $B(\rho)$ defined in (\ref{A_B_tortoise}) is 
$B(\rho)\approx 1/(1+\rho^2)$ and (\ref{tortoise_coord}) is easily integrated to give 
$\rho_*=\arctan y$, as claimed.}
\beq
y\,=\,\arctan (\rho)\,\,,
\eeq
and let $y_c$ be the $y$ coordinate of the pseudohorizon
$y_c\,=\,\arctan (\rho_c)$.
In terms of the variable $y$, the initial condition (\ref{chi_prime_initial_linear}),  for $w_0=1$,   takes the form
\beq
{d\chi\over dy}\Big|_{y=y_c}\,=\,\Omega\,\,\,.
\label{initial_chi}
\eeq
We will impose the following  infalling condition at the pseudohorizon
\beq
{dc\over dy}\Big|_{y=y_c}\,=\,i\,\Omega\,c(y_c)\,\,.
\label{horizon_infalling}
\eeq
Taking into account that, in general
\beq
{dc\over dy}\,=\,\Big[\,i{d\chi\over dy}\,+\,{1\over b}\,{db\over dy}\Big]\,c\,\,,
\eeq
this is equivalent to requiring that
\beq
{db\over dy}\Big|_{y=y_c}\,=\,0\,\,
\eeq
or, equivalently, that the infalling frequency is real. 
To simplify the calculation, let us now introduce a new complex field  
\beq
\psi(y)\,=\,{\rho\over \sqrt{1+\rho^2}}\,c(\rho)\,\,.
\eeq
As $c=\psi(y)/\sin y$, the infalling boundary condition (\ref{horizon_infalling}) takes the following  form in terms of $\psi$
\beq
{d\psi\over dy}\Big|_{y=y_c}\,=\,\Big(i\,\Omega\,+\cot y_c\Big)\,\psi(y_c)\,\,.
\eeq
Taking into account that $\cot y_c=1/\rho_c$, we finally get
\beq
{d\psi\over dy}\Big|_{y=y_c}\,=\,\Big(i\,\Omega\,+{1\over \rho_c}\,\Big)\,\psi(y_c)\,\,.
\eeq
The  general solution (\ref{general_c_linear}) can be written as
\beq
\psi(y)\,=\,d_1\,e^{i\barO\,y}\,+\,d_2\,e^{-i\barO\,y}\,\,,
\eeq
where  $d_1$ and $d_2$ are complex constants and $\barO$ is defined as
\beq
\barO\,\equiv \sqrt{1+\Omega^2}\,\,.
\eeq
We now impose the pseudohorizon boundary condition (\ref{horizon_infalling}). It is easy to demonstrate that (\ref{horizon_infalling}) is satisfied if the ratio of the constants $d_1$ and $d_2$ is given by
\beq
{d_1\over d_2}\,=\,{(\barO+\Omega)\rho_c\,-\,i\over (\barO-\Omega)\rho_c\,+\,i}\,
e^{-2i\,\barO\,y_c}\,\,.
\eeq
To rewrite this condition in a more convenient form, define the phases $\Lambda_c$ and $\overline\Lambda_c$ 
\beq
\Lambda_c\,\equiv\,\arctan\Big[{\barO-\Omega\over \rho_c}\Big]\,\,,
\qquad\qquad
\overline\Lambda_c\,\equiv\,\arctan\Big[{\barO+\Omega\over \rho_c}\Big]\,\,,
\eeq
in terms of which we can write
\bear
&&(\barO+\Omega) \rho_c+i\,=\,
\sqrt{(\barO+\Omega)^2 \rho_c^2+1}\,\,
e^{i\Lambda_c}\,\,,\rc\rc
&&(\barO-\Omega) \rho_c+i\,=\,
\sqrt{(\barO-\Omega)^2 \rho_c^2+1}\,\,
e^{i\overline\Lambda_c}\,\,,
\eear
and $d_1/d_2$ takes the form
\beq
{d_1\over d_2}\,=\,
{\sqrt{(\barO+\Omega)^2 \rho_c^2+1}\over \sqrt{(\barO-\Omega)^2 \rho_c^2+1}}\,\,\,\,
e^{-2i\,\barO\,y_c-i\Lambda_c-i\overline\Lambda_c}\,\,.
\label{ratio_d1_d2_with_Lambdas}
\eeq
It is now straightforward to relate the value of $\psi$ at $y=y_c$  with the constant $d_1$ and to obtain the value 
of $c$ at the pseudohorizon. We get
\beq
c(y_c)\,=\,{2\barO\,\sqrt{1+\rho_c^2}\over \sqrt{(\barO+\Omega)^2 \rho_c^2+1}}\,\,\,
d_1\,\,e^{i\barO\,y_c\,+\,i\Lambda_c}
\eeq
We choose  the phase of $c(y_c)$  to vanish at $\rho=\rho_c$.  This requirement determines the phase of $d_1$, which must be of the form
\beq
d_1\,=\,|d_1|\,e^{-i\barO\,y_c\,-\,i\Lambda_c}\,\,.
\eeq
Moreover, since
\beq
|c(y_c)|\,=\,{2\barO\,\sqrt{1+\rho_c^2}\over \sqrt{(\barO+\Omega)^2 \rho_c^2+1}}\,\,
|d_1|\,\,,
\eeq
we can fulfill (\ref{b0_w0_rho_c})  with $w_0=1$ by choosing $|d_1|$ to be
\beq
|d_1|\,=\,{\sqrt{1+\rho_c^2}\,\over 2\,\barO\,\Omega}\,\sqrt{(\barO+\Omega)^2 \rho_c^2+1}\,\,.
\eeq
The value of $|d_2|$ can be obtained from (\ref{ratio_d1_d2_with_Lambdas})
\beq
|d_2|\,=\,{\sqrt{1+\rho_c^2}\,\over 2\,\barO\,\Omega}\,\sqrt{(\barO-\Omega)^2 \rho_c^2+1}\,\,.
\eeq
From these values of $|d_1|$ and $|d_2|$ one can  readily check that the Joule heating
$q$ is given by 
\beq
q\,=\,\Omega\,\bar\Omega\big(\,
|d_1|^2\,-\,|d_2|^2\,\big)=\,(1+\rho_c^2)\,\rho_c^2\,\,,
\eeq
in agreement  with (\ref{q_w0_rho_c}), as it should. 
Taking into account these results, we can write $d_1$ and $d_2$ as
\bear
&&d_1\,=\,{\sqrt{1+\rho_c^2}\,\over 2\,\barO\,\Omega}\,\sqrt{(\barO+\Omega)^2 \rho_c^2+1}\,\,\,
e^{-i\barO\,y_c\,-\,i\Lambda_c}\,=\,
{\sqrt{1+\rho_c^2}\,\over 2\,\barO\,\Omega}\,
\Big[(\barO+\Omega) \rho_c-i\Big]\,e^{-i\barO\,y_c}
\,\,,\rc\rc
&&d_2\,=\,{\sqrt{1+\rho_c^2}\,\over 2\,\barO\,\Omega}\,\sqrt{(\barO-\Omega)^2 \rho_c^2+1}\,\,\,
e^{i\barO\,y_c\,+\,i\overline\Lambda_c}\,=\,
{\sqrt{1+\rho_c^2}\,\over 2\,\barO\,\Omega}\,
\Big[(\barO-\Omega) \rho_c+i\Big]\,e^{i\barO\,y_c}\,\,,
\qquad\qquad
\eear
and the complexified gauge potential $c(y)$ is given by
\bear
&&c(y)\,=\,{\sqrt{1+\rho_c^2}\,\over 2\,\barO\,\Omega\,\sin y}\Bigg[
\sqrt{(\barO+\Omega)^2 \rho_c^2+1}\,\,e^{i\barO\,(y-y_c)\,-\,i\Lambda_c}\,+\,\rc\rc
&&\qquad\qquad\qquad\qquad\qquad\qquad
+\sqrt{(\barO-\Omega)^2 \rho_c^2+1}\,\,e^{-i\barO\,(y-y_c)\,+\,i\overline\Lambda_c}\Bigg]\,\,.
\label{c_y_sol}
\eear
This expression can also be rewritten as
\bear
c(y)\,=\,{\sqrt{1+\rho_c^2}\,\over 2\,\barO\,\Omega\,\sin y}\Bigg[
\Big[(\barO+\Omega) \rho_c-i\Big]e^{i\barO\,(y-y_c)}\,+\,
\Big[(\barO-\Omega) \rho_c+i\Big]e^{-i\barO\,(y-y_c)}\Bigg]\,\,.
\label{c_y_rwritten}
\eear
From (\ref{c_y_rwritten}) we can   obtain the phase $\chi$
\beq
\tan\chi(y)\,=\,{\rho_c\,\Omega\,\sin\big[\barO(y-y_c)\big]\over
\rho_c\,\barO \cos\big[\barO(y-y_c)\big]\,+\,\sin\big[\barO(y-y_c)\big]}\,\,.
\eeq
Finally, we can expand this result near $\rho=\infty$  to obtain the electric field $E$
\bear
&&E\,=\,-i\,{\sqrt{1+\rho_c^2}\,\over 2\,\barO}\Bigg[
\sqrt{(\barO+\Omega)^2 \rho_c^2+1}\,\,e^{-i\barO\,(y_c-{\pi\over 2})\,-\,i\Lambda_c}\,+\,\rc\rc
&&\qquad\qquad\qquad\qquad\qquad\qquad
+\sqrt{(\barO-\Omega)^2 \rho_c^2+1}\,\,e^{i\barO\,(y_c-{\pi\over 2})\,+\,i\bar\Lambda_c}\Bigg]\,\,,
\label{E_linear_BH}
\eear
and the current $j$
\bear
&&j\,=\,-i\,{\sqrt{1+\rho_c^2}\,\over 2\,\Omega}\Bigg[
\sqrt{(\barO+\Omega)^2 \rho_c^2+1}\,\,e^{-i\barO\,(y_c-{\pi\over 2})\,-\,i\Lambda_c}\,-\,\rc\rc
&&\qquad\qquad\qquad\qquad\qquad\qquad
-\sqrt{(\barO-\Omega)^2 \rho_c^2+1}\,\,e^{i\barO\,(y_c-{\pi\over 2})\,+\,i\bar\Lambda_c}\Bigg]\,\,.
\label{j_linear_BH}
\eear
As a check one can verify that these results reduce to the ones obtained for Minkowski embeddings when we take 
$\rho_c=y_c=0$.

\begin{figure}[ht]
\center
 \includegraphics[width=0.5\textwidth]{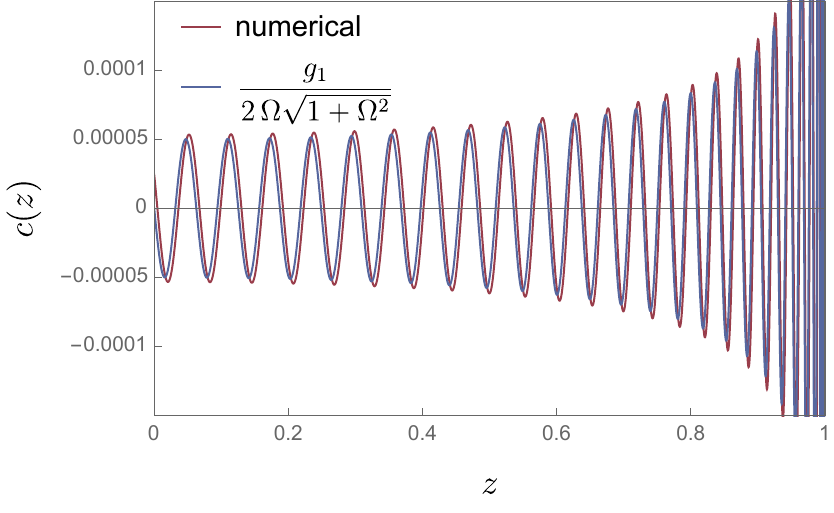}
  \caption{Comparison of the critical solution for $c$ obtained by numerically solving the non-linear full equation of motion and \eqref{c_linear_critical} for $\Omega=100$, $m=1$ and $\epsilon={1\over2}$, using $z^2={1\over 1+\rho^2}$. }
\label{c_vs_z}
\end{figure}

\subsection{Critical embeddings}\label{sec:critembed}

According to our general analysis of section \ref{Critical_bc}, the critical solutions satisfy $\Omega\,b(\rho=0)\,=\,w_0^2$. 
Therefore, if we define the parameter $\epsilon$ as
\beq
\epsilon\,\equiv\,\Omega\,b(\rho=0)\,\,,
\eeq
these embeddings must satisfy $\epsilon=w_0^2$. In our linearized analysis we have $w_0=1$. Thus, we are tempted to describe the critical embeddings by means of the solution $g_1(\rho)$ of (\ref{delta_c_sol_Min}) with $\epsilon=1$. Indeed, contrary to the other solution $g_2(\rho)$, the function $g_1(\rho)$ is regular at $\rho=0$ and one can easily find $\epsilon$ as a function of  $\Omega$ and the constant $c_1$
\beq
\epsilon\,=\,\Omega \,c_1\,\sqrt{1+\Omega^2}\,\,.
\eeq
Plugging this relation in (\ref{delta_c_sol_Min}), we get $c(\rho)$ as
\beq
c(\rho)\,=\,{\epsilon\over \Omega\sqrt{1+\Omega^2}}\,g_1(\rho)\,\,.
\label{c_linear_critical}
\eeq
It turns out, however, that this linearized solution cannot describe accurately the critical embedding near $\rho=0$. Indeed, we found in section \ref{Critical_bc} that the $b$ field behaves linearly  in $\rho$ around its value at $\rho=0$, whereas the function $g_1(\rho)$ behaves quadratically (see (\ref{g12_IR})). As argued in \cite{Kinoshita:2017uch} for the analogous D3-D7 case, the critical solutions behave highly non-linearly around $\rho=0$. Nevertheless the linear solution can describe rather accurately the true non-linear solution for $\rho\gg1$ if we change the value of $\epsilon$ from its naive value $\epsilon=1$. To illustrate this we compare the linear analytic solution for $\epsilon=1/2$ with the numerical results. In figure \ref{c_vs_z} we plot the values of $c(\rho)$ calculated by these two procedures.

\begin{figure}[!ht]
\center
 \includegraphics[width=0.47\textwidth]{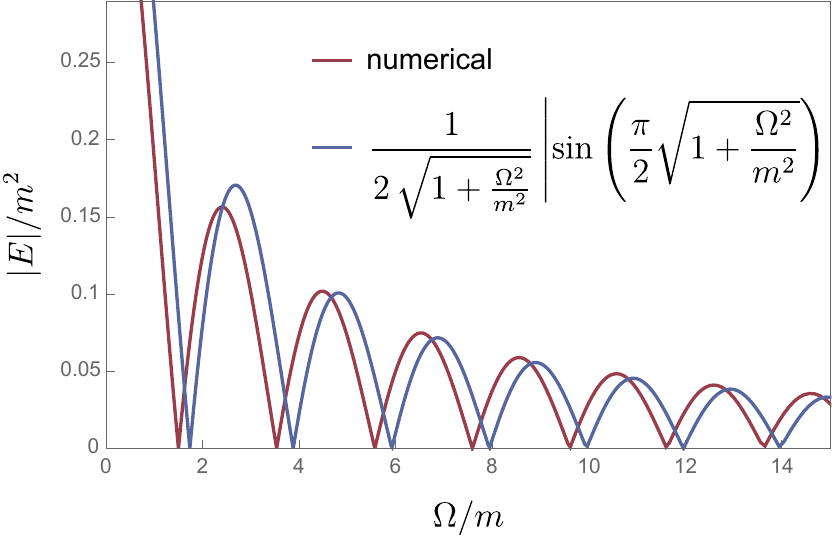}\qquad
 \includegraphics[width=0.47\textwidth]{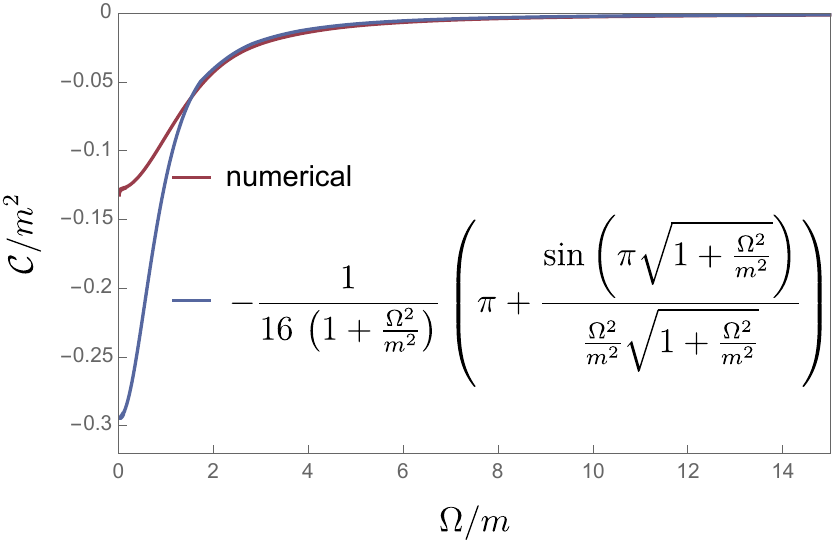}
  \caption{$E$ and $\mathcal{C}$ for critical solutions obtained numerically and using \eqref{ECMinkSol} with $\epsilon=1/2$. }
\label{EC_vs_Omega_Crit}
\end{figure}

For a general value of $\epsilon$, the electric field $E$, the current $j$ and the condensate $\mathcal{C}$ in the linear approximation are
\begin{flalign}\label{ECMinkSol}
&E=-i {\epsilon\over\sqrt{1+\Omega^2}} \sin \Bigg({\pi\over2}\sqrt{1+\Omega^2}\Bigg)\,\,,\\\nonumber
&j=-{\epsilon\over \Omega}\,\cos \Bigg({\pi\over2}\sqrt{1+\Omega^2}\Bigg)\\\nonumber
 &\mathcal{C} =-{1\over 4}\,{\epsilon^2\over1+\Omega^2}\left(\pi+{\sin\left(\pi\sqrt{1+\Omega^2}\right)\over\Omega^2\sqrt{1+\Omega^2}}\right)\,\,.
\end{flalign}
In figure \ref{EC_vs_Omega_Crit} we compare the approximate linear expressions  (\ref{ECMinkSol}) for $E$ and $\mathcal{C}$  (again for $\epsilon=1/2$) to the numerical result of the full non-linear calculation. A reasonable agreement for large $\Omega$ is found. The discrepancies can be attributed  to the intrinsic non-linearity of the critical solutions. In particular, eq. (\ref{ECMinkSol}) implies that the critical electric field decreases as $\left(\Omega/m\right)^{-1}$ for large $\Omega/m$.  We can also use (\ref{ECMinkSol}) to estimate the susceptibility parameter $\gamma_{xy}$ (defined in (\ref{gamma_def})) for critical embeddings:
\beq
\gamma_{xy}\,=\,{{\rm Im}\,\big(E\,j^{*}\big)\over |E|^2}\,=\,
{\sqrt{1+\Omega^2}\over \Omega}\,\cot \Bigg({\pi\over2}\sqrt{1+\Omega^2}\Bigg)\,\,.
\eeq
This expression explains qualitatively well the change of sign of  $\gamma_{xy}$ pointed out in section \ref{phase_diagram} and  reproduces quite accurately  the numerical results of $\gamma_{xy}$  for large values of the driving frequency $\Omega$.

\section{Conductivities in the massless case}
\label{app_C}

In the massless case the fluctuations $\delta w$ of the embedding function decouple from those of the gauge field
$\delta \vec c$.  Therefore, since we are interested in computing conductivities, we can concentrate in studying the equations for $\delta c_x$ and $\delta c_y$. In order to write these  equations in a more convenient form, let us define the following differential operators
\bear
&&{\cal O}_1\equiv \partial^2_t\,+\,{\rho^4(\rho_c^4-\rho^4)\over \rho_c^4+\rho^4}\partial^2_\rho-
{2\rho_c^4\rho^2\over \rho_c^4+\rho^4}\partial_t\partial_{\rho}+
{4\rho_c^4\rho\over \rho_c^4+\rho^4}\partial_t-2\rho^3\partial_\rho\,\,,
\rc\rc
&&{\cal O}_2\equiv-2\partial_t\,+\,{2\rho_c^4\rho\over \rho_c^4+\rho^4}\,\Big(
\rho\partial_\rho-2\Big)\,\,.
\eear
Then, $\delta c_x$ and $\delta c_y$ satisfy the following system of coupled second-order differential equations
\beq
\big({\cal O}_1-\Omega^2\big)\,\delta c_x\,+\,\Omega\,{\cal O}_2\,\delta c_y\,=\,0\,\,,
\qquad\qquad
\big({\cal O}_1-\Omega^2\big)\,\delta c_y\,-\,\Omega\,{\cal O}_2\,\delta c_x\,=\,0\,\,.
\eeq
To decouple these equations, let us consider the following  complex combinations of $\delta c_x$ and $\delta c_y$
\beq
\eta(t,\rho)\,\equiv\,\delta c_x(t,\rho)\,+\,i\delta c_y(t,\rho)\,\,,
\qquad\qquad
\tilde \eta(t,\rho)\,\equiv\,\delta c_x(t,\rho)\,-\,i\delta c_y(t,\rho)\,\,.
\eeq
Notice that $\tilde \eta$ is not the complex conjugate of $\eta$ since $\delta c_x$ and $\delta c_y$ are not necessarily real. It is straightforward to verify that the equations for $\eta$ and $\tilde\eta$ are indeed decoupled and given by
\beq
\big({\cal O}_1-\Omega^2\big)\,\eta\,-\,i\Omega\,{\cal O}_2\,\eta\,=\,0\,\,,
\qquad\qquad
\big({\cal O}_1-\Omega^2\big)\,\tilde\eta\,+\,i\Omega\,{\cal O}_2\,\tilde\eta\,=\,0\,\,.
\eeq
Let us now separate variables as
\beq
\eta(t,\rho)\,=\,\beta(\rho)\,e^{-i\omega\,t}\,\,,
\qquad\qquad
\tilde\eta(t,\rho)\,=\,\tilde\beta(\rho)\,e^{-i\omega\,t}\,\,,
\eeq
for some frequency $\omega$.  Then, remarkably, one can find the following general solutions 
\bear
&&\eta(t,\rho)\,=\,e^{-i\omega\,t}\, e^{i\,{\omega-\Omega\over \rho}}\Big[A+B\,
e^{i\,{\omega-\Omega\over \rho_c}\big(\arctan({\rho\over \rho_c})+{1\over 2}\,
\log{\rho-\rho_c\over \rho+\rho_c}\big)}\Big]\,\,,\rc\rc
&&\tilde\eta(t,\rho)\,=\,e^{-i\omega\,t}\, e^{i\,{\omega+\Omega\over \rho}}\Big[\tilde A+\tilde B\,
e^{i\,{\omega+\Omega\over \rho_c}\big(\arctan({\rho\over \rho_c})+{1\over 2}\,
\log{\rho-\rho_c\over \rho+\rho_c}\big)}\Big]\,\,,
\label{etas_general}
\eear
where $A$, $B$, $\tilde A$ and $\tilde B$ are complex constants which are determined by imposing boundary conditions both at the IR and UV. First of all, let us look at the regularity conditions at the pseudo-horizon $\rho=\rho_*$. These conditions are better studied by using the tortoise coordinates $(\tau, \rho_*)$, related to
$(t,\rho)$ in this massless case by the following differential relations
\beq
d\,\rho_*\,=\,{\rho^2\over \rho^4-\rho_c^4}\,d\rho\,\,,
\qquad\qquad
d\tau\,=\,dt\,-\,{\rho_c^4\over \rho^2(\rho^4-\rho_c^4)}\,d\rho\,\,,
\eeq
which can be integrated as
\bear
&&\rho_*\,=\,{1\over 2\rho_c}\Big[\arctan({\rho\over \rho_c})+{1\over 2}\,
\log{\rho-\rho_c\over \rho+\rho_c}\,-\,{\pi\over 2}\Big]\,\,,\rc\rc
&&\tau\,=\,t\,-\,{1\over \rho}\,-\,{1\over 2\rho_c}\Big[\arctan({\rho\over \rho_c})+{1\over 2}\,
\log{\rho-\rho_c\over \rho+\rho_c}\,-\,{\pi\over 2}\Big]\,\,.
\eear
Notice that the second of these equations can be simply rewritten as
\beq
\tau=t\,-\,{1\over \rho}\,-\,\rho_*\,\,.
\eeq
The new radial coordinate $\rho_*$ varies from $\rho_*=-\infty$ at the pseudohorizon to $\rho_*=0$ at the UV boundary. Actually, one can prove that in these regions it can be related to $\rho$ as
\beq
\rho_*\,=\,-{1\over \rho}+{\cal O}(\rho^{-5})\,\,,
\qquad (\rho\to\infty)\,\,,
\qquad\qquad
\rho_*\,\sim {1\over 4\rho_c}\,\log(\rho-\rho_c) \,\,,
\qquad (\rho\to\rho_c)\,\,.
\eeq
Inspecting the expression of $\eta$ in (\ref{etas_general}) one easily demonstrates that, in terms of the tortoise variable, it can be simply written as
\beq
\eta(\tau, \rho_*)\,=\,e^{-i{\Omega\over \rho}}\,
\Big[A\,e^{-i\omega (\tau+\rho_*)}\,+\,B_*\,e^{-2i\Omega\,\rho_*}\,
e^{-i\omega (\tau-\rho_*)}\Big]
\eeq
where $B_*$ is a new constant, related to $B$ as  $B_*\,=\,\exp[{i\pi(\omega-\Omega)\over 2\rho_c}]\,B$. 
It is now clear that $\eta(\tau, \rho_*)$ is the superposition of ingoing and outgoing waves at the pseudohorizon. 
 The infalling regularity condition requires that $B_*$ (and thus $B$) vanishes. Then, writing $\eta$ in our original $(t, \rho)$  coordinates, we have
\beq
\eta(t,\rho)\,=\,A\, e^{i\,{\omega-\Omega\over \rho}}\,e^{-i\omega\,t}\,\,.
\eeq
We can proceed similarly with $\tilde\eta$ and conclude that we should require that $\tilde B=0$. Therefore
\beq
\tilde\eta(t,\rho)\,=\,\tilde A\, e^{i\,{\omega+\Omega\over \rho}}\,e^{-i\omega\,t}\,\,.
\eeq
Therefore, we obtain that the fluctuations $\delta c_x$ and $\delta c_y$ regular at the pseudohorizon are
\beq
\delta c_x(t,\rho)\,=\,{1\over 2}\Big[A\,e^{i\,{\omega-\Omega\over \rho}}\,+
\,\tilde A\,e^{i\,{\omega+\Omega\over \rho}}\,\Big]\,e^{-i\omega\,t}\,\,,
\qquad\qquad
\delta c_y(t,\rho)\,=\,{1\over 2i}\Big[A\,e^{i\,{\omega-\Omega\over \rho}}\,-
\,\tilde A\,e^{i\,{\omega+\Omega\over \rho}}\,\Big]\,e^{-i\omega\,t}\,\,,
\label{delta_c_massless_fluct}
\eeq

Let us now impose the boundary conditions at the UV. These conditions are those written in (\ref{delta_c_bd})
and can be fulfilled if we add two solutions of the form (\ref{delta_c_massless_fluct})  with frequencies $\omega_+=\omega+\Omega$ and $\omega_-=\omega-\Omega$ with amplitudes $\vec \beta^{(0)}_{\pm}$ 
 at the UV given by  (\ref{beta_0_M_epsilon}), which can be explictely written as follows
\beq
\vec  \beta^{(0)}_{+}\,=\bemat{c} {\epsilon_x+i\epsilon_y\over 2\,i\omega}
 \\ \\- {\epsilon_x+i\epsilon_y\over 2\omega}
 \enmat \,\, ~~~ ,
 \qquad
 \vec  \beta^{(0)}_{-}\,=\bemat{c} {\epsilon_x-i\epsilon_y\over 2\,i\omega}
 \\ \\{\epsilon_x-i\epsilon_y\over 2\omega} \enmat \,\,.
 \label{beta_0_s}
\eeq

  Let $A_{\pm}$ and $\tilde A_{\pm}$ denote the constants in (\ref{delta_c_massless_fluct}) for the frequency $\omega_{\pm}$. From the leading UV term of our solution, we get that
\beq
\vec\beta^{(0)}_{\pm}\,=\bemat{c} {A_{\pm}+\tilde A_{\pm} \over 2}
 \\ \\{A_{\pm}-\tilde A_{\pm} \over 2i}
 \enmat \,\,.
 \label{beta_0_+_As}
\eeq
Let us now compare the expressions  of $\vec\beta^{(0)}_{+}$  in (\ref{beta_0_+_As}) and (\ref{beta_0_s}). They only match if the constants $A_+$ and $\tilde A_+$ are given by
\beq
A_+\,=\,-{i\over \omega} (\epsilon_x+i\epsilon_y)\,\,,
\qquad\qquad
\tilde A_+\,=\,0\,\,.
\label{A_+}
\eeq
By expanding around $\rho=\infty$ in (\ref{delta_c_massless_fluct}) we can now obtain the amplitude $\vec \beta^{(1)}_+$ of the subleading term
\beq
\beta^{(1)}_{+,x}\,=\,{i\over 2}\,(\omega_+-\Omega)\,A_+\,=\,{i\over 2}\,\omega\,A_+\,\,,
\qquad\qquad
\beta^{(1)}_{+,y}\,=\,{1\over 2}\,(\omega_+-\Omega)\,A_+\,=\,{1\over 2}\,\omega\,A_+\,\,.
\eeq
Using (\ref{A_+}) we can obtain $\vec \beta^{(1)}_{+,x}$ in terms of $\vec \epsilon$
\beq
\vec \beta^{(1)}_{+}\,=\,{\bf M}_+\,
\vec\epsilon\,\,,
\label{beta1_+_epsilon}
\eeq
where ${\bf M}_+$ is the matrix defined in (\ref{M_pm_def}). 
Thus, the matrix ${\bf X}_+$ introduced in (\ref{X_pm})  in this massless case is given by
\beq
{\bf X}_+\,=\,i\,\omega\,{\bf I}\,\,.
\eeq
Proceeding similarly for the frequency $\omega_-$,  we obtain that the constants $A_-$ and $\tilde A_-$ must be
\beq
A_-\,=\,0\,\,,
\qquad\qquad
\tilde A_-\,=\,-{i\over \omega} (\epsilon_x-i\epsilon_y)\,\,,
\label{A_-}
\eeq
and the corresponding amplitude for the subleading term takes the form
\beq
\beta^{(1)}_{-,x}\,=\,{i\over 2}\,(\omega_- + \Omega)\,\tilde A_-\,=\,{i\over 2}\,\omega\,\tilde A_-\,\,,
\qquad\qquad
\beta^{(1)}_{-,y}\,=\,-{1\over 2}\,(\omega_-+\Omega)\,\tilde A_-\,=\,-{1\over 2}\,\omega\,\tilde A_-\,\,,
\eeq
which, after using (\ref{A_-}), can be related to $\vec \epsilon$ as
\beq
\vec \beta^{(1)}_-\,=\,{\bf M}_-\,
\vec\epsilon\,\,.
\label{beta1_-_epsilon}
\eeq
Comparing (\ref{X_pm}) and (\ref{beta1_-_epsilon}) we immediately conclude that
\beq
{\bf X}_-\,=\,i\,\omega\,{\bf I}\,\,.
\eeq
Using these values of ${\bf X}_+$ and ${\bf X}_-$ we can now obtain the conductivities from (\ref{conduct}). As
\bear
&&{\bf M}_+ {\bf X}_+ {\bf M}_+   + {\bf M}_-  {\bf X}_- {\bf M}_- \,=\,i\omega\,\big[{\bf M}_+^2+{\bf M}_-^2\big]\,=\,
i\omega\,\big[{\bf M}_++{\bf M}_-\big]\,=\,i\omega\,{\bf I}\,\,,\rc\rc
&&{\bf M}_{\mp} {\bf X}_{\pm} {\bf M}_{\pm}\,=\,i\omega\,{\bf M}_{\mp}  {\bf M}_{\pm}\,=\,0\,\,.
\eear
This can be summarized as follows:
\beq
\boldsig(\omega)\,=\,{\bf I}\,\, ~~~~ ,
\qquad
\boldsig^+(\omega)\,=\,\boldsig^-(\omega)\,=\,0\,\,.
\eeq

\begin{figure}[!ht]
\center
\includegraphics[width=0.4\textwidth]{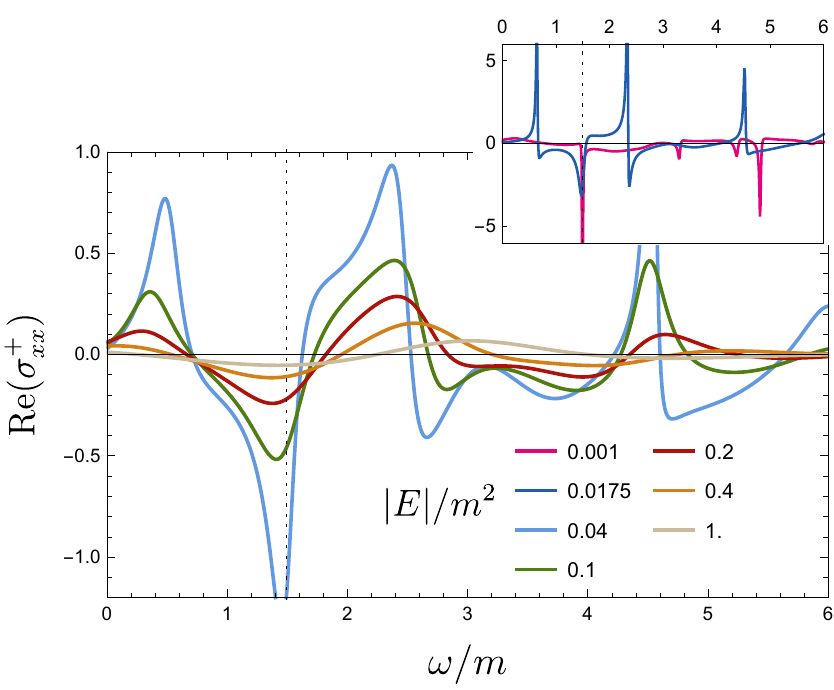} ~ \includegraphics[width=0.4\textwidth]{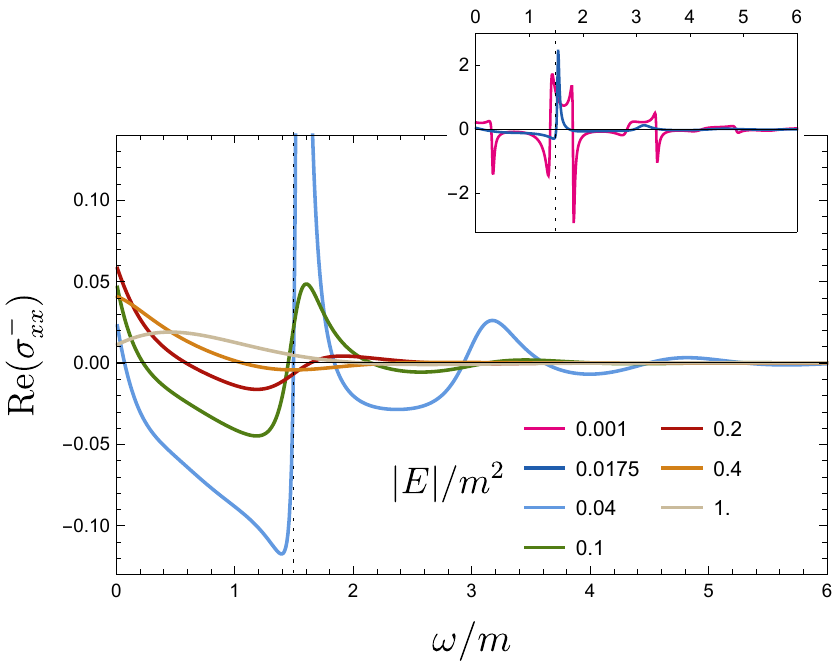} 
\includegraphics[width=0.4\textwidth]{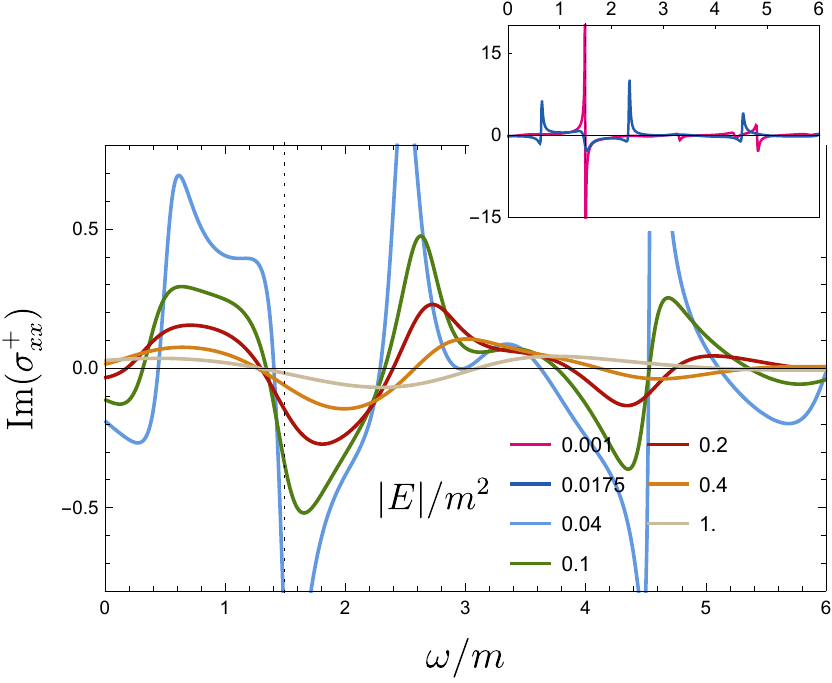} ~ \includegraphics[width=0.4\textwidth]{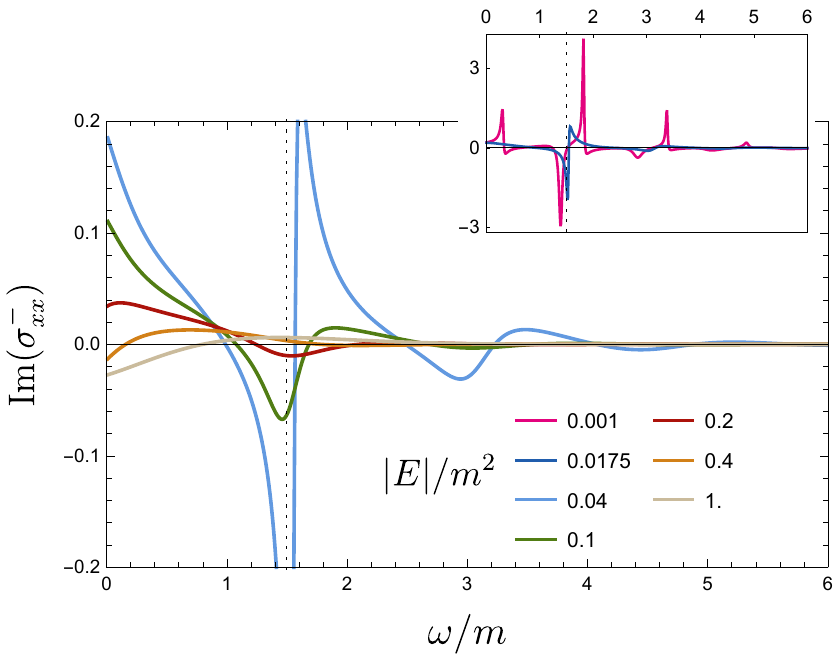} 
\caption{AC Conductivities for fixed value of $\Omega/m = 1.496$ and different values of $|E|/m^2$. They correspond to the vertically aligned dots in Fig.\,\ref{fig:PhSpMap} with the same color code.    In the insets, the curves for low values of $|E|/m^2\to 0$ deep into the critical wedge, where they develop peak resonances.  The value $\omega/m \to \Omega/m = 1.496$ is signalled with a dashed vertical line. }
\label{fig:AC_heterodyning}
\end{figure}

\section{More on optical conductivities}
\label{app_E}

In Fig.\,\ref{fig:AC_heterodyning} we show some plots for the heterodyning optical conductivities $\boldsig^\pm(\omega)$. As proven in \cite{Hashimoto:2016ize},
a set of non-trivial relations among the components of $\boldsig$ and $\boldsig^\pm$ allows to take as independent components $\sigma_{xx}, \sigma_{xy}$ and  $\sigma^\pm_{xx}$.
The relation  $\boldsig(-\omega)=\boldsig(\omega)^*$ implies that the DC conductivity will be real. For the heterodyning conductivities, in contrast, $\boldsig^\pm(-\omega)=\boldsig^\mp(\omega)^*$ and, hence, the associated  DC limits will be complex.

They exhibit a similar pattern of oscillations whose amplitude increases as we approach the critical point. Eventually they sharpen into peaks as well.  Comparing for example the peaks in $\sigma^{\pm}_{xx}$ with those in $\sigma_{xx}$ in Fig.\,\ref{fig:AC_conductivities} it seems apparent that they share half of them each. 
This is apparent in the heterodyning version of Figs.\ref{fig:criticalcond} shown below in Fig.\ref{fig:criticalcondpm}.
 Namely, while $\sigma_{xx}^-$ has only the central stable peaks,  $\sigma_{xx}^+$  has the external peaks whose average value is $\Omega$.

\begin{figure}[!ht]
\center
\includegraphics[width=0.45\textwidth]{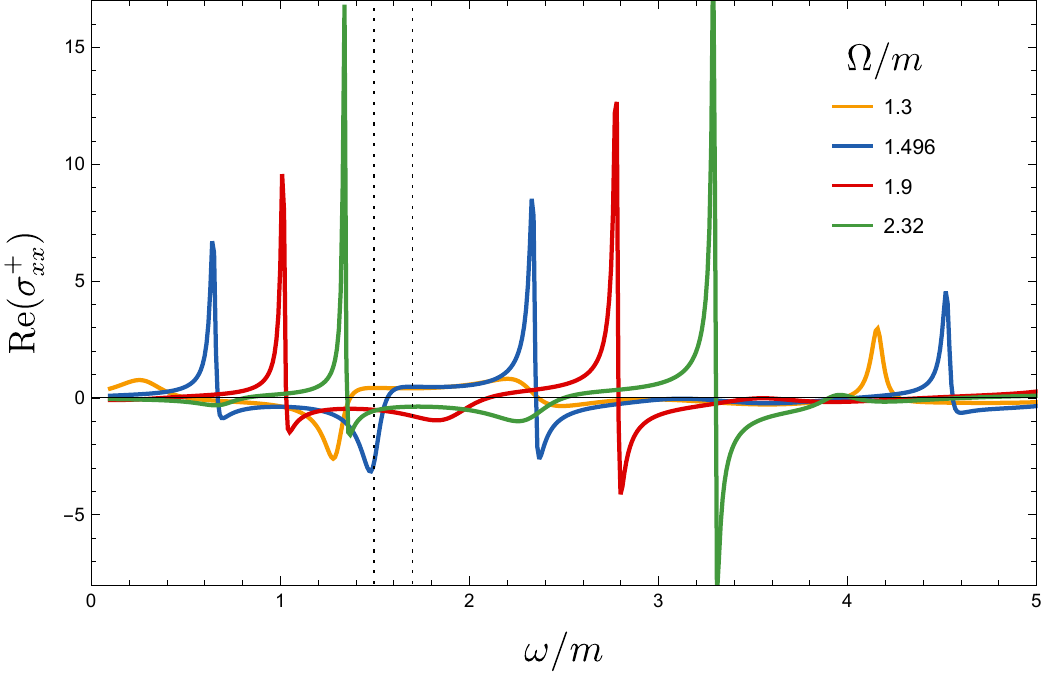} ~~~ \includegraphics[width=0.45\textwidth]{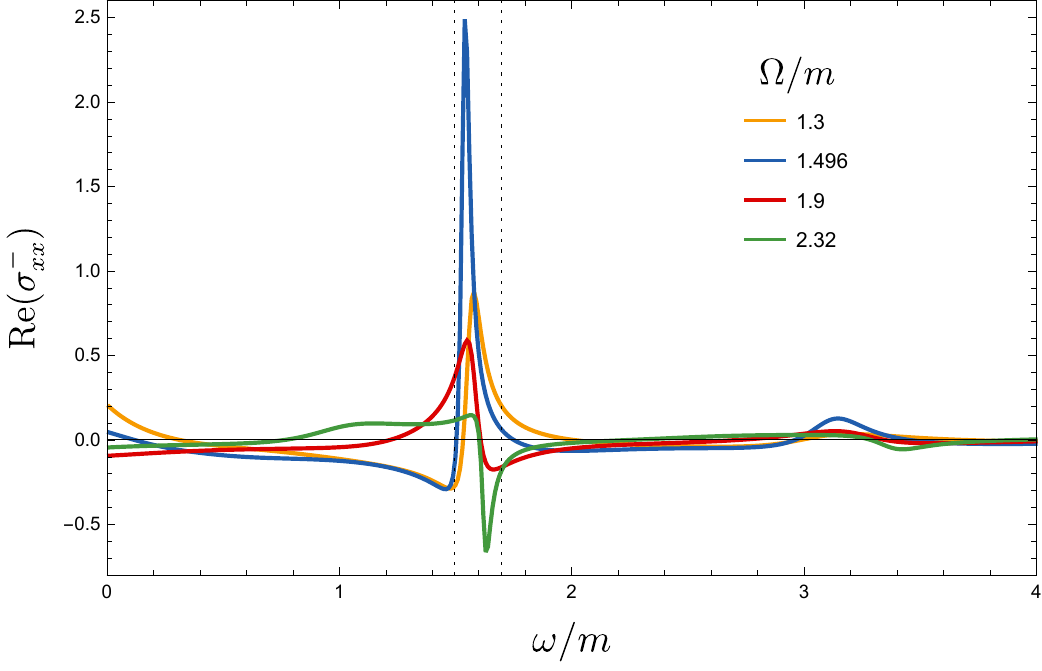} \\
\caption{Peaks in the AC conductivities along the critical line to the right of the first Floquet condensate (see Fig.\ref{fig:PhSpMap}). The vertical dashed lines signal the critical  Floquet and the  meson  mass frequencies respectively.  The central peaks bunch in this region whereas the two lateral peaks move with $\omega$ while keeping their inter-spacing almost constant.}
\label{fig:criticalcondpm}
\end{figure}

\begin{figure}[!ht]
\center
\includegraphics[width=0.45\textwidth]{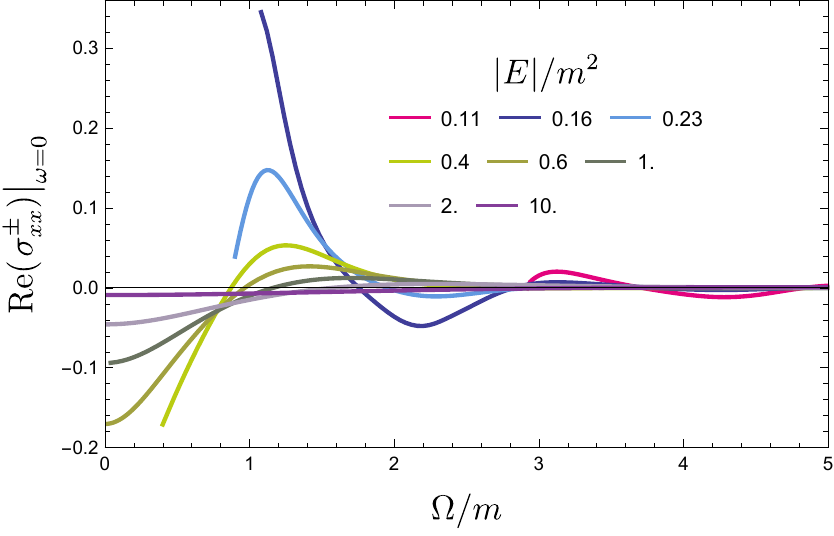} ~ \includegraphics[width=0.45\textwidth]{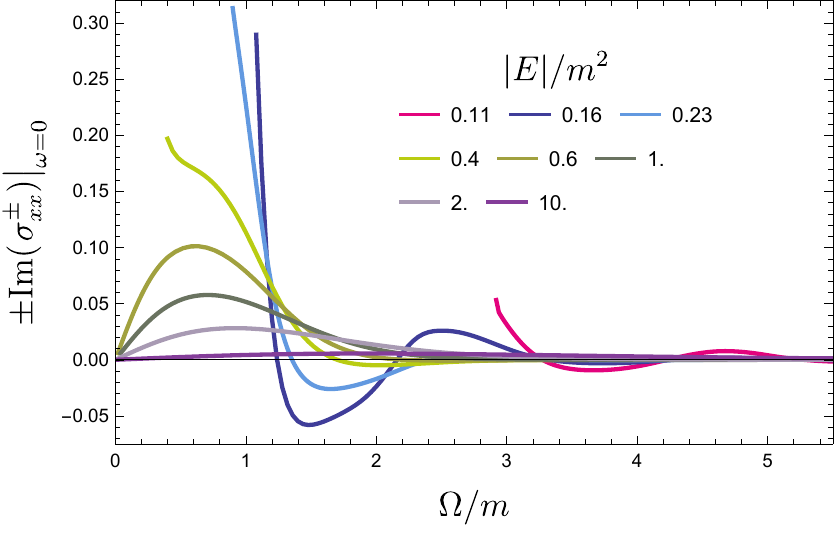} \\ ~\\
\includegraphics[width=0.45\textwidth]{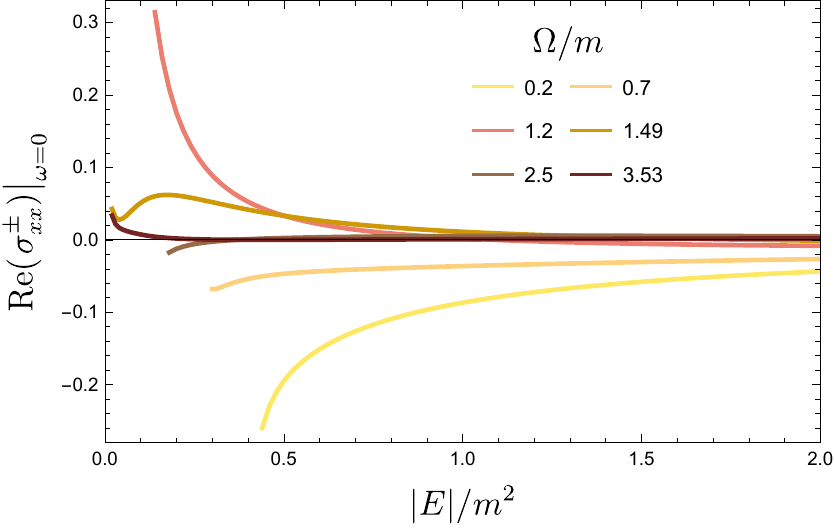} ~ \includegraphics[width=0.45\textwidth]{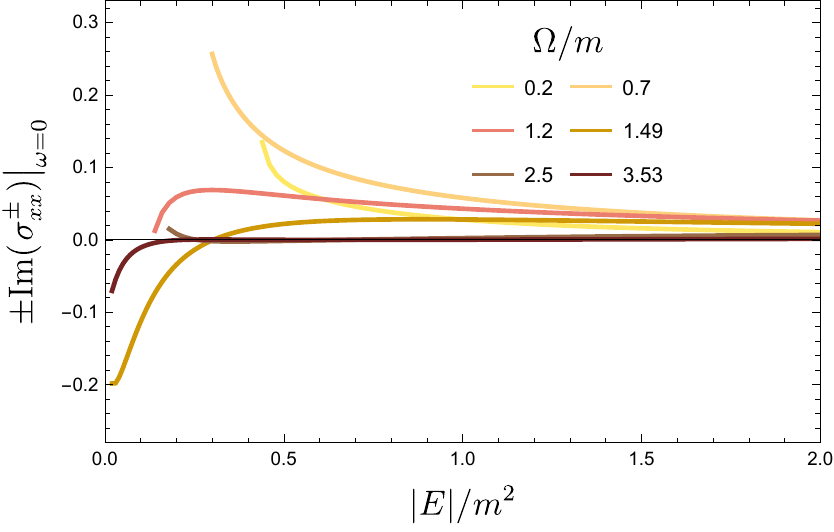} 
\caption{DC conductivities $\sigma^\pm_{xx}$  as function of either $\Omega/m$ or $|E|/m^2$. }
\label{fig:sigp_DC_Eovm2}
\end{figure}

\end{document}